\documentclass[11pt,a4paper]{article}
\pdfoutput=1

\usepackage{jheppub} 
\usepackage[utf8]{inputenc}
\usepackage[T1]{fontenc} 
\usepackage{float}
\usepackage{lmodern}
\usepackage{amsmath,amsfonts,amssymb}
\usepackage{caption}
\usepackage{hyperref}
\usepackage[all]{hypcap}
\usepackage{graphicx}
\usepackage{array}
\usepackage{slashed} 
\usepackage{dcolumn} 
\usepackage{paralist}
\usepackage{mathrsfs} 
\usepackage{subcaption}
\newif\ifonecol\onecolfalse

\newcommand{\mean}[1]{\ensuremath{\langle #1 \rangle}}
\newcommand{\nue}{\ensuremath{\nu_e}}
\newcommand{\numu}{\ensuremath{\nu_\mu}}
\newcommand{\nutau}{\ensuremath{\nu_\tau}}

\newcommand{\ie}{\textit{i.e.}}
\newcommand{\eg}{\textit{e.g.}}

\newcommand{\anti}[1]{\ensuremath{\bar{#1}}}

\newcommand{\dm}{DM}

\newcommand{\zp}{\ensuremath{Z^{\prime}}}

\newcommand{\ud}{\ensuremath{\mathrm{d}}}
\newcommand{\fdnde}{\ensuremath{\frac{\ud N_\chi}{\ud E_{\chi}}}}

\newcommand{\cmssr}{\ensuremath{\text{ cm}^{-2}\text{ s}^{-1}\text{ sr}^{-1}}}
\newcommand{\flxunit}{\ensuremath{\text{ GeV}\cmssr}}

\newcommand{\chisq}{\ensuremath{\chi^{2}}}

\newcommand{\nast}{\ensuremath{N_\text{Ast}}}
\newcommand{\tilnast}{\ensuremath{\tilde{N}_\text{Ast}}}
\newcommand{\ndm}{\ensuremath{N_\text{DM}}}
\newcommand{\gray}{$ \gamma $-ray}

\title{Boosted Dark Matter and  its implications for the features in  IceCube HESE data}

\author[a]{Atri Bhattacharya,}
\author[b,c]{Raj Gandhi,}
\author[b,c]{Aritra Gupta}  
\author[d]{and Satyanarayan Mukhopadhyay}
\affiliation[a]{Space sciences, Technologies and Astrophysics Research (STAR) Institute, Universit\'{e} de
Li\`{e}ge, B\^{a}t. B5a, 4000 Li\`{e}ge, Belgium}
\affiliation[b]{Harish-Chandra Research Institute, Chhatnag Road, Jhunsi, Allahabad-211019, India}
\affiliation[c]{Homi Bhabha National Institute, Training School Complex, Anushaktinagar, Mumbai - 400094, India}
\affiliation[d]{PITT-PACC, Department of Physics and Astronomy, University of Pittsburgh,
PA 15260, USA}

\emailAdd{A.Bhattacharya@ulg.ac.be}
\emailAdd{raj@hri.res.in}
\emailAdd{aritra@hri.res.in}
\emailAdd{satya@pitt.edu}

\preprint{PITT-PACC 1619}

\abstract{We study the implications of the premise that any new, relativistic, highly energetic neutral particle that  interacts with quarks and gluons would create cascade-like events in the IceCube (IC) detector. Such events  would be observationally indistinguishable from neutral current deep-inelastic (DIS) scattering events due to neutrinos. Consequently, one reason for deviations, breaks or excesses in the expected astrophysical power-law neutrino spectrum could be the flux of such a particle. Motivated by features in the recent 1347-day IceCube high energy starting event (HESE) data, we focus on particular boosted dark matter ($\chi$) related realizations of this premise.  Here, $\chi$ is assumed to be  much lighter than, and the result of, the slow decay of  a massive scalar ($\phi $) which constitutes a major fraction of the Universe's dark matter (DM). We show that this hypothesis, coupled with a standard power-law astrophysical neutrino flux is capable of providing very  good fits to the present data, along with a possible explanation of other features in the HESE sample. These features include  a) the paucity of events beyond $\sim 2$ PeV b) a spectral feature resembling a dip or a spectral change in the 400 TeV--1 PeV region and c) an excess in the $50-100$ TeV region.  We consider two different boosted  DM scenarios, and determine  the allowed mass ranges and couplings for four different types of mediators (scalar, pseudoscalar, vector and axial-vector) which could connect the standard and dark sectors.We consider constraints from gamma-ray observations and collider searches. We find that  the gamma-ray observations provide  the most restrictive constraints, disfavouring the $1\sigma$ allowed parameter space from IC fits, while still being consistent with the $3\sigma$ allowed region.  We also test  our proposal and its implications against the (statistically independent) sample of six year through-going muon track data recently released by IceCube.}


\begin{document}

\maketitle

\section{Introduction and Motivation}
In this section, we shall begin with a summary of the $1347$-day IceCube (IC) high-energy starting event (HESE) neutrino data, focussing on events with deposited energies greater than around $30$ TeV, and discuss some of its features, especially those that are of particular interest for this study. We shall then introduce two possible scenarios of boosted dark matter, which, in combination with a power-law astrophysical flux, can provide a good fit to these features. 

 \subsection{IceCube High Energy Starting  Events (HESE) and features of  the 1347-day data}

The observation of 54 HESE (\ie, events with their $ \nu N $ interaction vertices inside the detector) \cite{Aartsen:2015zva,1311.5238},
with deposited energies 
between 30 TeV to a maximum energy of 2.1 PeV  by the  IceCube experiment (IC) has opened an unprecedented window to the universe at  high energies.\footnote{
In addition to the analysis 
presented by the IceCube collaboration in \cite{Aartsen:2015zva,1311.5238}, a recent  analysis of the HESE data may be found in \cite{1605.01556}.} The data 
constitute an approximately $7\sigma$ signal in favour of  a non-atmospheric and extra-terrestrial origin of the events.\footnote{The statistical significance is dependent 
upon the largely theoretically modelled upper limits of the prompt neutrino flux from heavy meson decays. The $7\sigma $ value corresponds to the scenario where the 
prompt flux is assumed to be absent. Nonetheless, even with the highest upper limits from present computations, the statistical significance of a new signal over and 
above the atmospheric background is well above $ 5\sigma $.}
It is generally believed, but not conclusively known,  that the highest energy cosmic rays ($E\geq10^6$ GeV), for which observations now extend to $E \sim 10^{11}$ GeV, and ultra-high energy (UHE) neutrinos with energies greater than $\mathcal{O}(20)$ TeV, share common origins and are produced by the same cosmic accelerators. The specific nature of these accelerators, however, remains unknown, although over the years, anticipating their detection, several classes of highly energetic cosmic astrophysical sources have been studied as possible origins of these particles. For general discussions of this topic, we refer the reader to  \cite{0710.1557, astro-ph/0204527,1202.0466,1301.1703,1305.4123,1306.2309,1306.5021,1312.6587,1410.3680,1407.7536,1405.5487}.

 Subsequently,  based on the recent IC data, many authors have considered a host of source classes and possibilities for explaining both the origin and some emerging spectral features in the IC data. These efforts have been motivated, at least  in part, by evidence that the data, to an extent, diverge from expectations. The considered candidate sources  include gamma-ray bursts~\cite{1211.1974,1212.1260,1307.7596,1306.2274,1310.7061,1405.2091,
1405.5487,1411.7491,
1504.00107,1512.08513,1512.01559}, star-burst galaxies~\cite{1306.3417,1405.7648,1404.1189,1406.1099,1509.00983}, active galactic nuclei~\cite{1303.0300,1305.7404,1406.2633,1406.0645,1407.0907,1408.3664,1410.8124,
1411.2783,1411.3588,1501.07115},
remnants of hyper-novae~\cite{1310.1263} and of supernovae~\cite{1501.02615}, slow-jet supernovae~\cite{1407.2985}, microquasars~\cite{1410.0348}, neutron star mergers~\cite{1306.3006}, blackholes~\cite{1512.08596}, cosmic-ray interactions~\cite{1307.2158,1310.5123,1311.0287,1404.6237,1405.3797,1411.6457,1412.8590}, the galactic halo~\cite{1403.3206}, galaxy clusters~\cite{1410.8697}, dark matter decay~\cite{1303.7320,1308.1105,1312.3501,1410.5979,
1403.1862,1407.3280,1408.1745,
1411.1071,1503.02669,1503.04663,1505.06486,1506.08788,Boucenna:2015tra,1508.02500,
1601.02934,1606.04517,1607.05283}, and exotic particles,  processes or possibilities~\cite{1305.6907,1402.1681,1402.6678,1404.0622,
1404.2279,1404.2288,1404.2932,1404.7025,
1407.2848,1409.4180,1409.5896,1410.0408,
1410.3208,1411.5318,1411.5889,1507.03015,1507.03193,1606.06238,1606.07903,Dev:2016uxj}.

 It is generally accepted, however, that the charged particles in a 
source which link the acceleration of cosmic-rays to the acceleration of astrophysical neutrinos attain their high energies via Fermi shock acceleration~\cite{Fermi:1949ee}, and as a generic consequence, the neutrinos resulting from them are expected to follow a $E^{-2}$ spectrum~\cite{0710.1557, astro-ph/0204527}. Some variation from this general spectral behaviour may occur, however, depending on the details of the source, as discussed, for instance, in~\cite{2005PhRvL..95r1101K}.

IceCube is sensitive to high energy neutrinos via their electroweak charge and neutral current (CC and NC respectively) deep inelastic (DIS) interactions with nucleons in ice, which result in the deposition of detectable energy in the form of Cerenkov radiation.  An event may thus be classified as\footnote{This classification allows us to categorize most events. There are other, potentially important types of events, however, which have not yet been observed; \eg\  the {\it{double bang}}  events signalling the CC production of a highly energetic $\tau$ lepton~\cite{hep-ph/9405296}, and the {\it{pure muon}} and {\it {contained lollipop}}~\cite{Bhattacharya:2011qu} events which would unambiguously signal the detection of the Glashow resonance~\cite{PhysRev.118.316, Berezinsky:1981bt,Gandhi:1995tf}.}
\begin{itemize}
	\item \emph{a track}, produced by \numu\ CC  and a subset of \nutau\ CC interactions (where a produced $\tau$ decays to a  $\mu$) , characterized by a highly energetic charged lepton traversing a significant length of the detector, or 
	\item  \emph{a cascade}, produced by either i) \nue\ CC interactions, ii) a subset of \nutau\ CC interactions or iii)  NC interactions of all three flavours. Cascades  are  characterized by their light deposition originating from charged hadrons and leptons, distributed around the interaction vertex in an approximately spherically shaped signature.
\end{itemize} 
Additionally, because neutrino production in astrophysical sources stems from photohadronic interactions producing light mesons, such as pions and kaons, and to a smaller degree, some heavier charmed mesons, including $ D^{\pm}, D^{0} $, and their subsequent decays, the flux ratio at source is expected to be $ (\nue + \bar{\nu}_e: \numu + \bar{\nu}_\mu : \nutau + \bar{\nu}_\tau) = 1:2:0 $.
However, standard oscillations between the three flavours over cosmological distances renders this ratio close to $1:1:1$ \cite{0505017} by the time they arrive at earth.
In this situation, cascade events are expected to  constitute about 75--80\% of the total observed sample \cite{Beacom:2004jb}. The background to the HESE events is provided by the rapidly falling atmospheric neutrino flux and the muons created in cosmic-ray showers in the atmosphere.

We now describe the significant  features of the HESE data, some of which are fairly firm even at the present level of statistics, and others which, while  interesting and suggestive, are emergent and need further confirmation via more observations before they can be considered as established. (We note that the energies quoted below refer to those deposited by the primary in IC.)
\begin{itemize}
\item{The data, to a high level of significance (about $7\sigma$, as mentioned earlier), indicate that above a few tens of TeV,  the sources of the events are primarily non-atmospheric and extra-terrestrial in nature.}
\item{Due to the lack of multi-PeV events, including those from the Glashow Resonance \cite{Glashow:1960zz, Bhattacharya:2011qu, Barger:2014iua} in the range  6-10 PeV, a single power-law fit to the flux underlying the observed events now disfavours the expected spectral index from Fermi shock acceleration considerations, $ \gamma=-2$, by more than $4\sigma$. Indeed, for an assumed $E^{-2}$ spectrum, and with the corresponding best-fit normalization to the flux, about 3  additional cascade events are expected between 2 PeV and 10 PeV, largely due to the expected  presence of the Glashow resonance.  However, in spite of IceCube's high sensitivity at these energies, none have been observed thus far. The present best fit value of $\gamma$ is consequently significantly steeper, being around $ \gamma=-2.58$ \cite{Aartsen:2015zva, halzen_pheno16}.}
 \item{The data, when subjected to directional analyses \cite{Aartsen:2015zva,1309.2756,1309.4077,1311.5864,1311.7188,
1406.0376,1406.2160,1407.2243,
1410.5979,1501.05158,1507.05711,1509.00517,1509.03522,1510.00048,1511.09408,
1603.06733}, at its present level of statistics, is compatible with an isotropic diffuse flux, although several studies among the ones cited above indicate the presence of a small galactic bias. The accumulation of more data will be able to ascertain  whether the galactic bias is real, in which case it would imply important (and possibly new) underlying physics.}
\item{The three highest energy events \cite{Aartsen:2015zva}, with the estimated (central value) of the deposited energies of $1.04$ PeV, $1.14$ PeV and $2.0$ PeV are all cascade events from the southern hemisphere.
At these energies, \ie\ $ E_\nu \gtrsim 1$ PeV, the earth becomes opaque to neutrinos, thus filtering out neutrinos coming from the northern hemisphere.}
\item{Below $1$ PeV, there appears to be a dip in the spectrum, with no cascade events between roughly $400$ TeV and $1$ PeV.}\footnote{A recent analysis \cite{1611.07905} statistically reinforces the presence of a break in the spectrum in the region 200 - 500 TeV, which could have a bearing on this feature.}
\item{At lower energies, in the approximate range of $50-100$ TeV, there appears to be an excess, with a bump-like feature (compared to a simple power-law spectrum), which is primarily present in events from the southern hemisphere \cite{1410.1749}}. The maximum local significance of this excess is about $2.3 \sigma$, which is obtained when the lowest estimates for the conventional atmospheric neutrino background is adopted, with the prompt component of the background assumed to be negligible~\cite{MESE}.
\item{Finally, and importantly, the data when interpreted as being due to a single astrophysical power-law neutrino flux, appears to require an unusually high normalization for this flux, which is at the level of the Waxman-Bahcall (WB) bound ~\cite{Waxman:1998yy,Bahcall:1999yr} for neutrino fluxes from optically thin sources of high energy cosmic rays and neutrinos. This is  an aspect that is difficult to understand within the confines of the standard interpretive mechanism, which connects ultra-high energy neutrino fluxes to observations of the highest energy cosmic-rays \footnote{The WB bound is valid for sources which produce neutrinos as a result of $pp$ or $p\gamma$ interactions. It assumes that they  are optically thin to proton photo-meson and proton-nucleon interactions, allowing protons to escape. Such sources are characterized by an optical depth $\tau$ which is typically less than one.  As explained in \cite {Bahcall:1999yr}, the bound is conservative by a factor of $\sim 5/\tau$.}}.
\end{itemize}
 
\subsection{Deep Inelastic Scattering of Boosted Dark Matter in IceCube}
\label{intro2}
As proposed  in \cite{1407.3280}, if there is a source of long-lived, highly relativistic and energetic neutral particles in the present Universe which can interact with quarks or gluons, the signal produced by them in IceCube would, in all likelihood, be indistinguishable from the NC DIS cascade of a neutrino primary.  To the extent that the astrophysical neutrino flux is expected to follow a simple power-law behaviour, one could argue that features in the HESE data (as described in the previous subsection) which deviate from this, such as statistically significant excesses, spectral breaks or line-like features, could indicate the presence of  such a particle\footnote{Alternatively, such features could, of course, also indicate that the conventional neutrino astrophysical flux, while originating in standard physics,  is much less understood than we believe, and may have more than one component.}. Although there are strong constraints on the presence of additional relativistic degrees of freedom during the epochs of  recombination and big bang nucleosynthesis, such particles might be injected at later times  by the slow decay of a heavy particle, which, overall, is the approach we adopt here. 

 We consider the case  where this heavy particle constitutes a significant part of the dark matter (DM) density of the Universe. Its late-time decay produces a highly energetic flux of light dark matter (LDM) particles, which can then give rise to a subset of the NC DIS events at IC. We note that this is different from the  scenario where the heavy dark matter (HDM) particle directly decays to standard model particles, leading to a neutrino flux in IC, as discussed in, for instance~\cite{1303.7320,1308.1105,1312.3501,1410.5979,
1403.1862,1408.1745,
1411.1071,1503.04663,1505.06486,1506.08788,1507.01000,1508.02500,
1601.02934,1606.04517,1607.05283}. In the scenario(s) discussed here, in order to have NC DIS scattering with nuclei, the LDM particles need to couple to the SM quarks (or gluons) with appropriate strength. It is then possible that these interactions  could keep them in chemical equilibrium with the SM sector in the early Universe. Thus, the standard thermal freeze-out mechanism will give rise to a relic density of the LDM particles as well in the present Universe, though the exact value of their present-day density would in general depend upon all the annihilation modes open and the corresponding annihilation rates. It is important to note that the couplings relevant for the IC analysis  provide only a lower bound on the total annihilation rate. For our purpose, the precise relic density of LDM is not of particular relevance, and we simply need to ensure that it annihilates sufficiently fast in order not to overclose the Universe, while its relic abundance should not be too high, in order to allow
for a sufficient HDM presence in the universe.
The latter is needed to produce enough of the relativistic LDM flux from its late time decays.
In other words, scenarios where the LDM abundance is small are preferred but not required. Similarly,  for phenomenological analysis of the IC data, the production mechanism of the HDM particle does not play any essential role. Therefore, we abstain from discussing specific cosmological models for HDM production in this article, and instead refer the reader to possibilities discussed in Refs.~\cite{1402.2846, 1201.3696, hep-ph/0203118, hep-ph/0205246}. We further note that general considerations of partial-wave unitarity of scattering amplitudes imply an upper bound on the mass of any DM particle that participates in  standard thermal equilibrium production processes and then  freezes out. Such a particle should be lighter than a few hundred TeV, as discussed in \cite{PhysRevLett.64.615}. As we shall see,  the HDM under consideration here is necessarily non-thermal due to this reason\footnote{We note that a two-component ${\it{thermal}}$ WIMP-like DM scenario, with the lighter particle (of mass $O$(1 GeV)) being boosted 
after production (via annihilation in the galactic halo of its heavier partner of mass $O(100)$ GeV) and subsequently  detected in neutrino experiments has been discussed in \cite{1405.7370}. Boosted thermal DM detection from the sun and the galactic center due to annihilation of a heavier counterpart at similar masses and energies  has been discussed in \cite{1410.2246, 1411.6632, 1611.09866}.}.

In what follows,  we pursue two specific realizations (labelled Scenario I and II below) of such a dark matter sector, which, in combination with a power-law astrophysical component, provide a good description to the features in the IC data described in the previous subsection. For each realization, we perform a likelihood analysis to fit the IC HESE data and its observed features, in terms of a combination of four distinct fluxes. These fluxes are: 
\begin{enumerate}
\item {\bf Flux-1:} An underlying power-law flux of astrophysical neutrinos, $ \Phi_\text{Ast} = N_\text{Ast} E^{-\gamma} $, whose normalization ($N_\text{Ast}$) and index ($\gamma$) are left free.

\item {\bf Flux-2:} A flux of boosted light dark matter (LDM) particles ($\chi$), which results from the late-time decay of a heavy dark matter (HDM) particle ($\phi$). When $\chi$ is much lighter than $\phi$, its scattering in IC resembles the NC DIS scattering of an energetic neutrino, giving rise to cascade-like events.

\item  {\bf Flux-3:} The flux of secondary neutrinos resulting from three-body decay of the HDM, where a mediator particle is radiated off a daughter LDM particle. The mediator then subsequently decays to SM particles, producing neutrinos down the decay chain. Since the NC DIS scattering that results from  Flux-2 requires a mediator particle which couples to both the LDM and the SM quarks, such a secondary neutrino flux is always present.

\item {\bf Flux-4:} The conventional, fixed, and well-understood, atmospheric neutrino and muon background flux, which is adapted from IC analyses \cite{Aartsen:2015zva,1311.5238}.
\end{enumerate}

\subsubsection*{Scenario I : PeV events  originating from DIS scattering of boosted LDM at IC}
In Scenario I, the three highest energy PeV events, which are cascades characterized by energy depositions (central values) of $1.04$ PeV, $1.14$ PeV and $2.0$ PeV, are assumed to be due to Flux-2 above, requiring an HDM mass of $\mathcal{O}(5)$ PeV. Both Flux-1 and Flux-3 contribute to account for rest of the HESE events, including the small bump-like excess in the $30-100$ TeV range. This scenario, in a  natural manner,  allows for the presence of a gap, or break in the spectrum between 400 TeV to 1 PeV\footnote{The statistical significance of such a break has now increased due to the recent release of six-year muon track data~\cite{1607.08006}; see, for instance, the discussion in  ~\cite{1611.07905}. Additionally, as we shall see below, by providing a significant fraction of the events directly (via Flux 2) or indirectly (via Flux 3) from DM,   this scenario does not require the astrophysical neutrino flux to be pushed up uncomfortably close to the Waxman-Bahcall bound, unlike the standard single power-law interpretation. }.

A  similar scenario has previously been studied in Refs.~\cite{1407.3280, 1503.02669}, in which the 988-day HESE data were taken into account.  While Ref.~\cite{1407.3280} ascribed the events below a PeV upto tens of TeV  entirely to the astrophysical flux (Flux-1),  Ref.~\cite{1503.02669}, ascribed these as being generated by  the secondary neutrino flux from three-body HDM decay (Flux-3).  In this study we do not make any assumption regarding the specific origin of these sub-PeV events, and allow  any viable combination of Flux-1 and Flux-3 in the fitting procedure. As we shall see later, one of our main results from the fit to the HESE data within Scenario I is that with the current level of statistics, a broad range of combinations of Flux-1 and Flux-3 can fit the sub-PeV events, while the PeV events are explained by Flux-2. We note in passing that, in Ref.~\cite{1503.02669} the DM model parameter space was guided by the requirement that the LDM annihilation in the present Universe explain the diffuse gamma ray excess observed from the Galactic centre region~\cite{TheFermi-LAT:2015kwa} in the Fermi-LAT data.  In the present study, the  focus is entirely on satisfactorily fitting the IC events.

\subsubsection*{Scenario II : PeV events  from an astrophysical flux and  the $30-100$ TeV excess from LDM DIS scattering}

In Scenario II, we relax the assumption made regarding the origin of the three PeV events in Scenario I, and perform a completely general fit to both the PeV and the sub-PeV HESE data, with all four of the flux components taken together. Essentially, this implies that the mass of the HDM particle is now kept floating in the fit as well.  We find that both the best-fit scenario and the statistically favoured regions correspond to a case where the PeV events are explained by the astrophysical neutrino flux (Flux-1), while the excess in the $30-100$ TeV window primarily stems from the LDM scattering (Flux-2). Flux-3, which now populates the low 1--10 TeV bins becomes inconsequential to the fit, since the IC threshold for
the HESE events is 30 TeV. Expectedly, in order for the astrophysical flux to account for the PeV events, the slope of the underlying power-law spectrum in Scenario II is significantly flatter compared to that in Scenario I.

In addition to performing  general fits to the PeV and sub-PeV HESE data as described above, we also explore, for both Scenarios I and II, the extent to which different Lorentz structures of the LDM coupling with the SM quarks impact the results.  While a vector mediator coupling to the SM quarks and the LDM was considered in Ref.~\cite{1407.3280}, a pseudo-scalar mediator was employed in Ref.~\cite{1503.02669}. Adopting a more general approach, we consider scalar, pseudo-scalar, vector and axial-vector mediators. However, we find (expectedly) that  if the LDM relic density is appreciable,  strong limits on the spin-independent coherent elastic scattering cross-section with nuclei of the relic LDM component come into play and restrict  the available parameter space for scalar and vector mediators.  There are also interesting differences between the pseudo-scalar and axial-vector scenarios insofar as fitting the IC data, as we shall show in  later sections. 

Finally, as emphasized in Ref.~\cite{1503.02669}, the three-body decay of the HDM particles that gives rise to the secondary neutrino flux (Flux-3 above), also produces a flux of diffuse gamma-rays in a broad energy range, which is constrained from the measurements by the Fermi-LAT telescope~\cite{Ackermann:2014usa} at lower energies, and by the cosmic ray air shower experiments (KASCADE~\cite{Feng:2015dye} and GRAPES-3~\cite{Gupta:2009zz}) at higher energies~\cite{Ahlers:2013xia}. We find that the parameter space of the proposed dark matter scenarios that can fit the IC data is significantly constrained by the upper bounds on residual diffuse gamma ray fluxes\footnote{We note that stronger constraints, based on IC data and Fermi-LAT,  as discussed recently in \cite{1612.05638} are evaded in our work since they are derived assuming the two-body decay of dark matter directly to SM particles, \eg $\,\,b\bar{b}$.}.

The rest of this paper is organized as follows:   Sec.~\ref{LDM} examines the different ways the LDM particle can interact with SM quarks, and summarizes the current constraints on the effective couplings and the mass parameters, using  gamma ray and collider data. We also discuss the general method used to calculate the contribution made by the HDM three-body decay to galactic and extra-galactic gamma-ray fluxes.
 Sec.~\ref{Scenario1} focusses on Scenario I and  describes our procedure for deriving best-fits to the observed IC HESE data for it,  and the results obtained for different choices of the mediator. The validity of these results is then examined in the light of various constraints.
Similarly, Sec.\ \ref{sec:scenario2} repeats this  for Scenario II.
Although the focus of this work is on understanding the HESE data, IC has 
recently released a statistically independent sample of high energy muon track events
\cite{1607.08006} for the neutrino energies between $190$ TeV to $9$ PeV, 
where the interaction vertex is allowed to be outside the detector.
 Sec.~\ref{sec:track},  examines both the scenarios considered here in the light of this data sample.
Finally, our findings are recapitulated and summarized in Sec.~\ref{sec:summary}.

\section{LDM interaction with quarks: simplified models and current constraints}
\label{LDM}
  This section  provides further details on how we model the interaction of the LDM with the SM quarks. In what follows, we shall work with a representative  model where the HDM ($\phi$) is described by a real scalar field, and the LDM ($\chi$) is a neutral Dirac fermion, both of which are singlets under the standard model gauge interactions. The interaction of the heavy dark matter particle with the LDMs is described by an Yukawa term of the form $g_{\phi \chi \chi} \phi \overline{\chi} {\chi}$. 

We further assume that the LDM particles are stabilized on the cosmological scale by imposing a $Z_2$ symmetry, under which the LDM field is odd, and all other fields are even. The LDM can interact with the SM fermions (quarks in particular) via scalar, pseudo-scalar, vector, axial-vector or tensor effective interactions. To describe such effective interactions we introduce a simplified model, where the interactions are mediated by a $Z_2$ even spin-0 or spin-1 particle. The LDM can also couple to SM fermions via a $Z_2$ odd mediator, which carries the quantum numbers of the SM fermion it couples to. We do not consider the t-channel models or the tensor type interaction in this study. 

In the following sub-sections we shall describe the simplified model setup and mention the generic constraints on the couplings of a spin-0 or spin-1 mediator to the LDM and the SM fermions. Such constraints on the coupling and mass parameters can be modified within the context of a specific UV complete scenario, especially if it necessarily involves other light degrees of freedom not included in the simplified model. However, since the primary focus of this study is to determine the  combination of different fluxes which can fit the features  observed in the IC data, the simplified models chosen are sufficient for this purpose. Our approach allows us  to draw general conclusions regarding the possible contributions of  LDM scattering and the secondary neutrino fluxes, while being broadly consistent with constraints from  experiments and observations.

\subsection{Spin-0 mediators}
\label{spin0}
The parity-conserving effective interaction Lagrangian (after electroweak symmetry breaking) of the LDM ${\chi}$ with SM fermions $f$, involving a scalar mediator $S$ or a pseudo-scalar mediator $A$ can be written as follows:

\begin{equation}
\mathcal{L}_{\rm S} = \sum_{f} \frac {g_{Sf} m_f} {v} S \overline{f}  f + g_{S\chi} S \overline{{\chi}} \chi
\end{equation}

\begin{equation}
\mathcal{L}_{\rm P} = \sum_{f} \frac{i g_{Pf} m_f} {v} A \overline{f}  \gamma_5 f +  i g_{P\chi} A \overline{{\chi}}  \gamma_5  \chi
\end{equation}
Here $m_f$ is the mass of the SM fermion $f$, $g_{S\chi}$ ($g_{P\chi}$) represents the coupling of the LDM with the scalar (pseudoscalar) mediator, and $v ~(\approx 246$ GeV) stands for the vacuum expectation value of the SM Higgs doublet (in the presence of other sources of electroweak symmetry breaking the definition of $v$ will be appropriately  modified). The sum over fermion flavours can in principle include all SM quarks and leptons, although for our current study, the quark couplings are more relevant. We shall take the coupling factors $g_{Sf}$ and $g_{Pf}$, which appear in the coupling of fermion flavour $f$ with the scalar and the pseudo-scalar mediators respectively, to be independent of the quark flavour for simplicity.

A SM singlet spin-0 mediator cannot couple in a gauge-invariant way to SM fermion pairs via dimension-four operators. One way to introduce such a coupling is via mixing with the neutral SM-like Higgs boson after electroweak symmetry breaking. Such a mixing, if substantial, can however modify the SM-like Higgs properties leading to strong constraints from current LHC data. Other possible ways include introducing a two Higgs doublet model (and mixing of the singlet scalar with the additional neutral scalar boson(s)), or introducing new vector-like fermions to which the singlet scalar couples, and which in turn can mix with the SM fermions~\cite{Izaguirre:2014vva}. In all such cases  the couplings of the singlet-like scalar to SM fermions should be proportional to the fermion Yukawa couplings in order to be consistent with the assumption of minimal flavour violation, thus avoiding flavour-changing neutral current (FCNC) constraints~\cite{D'Ambrosio:2002ex}.  

\subsection{\label{spin1}Spin-1 mediators}

The effective interaction Lagrangian involving a spin-1 mediator, $Z^\prime$, to SM fermions $f$ and the LDM $\chi$ can be written as follows:

\begin{equation}
\mathcal{L} = \overline{\chi} \left (g_{V\chi} \gamma^\mu + g_{A\chi} \gamma^\mu \gamma_5 \right) \chi Z^{\prime}_\mu +\sum_{f} \overline{f} \gamma^\mu\left(g_{Lf}  P_L  + g_{Rf}  P_R  \right) f Z^{\prime}_\mu.
\end{equation}
Here the subscripts $V,A,L$, and $R$ refer to vector, axial-vector, left-chiral and right-chiral couplings respectively. The  left and right handed SM fermion currents are invariant under the SM ${\rm SU}(3)_C \times {\rm SU}(2)_L \times {\rm U}(1)_Y$ gauge transformations. Therefore, in general, both vector and axial-vector interactions are present with coefficients $g_{Vf}=g_{Rf} + g_{Lf} $ and $g_{Af}=g_{Rf} - g_{Lf} $. In order to obtain only vector or axial-vector SM fermion currents at a low energy scale, we need to set $g_{Rf}=g_{Lf} $ or $g_{Rf}=-g_{Lf} $, respectively.

If the  $Z^\prime$ couples to charged leptons, there are strong upper bounds on its mass from collider searches for dilepton resonances from the LHC. In order to avoid them, we assume the leptonic couplings to be absent. In a minimal scenario with only the SM Higgs doublet giving mass to all the SM fermions, we encounter further relations from ${\rm U}(1)^\prime$ gauge invariance (here, $Z^\prime$ is the gauge field corresponding to the ${\rm U}(1)^\prime$ gauge interaction) on the coupling coefficients to quarks and leptons~\cite{Kahlhoefer:2015bea}. This is because if left and right handed SM fermions have different charges under the new gauge group, the SM Higgs doublet needs to be charged under ${\rm U}(1)^\prime$ as well. Thus, when a single Higgs doublet gives rise to the mass of both SM quarks and charged leptons, if the quarks are charged under ${\rm U}(1)^\prime$, so would be the leptons. However, such constraints can be avoided in a non-minimal scenario, for example in a two Higgs doublet model, where different Higgs bosons are responsible for giving mass to quarks and leptons, thereby making their ${\rm U}(1)^\prime$ charges uncorrelated.  We  keep in view such considerations related to  ultra-violet completion for this study, although we do not fully flesh out their consequences.

\subsection{\label{ssec:gen-constraints}Constraints on the couplings and the mass parameters}
\begin{figure}[htb]
  \centering
  \begin{subfigure}{0.6\textwidth}
    \centering
    \includegraphics[width=\textwidth]{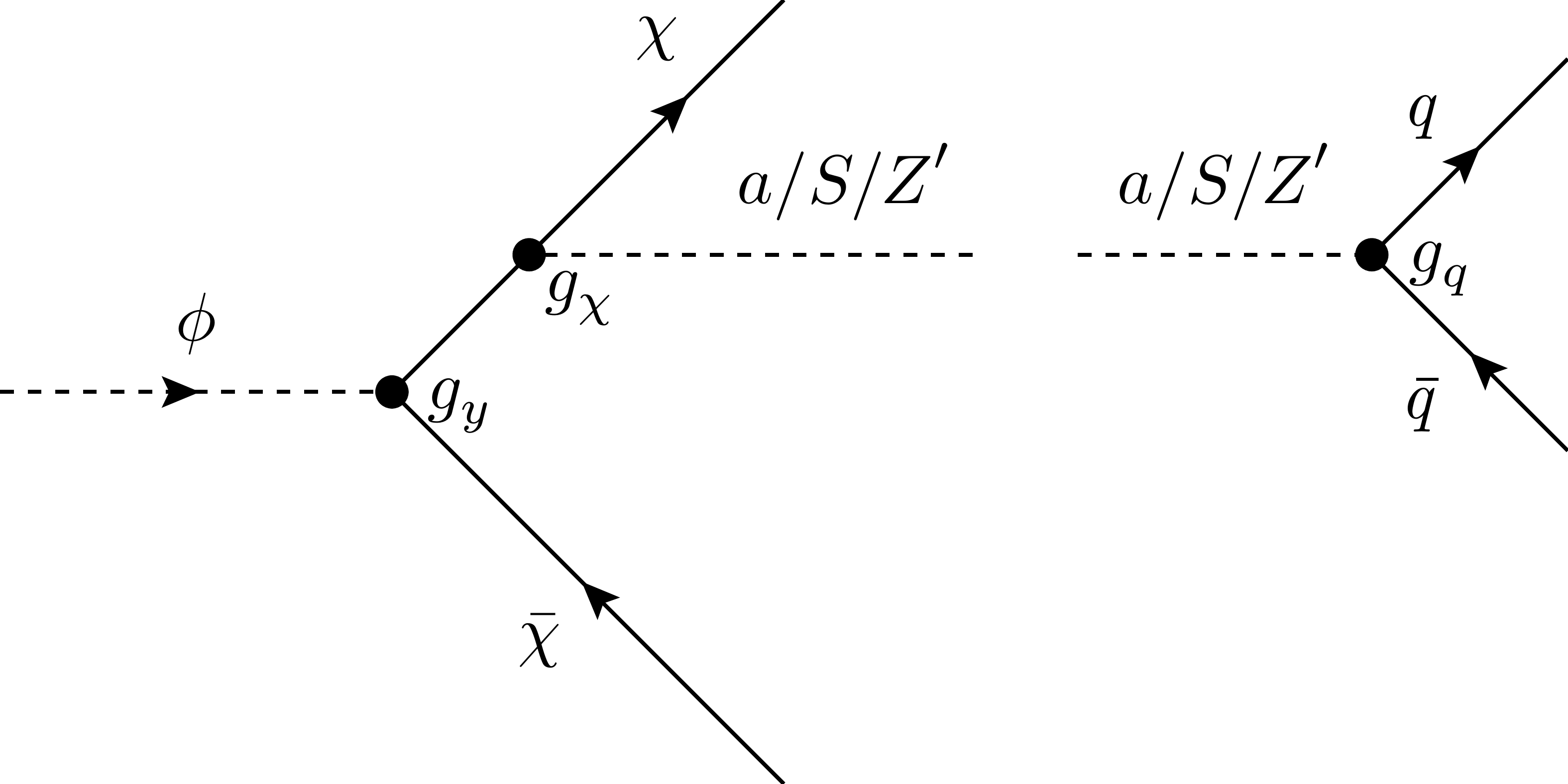}
  \end{subfigure}
  \begin{subfigure}{0.39\textwidth}
    \centering
    \includegraphics[width=0.5\textwidth]{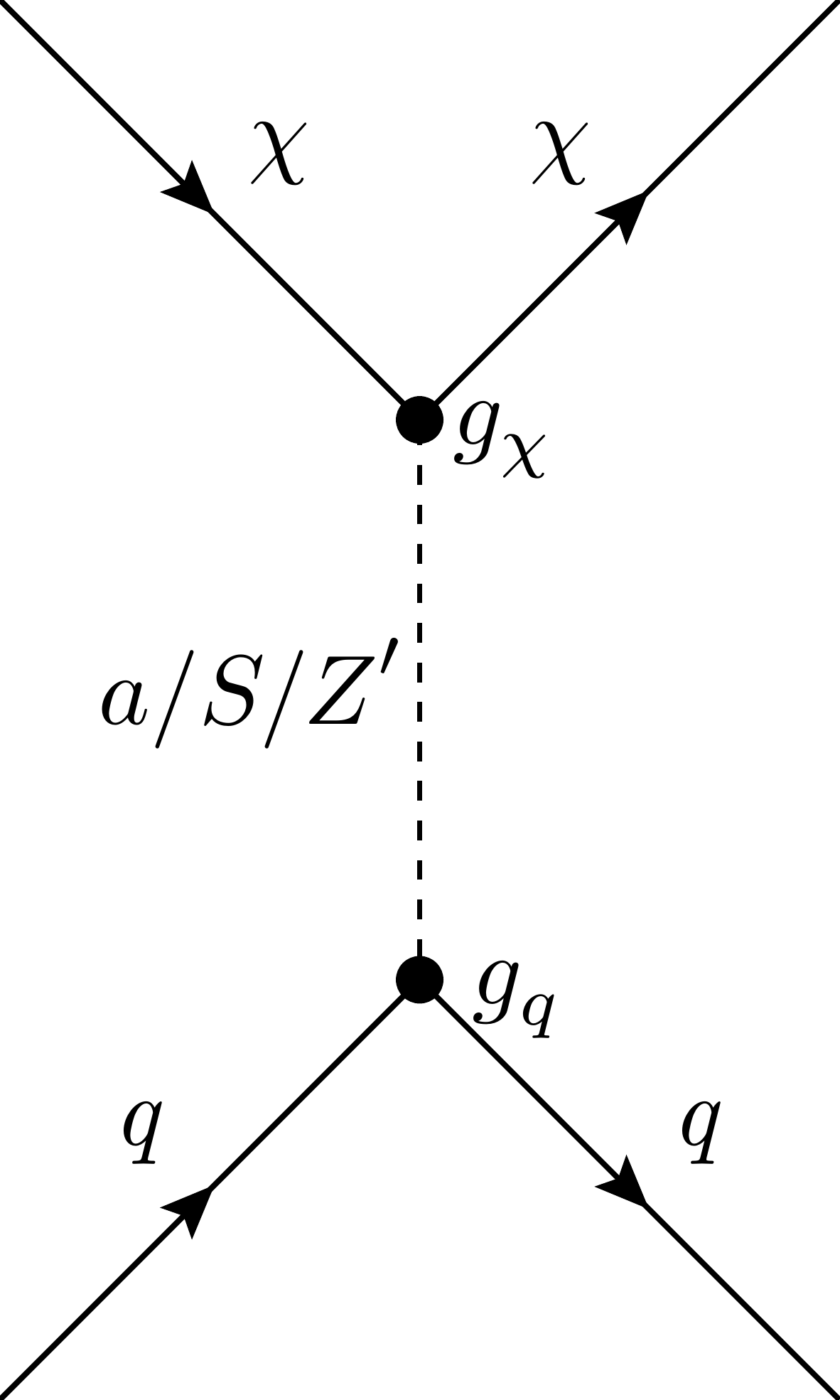}
  \end{subfigure}
  \caption{\label{fig:feyn-spin0}The interactions corresponding to $ \phi $ decay (left), mediator decay (centre) and $ \chi q $
           scattering (right) involving a generic mediator, along with relevant coupling constants.}
\end{figure}

Figure 1 shows the main interaction vertices which are relevant for both Scenario I and II. $g_y$ represents the coupling between the HDM and LDM leading to the slow decay of the former, with lifetime $\tau_\phi$. The other couplings shown correspond to the vertices of either (a) SM quarks or  (b) the LDM interacting with  a generic mediator, which can be a pseudo-scalar ($a$) or a scalar ($S$) or a spin-1
boson ($Z^\prime$) which couples via vector and/or axial-vector couplings, as discussed in the previous section.

The rate of LDM DIS scattering at IC is proportional to $(g_q g_\chi)^2$, where $g_q$ and $g_\chi$ are the mediator-quark pair and mediator-LDM pair couplings, respectively.  It is also proportional to $g_y^2$, or, equivalently, inversely proportional to  $\tau_\phi$\footnote{This assumes that the two-body decay to $\chi$ is the dominant mode.}. Finally, the IC event rates are also proportional to the fractional contribution of the HDM to the total DM density, $f_{\phi} = \Omega_{\phi}/\Omega_{\rm DM}$. Here, $\Omega_{\rm DM} = 0.1198/h^2$ (with $h$ being the normalized Hubble constant) from recent PLANCK results\cite{1502.01589}. 

From the above considerations,  the IC event rate from LDM DIS scattering, for a given choice of mediator mass $m_{\rm M}$, is determined by the quantity $F=f_\phi g_q^2  g_\chi^2 / \tau_\phi$. It is useful to determine its maximum allowed value. In order to keep the couplings perturbative, we require $g_{\chi,q} < 4 \pi$.  We also require the lifetime of the HDM to be longer than the age of the Universe $\tau_\phi \gtrsim 4.35 \times 10^{17}$ seconds. And since $f_\phi < 1$, we obtain the upper bound, $F \lesssim 5.7 \times 10^{-14} {~\rm s}^{-1}$.  If the  value of $F$ exceeds this maximum, the couplings will not be perturbative, or the HDM would have decayed too quickly to have an appreciable density in the present Universe.

The secondary neutrino flux from the three-body decay of $\phi$ (Flux-3 in Sec.~\ref{intro2}), is proportional to $g_\chi^2$ (again, in the limit where the two-body decay width is much larger than the three-body width). It is also  inversely proportional to the life-time of the HDM, $\tau_\phi$.
In addition to the mass of the $\phi$, $ \tau_\phi $ is determined by $g_{\phi \chi \chi}$ when the two body decay to LDM pairs dominate. Thus, the parameters relevant for fitting the features in the IC data in our work are $g_q, g_\chi$, mass of the mediator particle ($m_{M}$), and $\tau_\phi$. The results do not depend on $m_\chi$, as long as it is significantly lower than $m_\phi$. 

It is useful to examine the ball-park  numerical values of some of the quantities which are used to fit the IC events using DIS $\chi$-nucleon scattering. The cross section depends essentially on $F$ and the mediator mass, $m_{\rm M}$. Hence, given a  certain value of $m_{\rm M}$, and a value for the factor $F$, one could obtain minimum value of the couplings needed to fit an observed number of cascade events.  This is given by $g_q g_\chi \gtrsim (F \times 4.35 \times 10^{17})^{(1/2)}$, assuming  $f_\phi < 1$, and $\tau_\phi \gtrsim 4.35 \times 10^{17}$ seconds. A typical value that occurs in the fits is, for instance,  $F \sim 10^{-26} {~\rm s}^{-1}$, and using  this leads to a lower bound $g_q g_\chi \gtrsim 6.6 \times 10^{-5}$.  Assuming, for simplicity, $g_q \sim g_\chi =g$, each coupling should thus be greater than about $8 \times 10^{-3}$. 

As mentioned earlier, (in Sec.~\ref{intro2}), the most restrictive constraint on the value of $F$ comes from the upper bound on the flux of diffuse gamma rays. We defer a detailed discussion of our computation of the gamma ray flux from the three-body decay of the HDM, and the resulting constraints to Sec.~\ref{sssec:gamma}. Significant constraints also arise  from collider experiments, where the mediator and the LDM particles can be directly produced, and we discuss these in the next sub-section.

 The relic density of $\chi$, which we denote as $f_{\chi} = \Omega_{\chi}/\Omega_{\rm DM}$, is not of direct relevance to our study, which focusses  on the IC events coming either from DIS scattering of the LDM in IC, and on the flux of secondary neutrinos from the three-body decay of the HDM. However, direct detection constraints can be important  if there is a significant density of the LDM in the current Universe. It is well-known that if $f_\chi$ is significant, the spin-independent direct detection bounds on the scalar and vector interactions are very strong, and thus would force us to focus on either pseudo-scalar or axial-vector couplings (or relegate us to corners of $m_\chi$ values which are not yet probed by the direct detection experiments). For our purpose, we could either assume that this is the case, or, equivalently, that the $\chi$  density is indeed small. If the latter,  within the simplified model setup discussed above, the relic density of $\chi$ can be diluted to very small values in two ways. The first is by increasing $g_\chi$, and restricting to values of $m_\chi > m_M$, such that the dominant annihilation mode of $\chi$ is to the mediator pair, which can then decay to the SM fermions even via a small $g_q$. The second way (albeit  fine-tuned), is by setting $m_\chi$ close to $m_M/2$, thereby allowing for a resonant annihilation of LDM pairs to SM quarks. Since the IC event rates do not depend upon $m_\chi$ as long as it is significantly smaller than the HDM mass, both these approaches do not affect the IC event rates. Finally, there can always be additional annihilation modes of the LDM not described by the simplified models which do not affect the IC computations, but help make $f_\chi$ small. 

With respect to the choices of mediators, we note that as far as the IC DIS scattering cross-sections are concerned, the exact Lorentz structure of the couplings is not important. However, as we shall see later, the two-body branching ratio of the HDM to LDM pair is sensitive to the Lorentz structure.

\subsubsection{\label{sssec:collider-cons}Collider constraints}
The collider constraints are sensitive to the interplay of several  couplings and mass parameters relevant to our study, specifically, $g_q,g_\chi,m_\chi$ and $m_M$. A scalar or pseudo-scalar mediator particle which dominantly couples to heavy fermions  can be produced in association with one or two b-quarks (involving the parton level processes $g ~b (~\overline{b}) \rightarrow ~b (~\overline{b}) ~S/A$ and $g ~g \rightarrow ~b ~\overline{b} ~S/A$ respectively). Such a final state may be accessible to LHC searches if the (pseudo-)scalar decays further to an LDM pair $S/A \rightarrow \chi \overline{\chi}$. However, in case, $m_\chi > m_{S/A}$, the (pseudo-)scalar would decay back to the SM  fermion pairs, thereby making the search considerably harder due to large SM backgrounds. On the other hand, off-shell $S/A$ production does lead to a cross-section in the one or two b-jet(s) and missing transverse momentum (MET) channel. Furthermore, an effective coupling of $S/A$ to gluon pairs is also generated by the top quark loop, and therefore, mono-jet and missing energy searches are also relevant. These bounds have been computed in, for example, Ref.~\cite{Buckley:2014fba}. The current bounds from these searches are weaker than $g_q g_\chi \lesssim \mathcal{O}(0.1)$, across the range of $m_M$ and $m_\chi$ of our interest~\cite{Buckley:2014fba}. As we shall see later, the coupling values required in our study are well within the current collider limits. For individual couplings, values of $\mathcal{O}(0.3)$ should be allowed, although the LHC bounds are very sensitive to the ratio $g_\chi/g_q$, which determines the rate of events with MET. 

In the case of a spin-1 mediator with either vector or axial vector couplings to SM quarks, the strongest collider constraints come from dijet resonance searches, where the mediator is produced on-shell, and decays back to the SM quarks. Depending upon the values of $g_\chi$ and $g_q$, monojet and MET searches could  also be important, especially if a) the mediator width is  large, making the resonance searches harder, or if b)  $g_\chi > g_q$ for a given value of the product $g_\chi g_q$, such that the branching ratio to LDM pairs dominates the on-shell mediator decay (when $m_M > 2m_\chi$). Bounds on couplings in the axial-vector case   have been discussed  in Ref.~\cite{Chala:2015ama}, which combines the results of different experiments spanning a range of centre of mass energies, including (8 TeV) LHC (ATLAS and CMS), Tevatron and UA2. Similar considerations and bounds would apply to the vector mediator case. For $\mathcal{O}(1)$ values of $g_q g_\chi$,  bounds from dijet searches  cover $M_{Z^\prime}$ masses in the range of $100$ GeV to $2-3$ TeV, depending upon the ratio $g_\chi/g_q$, across the range of $m_\chi$ values. For a detailed discussion of these  bounds for different values of $g_q g_\chi$ and  $g_\chi/g_q$, we refer the reader to Ref.~\cite{Chala:2015ama}. With the recent $13$ TeV $15.7 \text{ fb}^{-1}$ LHC data, ATLAS limits on the $Z^\prime$ coupling to quarks vary in the range of $0.1$ to $0.33$, as $M_{Z^\prime}$ is varied in the range $1.5$ to $3.5$ TeV, when the mediator decay to LDM pairs is absent~\cite{ATLAS:2016lvi}. Thus, we conclude that the collider bounds on the spin-1 boson couplings are in the range of $\mathcal{O}(0.1)$, and the  values required to fit  the IC event rates are very much allowed by collider constraints. 

\subsubsection{\label{sssec:gamma}Contributions to Galactic and Extra-Galactic  Gamma-Ray Fluxes from HDM Decay}
The three-body decay of the HDM to a pair of LDMs and a mediator particle (where the mediator particle is radiated by an LDM in the final state), will necessarily contribute to a diffuse gamma ray flux spanning a wide range of energies. This sub-section describes the general method we use to calculate these contributions. The mediator particles lead to hadronic final states via their decays to quark pairs or to hadronically decaying tau pairs, with  gamma rays  originating from the decays of neutral pions produced in the cascade.  Leptonic decays of the mediator can also give rise to high-energy photons via bremsstrahlung and inverse Compton scattering. In the computation of the gamma ray constraints, we only consider the hadronic decay modes of the mediator via quark final states, since the coupling of the mediator to quarks is essential to explaining the IC events in our scenario. For the case of a  (pseudo)scalar mediator,  the leptonic couplings are expected to be small due to the  smaller Yukawa couplings of the charged leptons, while for the case of (axial-)vector mediators, as discussed in Sec.~\ref{spin1},  consistency with dilepton resonance search constraints  favour a setup in which the leptonic couplings are absent. We note in passing that the same three body decays would also lead to signatures in cosmic rays, and there can be additional constraints from measurements of positron and anti-proton fluxes. Due to the large uncertainties in  diffusion and propagation models of cosmic rays, we do not include these constraints in our analysis.

The gamma ray flux, like the secondary neutrino flux which we calculate below in Section 3, has a galactic and an extra-galactic component~\cite{Cirelli:2010xx}:

\begin{equation}
\frac{d\Phi_{\rm Isotropic}}{dE_{\gamma}} = \frac{d\Phi_{\rm ExGal}}{dE_{\gamma}} + 4 \pi \frac{d\Phi_{\rm Gal}}{dE_{\gamma} d\Omega} \bigg |_{\rm Min} 
\end{equation}
The extra-galactic flux is isotropic and diffuse (after subtracting out contributions from known astrophysical sources), while the minimum of the galactic flux  is an irreducible isotropic contribution to the diffuse flux \cite{Cirelli:2010xx}. Since the most important constraints on very high-energy gamma-rays come from air-shower experiments, observations of which are confined to the direction opposite to the Galactic center, we take this minimum to be the flux from the anti-Galactic center, following  Refs.~\cite{Cirelli:2010xx, Cirelli:2012ut}. 

Unlike the neutrino flux, the extra-galactic gamma-ray component suffers significant attenuation due to pair creation processes, and consequently in the energy region of interest here,  one finds the galactic component to be the dominant one from any given direction in the sky. This  is given by
  
\begin{equation}
\frac{d\Phi_{\rm Gal}}{dE_{\gamma} d\Omega} = \frac{1}{4 \pi} \frac{\Gamma_{\rm dec}}{M_{\rm DM}} \int_{\rm los} ds \rho_{\rm halo} [r(s,\psi)] \frac{dN} {dE_{\gamma}}
\end{equation}
where, $\Gamma_{\rm dec}$ is the total decay width of the HDM, $M_{\rm DM}$ is its mass, and the line of sight integral over the DM halo density $\rho_{\rm halo} [r(s,\psi)] $ is performed along the direction of the anti-GC. We take the DM density profile in our galaxy to be described by a Navarro-Frenk-White distribution~\cite{astro-ph/9508025}:

\begin{equation}
\rho_{\rm NFW} (r) = \rho_s \frac{r_s}{r} \left(1+ \frac{r}{r_s} \right)^{-2}
\end{equation}
with the standard parameter choices, $\rho_s=0.18 \text{ GeV cm}^{-3}$ and $r_s=24$ kpc.  Here, $dN/dE_{\gamma}$ represents the gamma-ray spectra per decay of the HDM in the HDM rest frame. We take the prompt gamma ray energy distribution in the rest frame of the mediator from {\tt PPPC4}~\cite{1012.4515}, and then subsequently fold it with the three-body differential energy distribution of the mediator obtained using {\tt CalcHEP}~\cite{1207.6082}, and finally boost the resulting gamma ray spectra to the rest frame of the decaying HDM. 

The extra-galactic component of the flux is given by~\cite{Cirelli:2010xx}
\begin{equation}
\frac{d\Phi_{\rm ExGal}}{dE_{\gamma}} = \frac{\Omega_{\rm DM} \rho_{\rm c, 0}}{M_{\rm DM} \tau_{\rm DM}} \int_0^{\inf} dz \frac{e^{-\tau(E_{\gamma}(z),z)}}{H(z)} \frac{dN} {dE_{\gamma}}(E_{\gamma}(z),z)
\end{equation}
where, the Hubble constant is given by $H(z)=H_0 \sqrt{\Omega_M (1+z)^3+\Omega_{\Lambda}}$, with $H_0$ being the present Hubble expansion rate, and $\Omega_M,\Omega_{\rm DM}$ and $\Omega_\Lambda$ are the matter, DM and dark energy densities respectively, in terms of the present critical density,  $\rho_{\rm c, 0}$. We take the values of all relevant cosmological parameters from recent Planck best fits~\cite{1303.5076,1502.01589}. The attenuation factor $e^{-\tau(E_{\gamma}(z),z)}$ describes the absorption of gamma rays described above, as a function of the redshift $z$ and observed gamma-ray energy $E_\gamma$, which we take from {\tt PPPC4} tables~\cite{1012.4515}.

Having established the framework and general considerations  for our study, and outlined the constraints to which it is subject,  in  the sections to follow we proceed with the specific calculations necessary to demonstrate how IC data may be understood in scenarios combining boosted dark matter and astrophysical neutrinos.

\section{Scenario I: PeV events caused by LDM scattering on Ice and its implications}
\label{Scenario1}

In this section we consider a  scenario where  boosted DM scattering off ice-nuclei leads to the three events at energies above a PeV seen in the 1347-day HESE sample. In the present data-set, these events are somewhat separated from the others, since there appear to be no  HESE events in the region $ 400$ TeV$\leq E_{dep}\leq1$ PeV, providing some justification for considering them as disparate from the rest. 

Both a) the details of the scattering cross-section of the LDM with ice-nuclei,
and b) the three-body spectrum leading to the secondary neutrino flux in sub-PeV energies
 depend  on the particle mediating  the $ \chi N $ interaction.
Thus we first examine different mediator candidates --- pseudo-scalar, scalar,
vector and axial vector --- and determine how the corresponding 
fits and parameters change when a specific choice is made.

As discussed in Sec.~\ref{ssec:gen-constraints}, for the (dominant) two-body decay of the HDM ($ \phi $)
 into a pair of LDM ($ \chi\bar{\chi} $), the corresponding event rate for $ \chi N $ scattering is proportional to $F= f_\phi \,(g_\chi \, g_q)^2 / \tau_\phi$.
The observed rate of the PeV events in IC, along with their deposited energies, then determines a) 
the ratio of couplings and lifetime $F$, and b) the mass ($m_\phi$) of the HDM ($ \phi $), using the usual two-body decay kinematics \cite{1407.3280}. 
Specifically, if the mean inelasticity of the interaction of the LDM with the
ice-nuclei, mediated by a particle $a$ is given
by $ \mean{y_a} $, then we require the LDM flux from HDM decay  to peak
around energies $ E_\text{PeV} / \mean{y_a} $,  where $ E_\text{PeV} $
represents an estimated average  deposited energy at IC for such events.

In this scenario, events in the sub-PeV energy range are then explained by a combination of events from Flux-1 (an astrophysical power-law neutrino flux),  Flux-3 (the secondary flux of neutrinos from three-body HDM decay) and Flux-4 (the standard atmospheric neutrino and muon flux), as outlined in Sec.~\ref{intro2}.
For Flux-4, we  use the best-fit background estimates from the IC analysis.
 We determine the best-fit combination of Flux-1 and Flux-3, which, when folded in with the IC-determined best-fit Flux-4 will explain all the sub-PeV observed events in the 1347-days HESE sample.
The parameters relevant to this sub-PeV best-fit  are $ m_a $, $(f_\phi\,g^2_\chi/{\tau_\phi})$,
$ N_\text{Ast} $ (the number of sub-PeV events from Flux-1), and $ \gamma $ (the power-law index for Flux-1).

The total number of shower events within each IC energy bin is given by \cite{1404.0017}:
\begin{equation}\label{eq:chi-event-rate}
{\rm N}^{\text{cascade},NC}_{\chi} = T \, N_A \, \int_{E_{min}}^{m_{\phi}/2}
dE_{\chi}\,{\rm M}^{NC}(E_\chi)\,\dfrac{d\Phi_\chi}{dE_\chi} 
\int_{y_{min}}^{y_{max}}dy \, \dfrac{d\sigma^{NC}(E_{\chi},y)}{dy}
\end{equation}
Here $y$ is the inelasticity parameter, defined in the laboratory frame  by $y = {E_{dep}}/{E_{\chi}}$, with $E_{dep}$ being the energy deposited in the detector and  $E_\chi$  denotes the energy of the incident dark matter, $T$ the runtime of the detector (1347 days) and $N_A$ is the Avogadro number. The limits of the integration are given by $y_{min} = {E_{min}^{dep}}/{E_\chi}$ and $y_{max} = \rm min \left(1,{E_{max}^{dep}}/{E_\chi}\right)$. $E_{min}^{dep}$ and
$E_{max}^{dep}$ are the minimum and maximum deposited energies for an IC energy-bin. $M^{NC}\,(E_\chi)$ is the energy dependent effective detector mass for neutral current interactions obtained from \cite{1311.5238}. ${d\sigma^{NC}(E_{\chi},y)}/{dy}$ is the differential $\chi N$ scattering cross-section, which we  quantify  below. 

The total flux ${d\Phi_\chi}/{dE_\chi}$ is composed of two parts, the Galactic component ${d\Phi_\chi^{\rm GC}}/{dE_\chi}$ and the red-shift ($z$) dependant extra-Galactic component ${d\Phi_{\chi}^{\rm EG}}/{dE_\chi}$. They are given by\cite{Esmaili:2012us,1308.1105} :

\begin{eqnarray}\label{difffgal}
\dfrac{d\Phi_\chi^{\rm GC}}{dE_\chi} &=& D_\text{G} \fdnde \nonumber \\
\dfrac{d\Phi_{\chi}^{\rm EG}}{dE_\chi} &=& D_\text{EG}\! \int_{0}^{\infty}{dz} \frac{1}{H(z)} \fdnde \left[(1+z)E_{\chi}\right],
\end{eqnarray}
where, 
\[D_\text{G} = 1.7\times 10^{-8} \left( \frac{1\text{ TeV}}{m_{\phi}} \right)
\left( \frac{10^{26}\text{ s}}{\tau_{\phi}} \right) \cmssr\]
and
\[D_\text{EG} = 1.4 \times 10^{-8} \left( \frac{1\text{ TeV}}{m_{\phi}} \right)
\left( \frac{10^{26}\text{ s}}{\tau_{\phi}} \right) \cmssr.\]
For the two-body decay $\phi \to \anti{\chi}\chi$, the flux at source is  given by :
\begin{equation}
\fdnde = 2\delta\left(E_\chi - \frac{1}{2}m_{\phi}\right)\,,
\label{diracdelta}
\end{equation}
where, $E_\chi$ denotes the incident energy at IC for each $\chi$ particle.

\paragraph*{}
We next describe the computation of  the secondary neutrino flux due to the $\phi \to \chi \bar{\chi} a$ three-body decay mode, where one of the daughters ($ a $) is the mediating particle in  $ \chi N $ scattering.
The general procedure is the same as outlined in \cite{1503.02669}. In our representative calculation here, $a$ is assumed to decay to a $q\bar{q}$ pair, which by further hadronisation and decays leads to the secondary neutrino spectrum.
It is straightforward to  obtain the resulting neutrino flux in the rest frame of $a$ (see, \eg, \cite{1012.4515}), using event generators that implement the necessary showering and hadronisation algorithms, such as \verb+PYTHIA8+ \cite{0710.3820}. This flux is then boosted to the  lab-frame, which is, approximately, the $\phi$ rest frame.

This boosted flux  in the $\phi$ rest frame is used in conjunction with  Eq.~\ref{difffgal} to get the final flux of the secondary neutrinos. The neutrino event rates from this source are determined by folding this flux with the effective area and the exposure time of the detector \cite{1311.5238}.

Having obtained the event rates for the secondary neutrinos, one defines the $ \chisq $ 
necessary to quantify our goodness of fit to the observed data:
\begin{multline}
  \chisq \equiv \chisq(m_a, f_\phi g^2_\chi / \tau_\phi, \nast, \gamma)\\
  = \left[N^\text{sub-PeV}(m_a, f_\phi g^2_\chi / \tau_\phi, \nast, \gamma) - N^\text{sub-PeV}_\text{obs}\right]^2/N^\text{sub-PeV}(m_a, f_\phi g^2_\chi / \tau_\phi, \nast, \gamma)
\end{multline}
Minimizing this \chisq\ determines the best-fit point in the parameter space of
$ \lbrace m_a, f_\phi g^2_\chi /\tau_\phi$, $\nast, \gamma \rbrace $.
It should be noted that the sub-PeV events in Scenario I are due both to the decay of the mediator and 
a uniform power-law spectrum typical of diffuse astrophysical sources, which is why the overall $ \chisq $ function
is dependent on all the four parameters shown above.

\paragraph*{}
We now turn to discussing the results for specific mediators.

\subsection{Pseudoscalar mediator}

When the mediator is a pseudo-scalar particle, the corresponding double differential cross-section is given by :
\begin{equation}\label{eq:ps-d2sigma}
\dfrac{d^2\sigma}{dxdy}=\sum_{q}\dfrac{1}{32\,\pi}\dfrac{E_\chi}{x \, M_N (E_\chi^2 - m_\chi^2)}
\dfrac{(g_\chi \, g_q)^2(Q^2)^2}{(Q^2 + m_a^2)^2}\,f_q(x, Q^2)
\end{equation}
where $x$ is the Bjorken scaling parameter, $M_N, m_\chi\text{ and } m_a$ are the masses of the nucleon, LDM,
and the mediator respectively, and $Q^2 = 2xyM_NE_\chi$. $f_q(x,Q^2)$ is the parton distribution function (PDF) of the quark $q$ in the nucleon.
We henceforth use the \verb+CT10+ PDFs \cite{Lai:2010vv} throughout our work.

\begin{figure}
\centering
\includegraphics[width=0.49\textwidth]{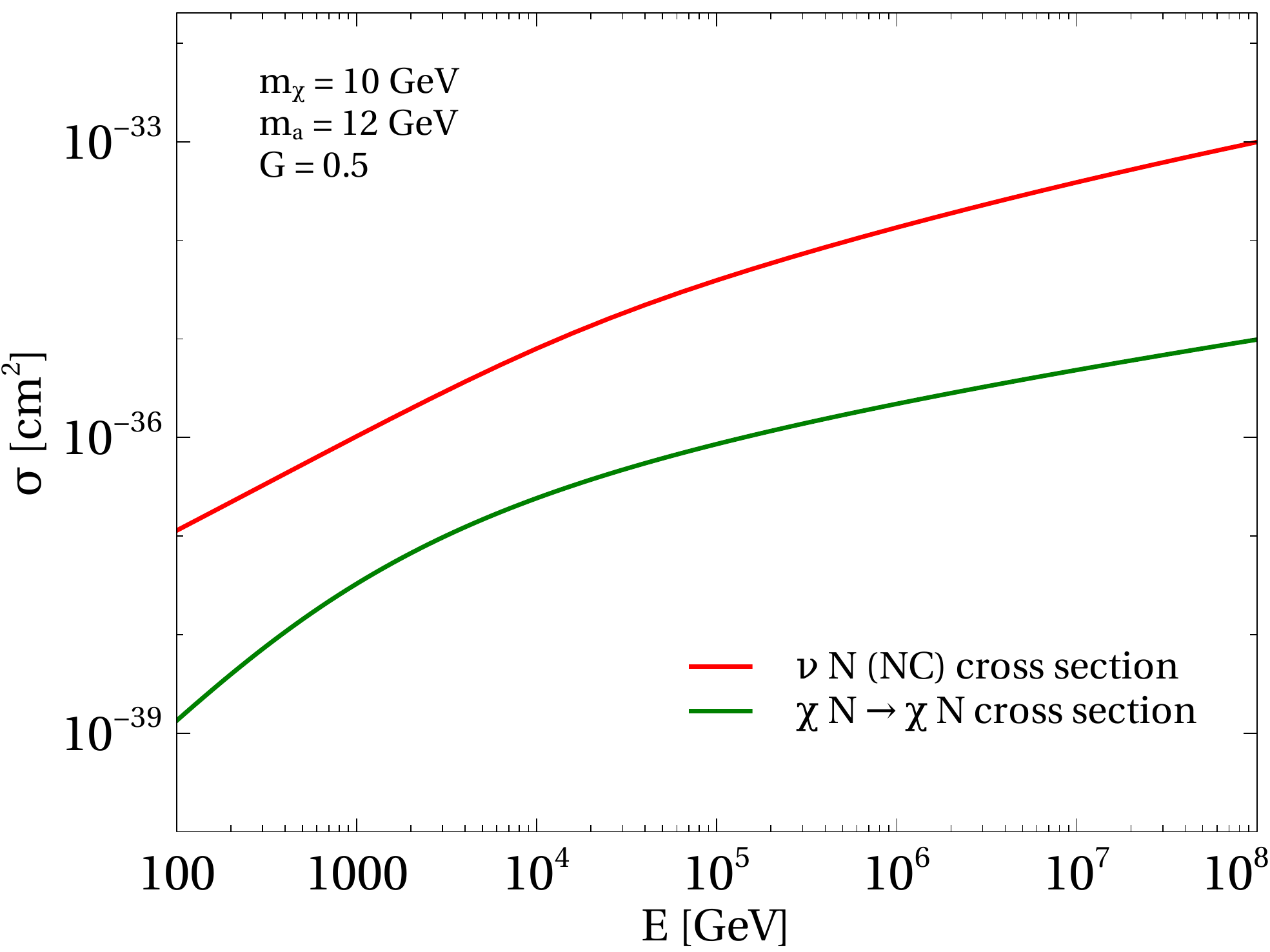}
\includegraphics[width=0.49\textwidth]{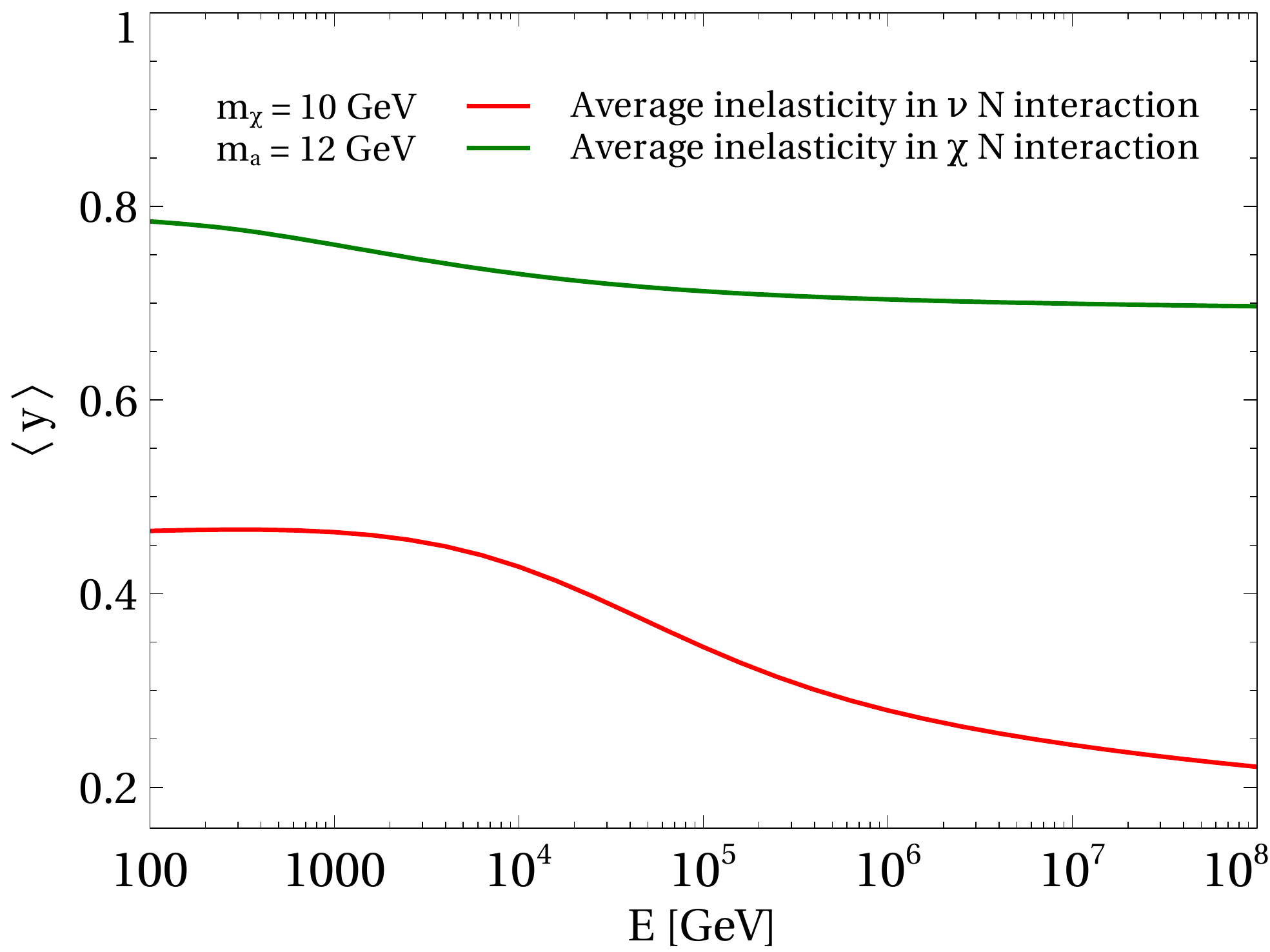}
\caption{Representative plots showing the relative behaviour of $\chi N$ and $\nu N$ neutral current cross sections (left).  Average inelasticities are also plotted for both cases (right).}
\label{xsec}
\end{figure}

Eq.\ \eqref{eq:ps-d2sigma} allows us to compute the event rates (using Eq.\ \eqref{eq:chi-event-rate})
and the mean inelasticity of the $ \chi N $ scattering process.
In Fig. \ref{xsec} we show the total deep inelastic $\chi N \rightarrow \chi N$ cross section and the average inelasticity ($\langle y \rangle$), and compare them with the $\nu N \rightarrow \nu N$ case \cite{Gandhi:1995tf,Gandhi:1998ri,1106.3723}.

\begin{figure}[htb!]
\centering
\includegraphics[scale=0.4]{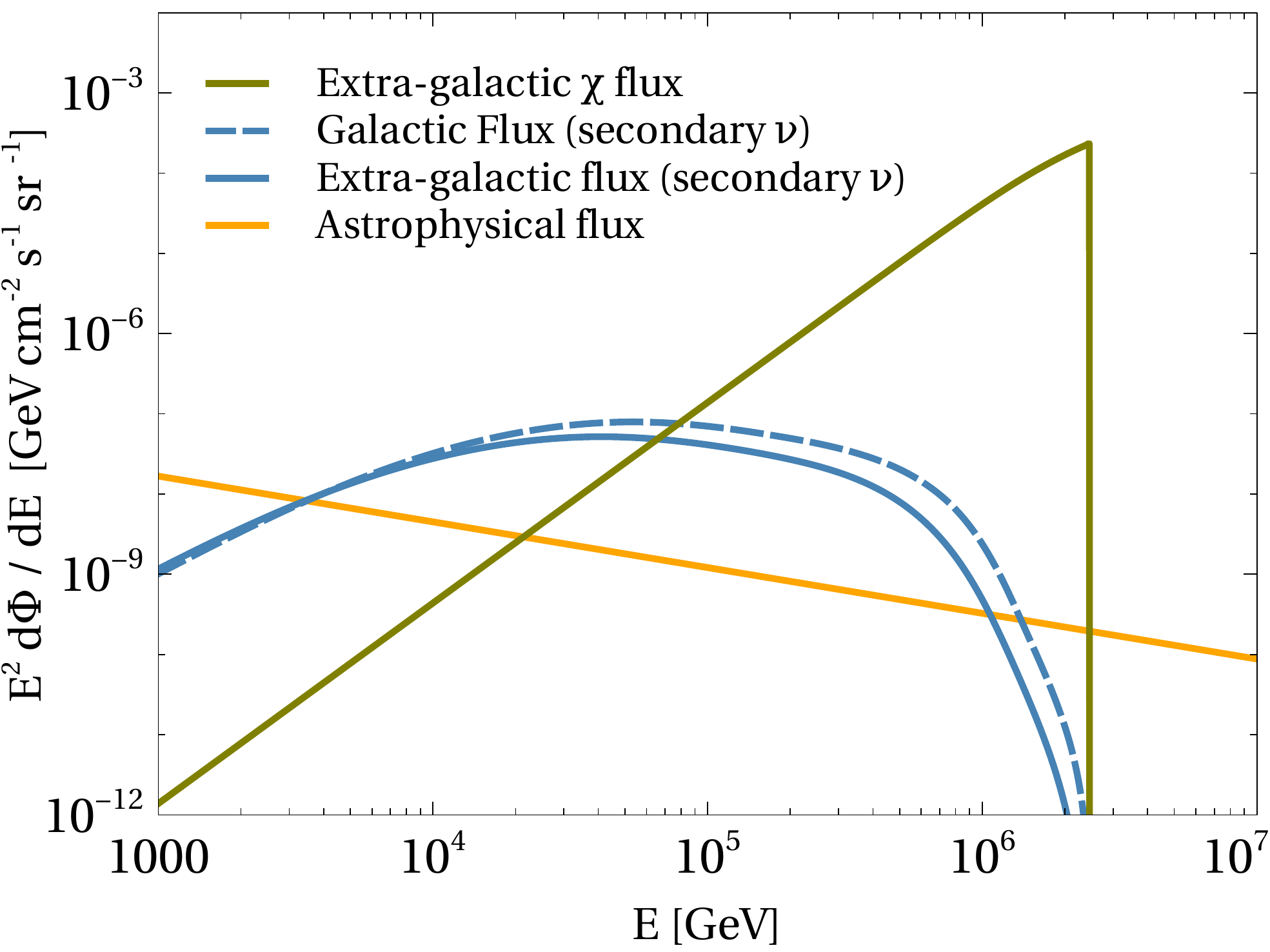}
\caption{Relevant fluxes that contribute towards the PeV and the sub-PeV events in Scenario I. The galactic $\chi$ flux is not shown since it originates from the two body decay of $\phi$, and is  given by the simple form in Eq. \ref{diracdelta}, unlike the extra-galactic  flux, which exhibits a $z$ dependance. The values of parameters used to calculate the fluxes are given in Table. \ref{tab1}.}
\label{flx}
\end{figure}

Fig.~\ref{flx} shows the individual flux components that contribute to  the PeV and the sub-PeV events  in Scenario I. This is a representative plot, and the parameters that were used while calculating the fluxes are the best-fit values shown in Table.~\ref{tab1}.

\begin{table}[htb]
\begin{center}
\begin{tabular} {c c c c c c}
\hline
Parameter & $m_a$ [GeV] & $g_{q}$ & $f_{\phi} g_\chi^2 / \tau_\phi\ [\text{s}^{-1}]$ & $\gamma$ & $\tilde{N}_{ast}$ (all flavour) \\
\hline
$a \rightarrow b \bar{b}$ & 12.0 & 0.32 & $1.23 \times 10^{-26}$ & 2.57 & $1.21 \times 10^{-9}$ \\
$a \rightarrow c \bar{c}$ & 5.3  & 0.50 & $5.02 \times 10^{-27}$ & 2.61 & $5.40 \times 10^{-9}$ \\
\hline
\end{tabular}
\end{center}
\caption{The best fit values of relevant parameters in case of a pseudoscalar mediator $a$, when it dominantly decays to $b\bar{b}$ and $c\bar{c}$ respectively. \tilnast\ is given in units of\flxunit.}
\label{tab1}
\end{table}

\begin{figure}[htb]
  \centering
  \includegraphics[width=\textwidth]{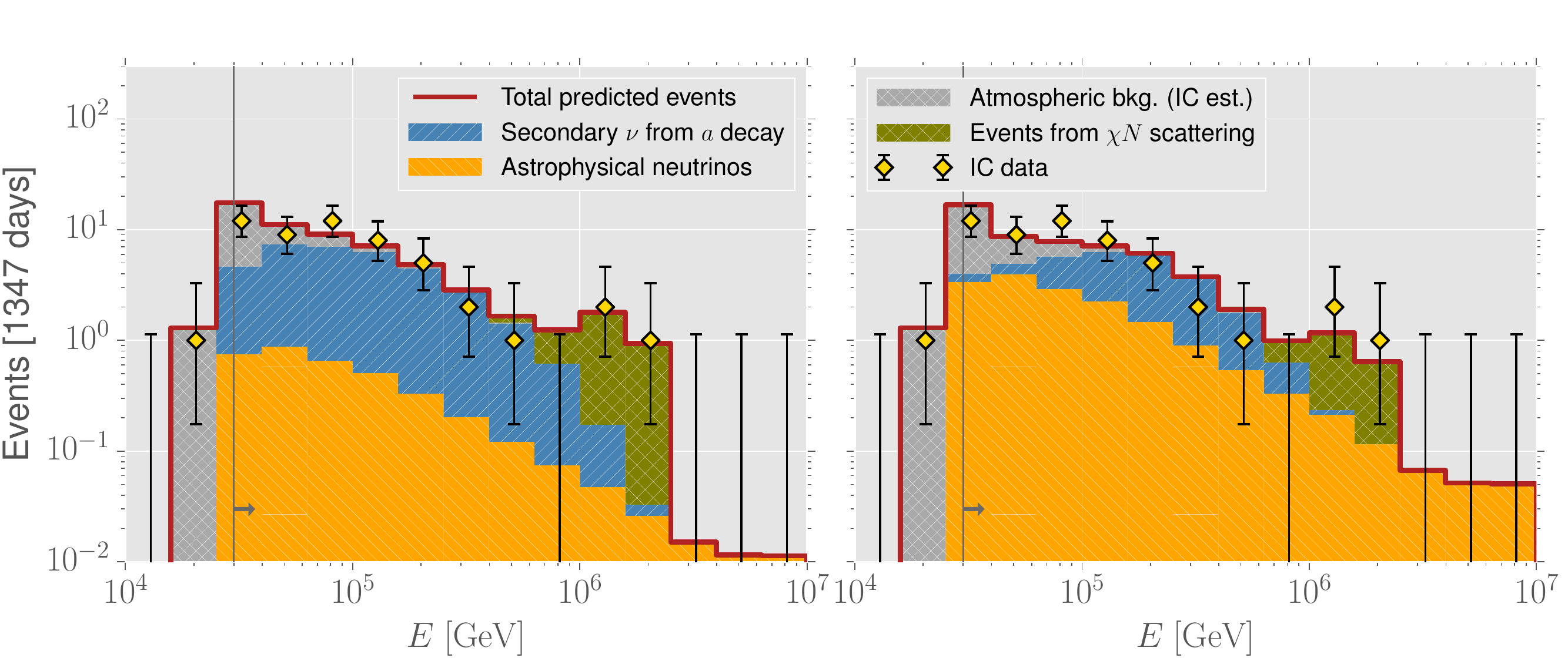}
  \caption{\label{fig:ps-fixedhdm}Best-fit events (stacked bars) from a combination of secondary $ \nu $'s, astrophysical $ \nu $'s and background in the sub-PeV energies, with LDM events explaining the PeV+ events. The best-fit value of $m_\phi = 5.34$ PeV. \textbf{Left}: Decays to $b\bar{b}$. \textbf{Right}: The mediator mass limited to below $ b\bar{b} $ production threshold, so that it can dominantly decay only to $c\bar{c}$ pairs.}
\end{figure}

As discussed previously, in Scenario I, the sub-PeV events depend on the mediator mass $m_a$, the ratio $f_\phi\,g_\chi^2 / \tau_\phi$  and  on the HDM mass $m_\phi$. The  three PeV events, on the other hand depend on $m_a$, the ratio $F = f_\phi g_\chi^2 g_q^2 / \tau_\phi$ and as well as on $m_\phi$.
Treating  the PeV events as arising from two-body decay of the $\phi$ to $\chi\bar{\chi}$ using  gives us $ m_\phi \simeq 5.3 $ PeV.
A major fraction of the sub-PeV events arise from the secondary neutrino flux, and for this  we carry out calculations in two different
kinematic regions:\begin{inparaenum}[\itshape a\upshape)]
\item where the mediator mass lies above the $ b\bar{b} $
      production threshold, and
\item where it  lies below this threshold, making $ c\bar{c} $ the main decay mode.
\end{inparaenum}
The results for best fits to the data using events from  all of the above fluxes,  and considering both kinematic regions,  are shown in Fig.\ \ref{fig:ps-fixedhdm}. The solid red line represents the total of the contributions from the various fluxes, and we find that it provides a good description to the data across the energy range of the sample. The 
best fit values of the parameters are given in Table \ref{tab1}.
The corresponding normalisation of the astrophysical flux is shown in
terms of the flux at the 100 TeV bin $ \tilde{N}_\text{Ast} = E^2\Phi_\text{Ast} \vert_{100\text{ TeV}}\!\flxunit$.

We note the following features of Fig \ref{fig:ps-fixedhdm}, which also conform to emergent features of IC data:
\begin{itemize}
\item{The secondary neutrino event spectrum has a shape that would allow it to account for a `bump', or excess, such as presently seen in the vicinity of 30--100 TeV.}
\item{The astrophysical neutrino contribution, especially in the $b\bar{b}$ case, is not a major component. This is unlike the standard situation where only astrophysical neutrinos account for events beyond  $30$ TeV, requiring a flux very close to the Waxman-Bahcall bound.}
\item{ A dip in the region 400--1000 TeV occurs naturally due to the presence of  fluxes of different origin in this region.}
\item{Over the present exposure period, no  HESE events are expected in the region beyond 2--3 PeV, since the only contributing flux here is the astrophysical flux, which is significantly lower in this scenario as opposed to the IC best-fits. With more exposure, some astrophysical events can be expected to show up in this region. }
\end{itemize}

\subsubsection{Parameter correlation analyses}

It is useful to examine the parameter space for Scenario I allowed by IC data. We use the case of a pseudo-scalar mediator as representative, and  examine the correlations and degeneracies between the parameters.
We give contour plots between pairs of parameters for
each of the LDM decay scenarios considered above, $\ie$ for decay to $b\bar{b}$ and to $c\bar{c}$. Noting that the  sub-PeV events in the HESE sample that do not have their origin in the atmosphere are, in our scenario, either from the secondary neutrino flux or from the astrophysical (power-law) neutrino flux, we denote the total number (in the 1347-day sample) of the former by  $N_\text{DM}$, and that of the latter by $ N_\text{Ast} $.

For each case we start with the best-fit values  obtained in the previous section for each of the parameters in the set:
$ \lbrace N_\text{DM}, m_a, N_\text{Ast}, \gamma, m_\phi, g_q \rbrace $.
 We note that $N_\text{DM}$  is proportional to $ (f_\phi \, g_\chi^2)/\tau_\phi $, whereas the primary 
DM component of the event spectrum, coming from $ \chi $ scattering off ice nuclei
at PeV energies is related to $ m_\phi $, $ f_\phi \, (g_\chi g_q)^2/\tau_\phi $ and $ m_a $.
For a fixed $ \gamma $, specifying the $ N_\text{Ast} $ is tantamount to
specifying the overall astrophysical flux normalisation $ A $ in the uniform power-law spectrum
$ \Phi_\text{Ast} = A E^{-\gamma}$.

\begin{figure}[htb]
	\centering
	\begin{subfigure}{0.49\textwidth}
		\centering
		\includegraphics[width=0.99\textwidth]{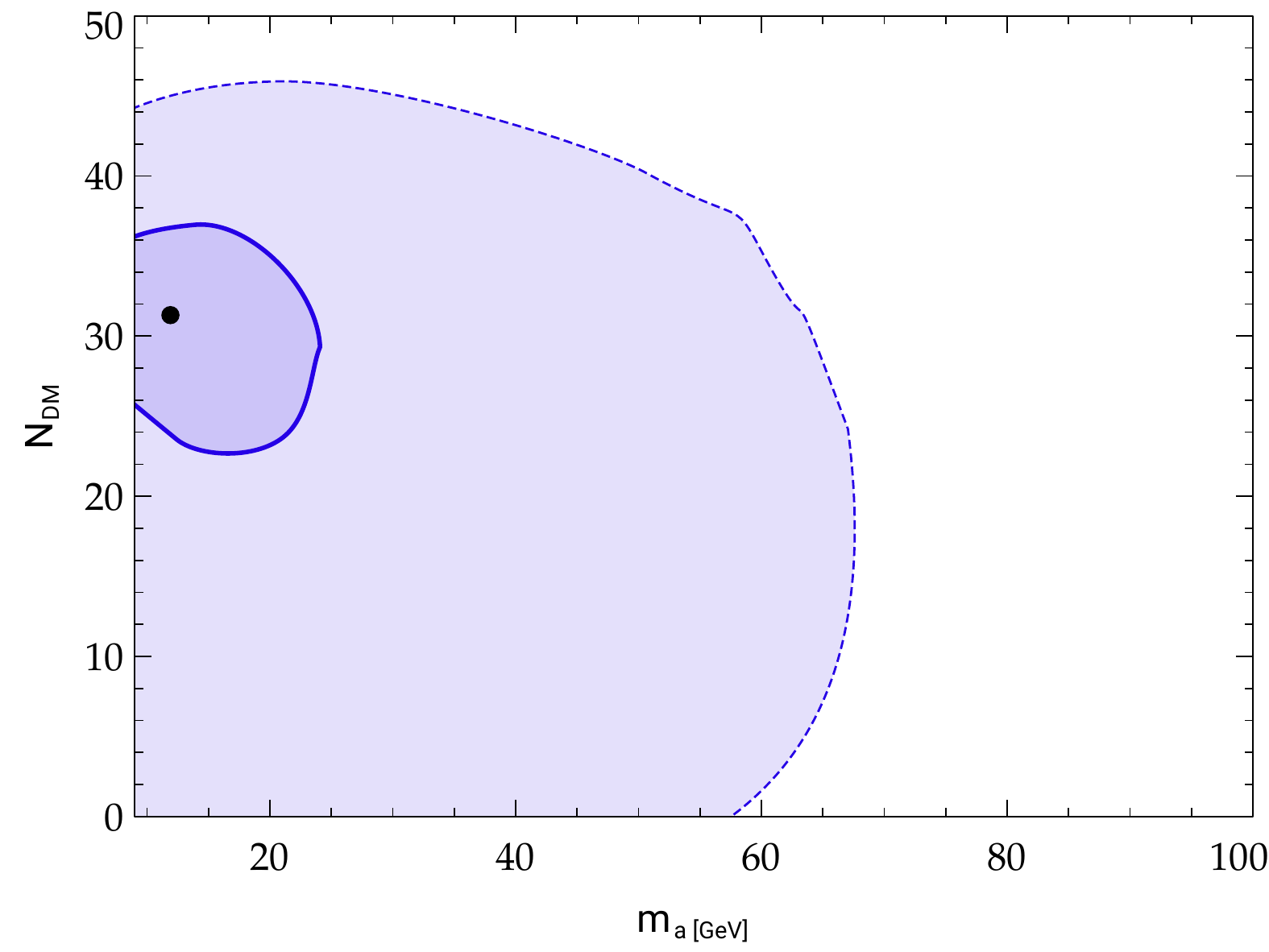}
	\end{subfigure}
	\begin{subfigure}{0.49\textwidth}
		\centering
		\includegraphics[width=0.99\textwidth]{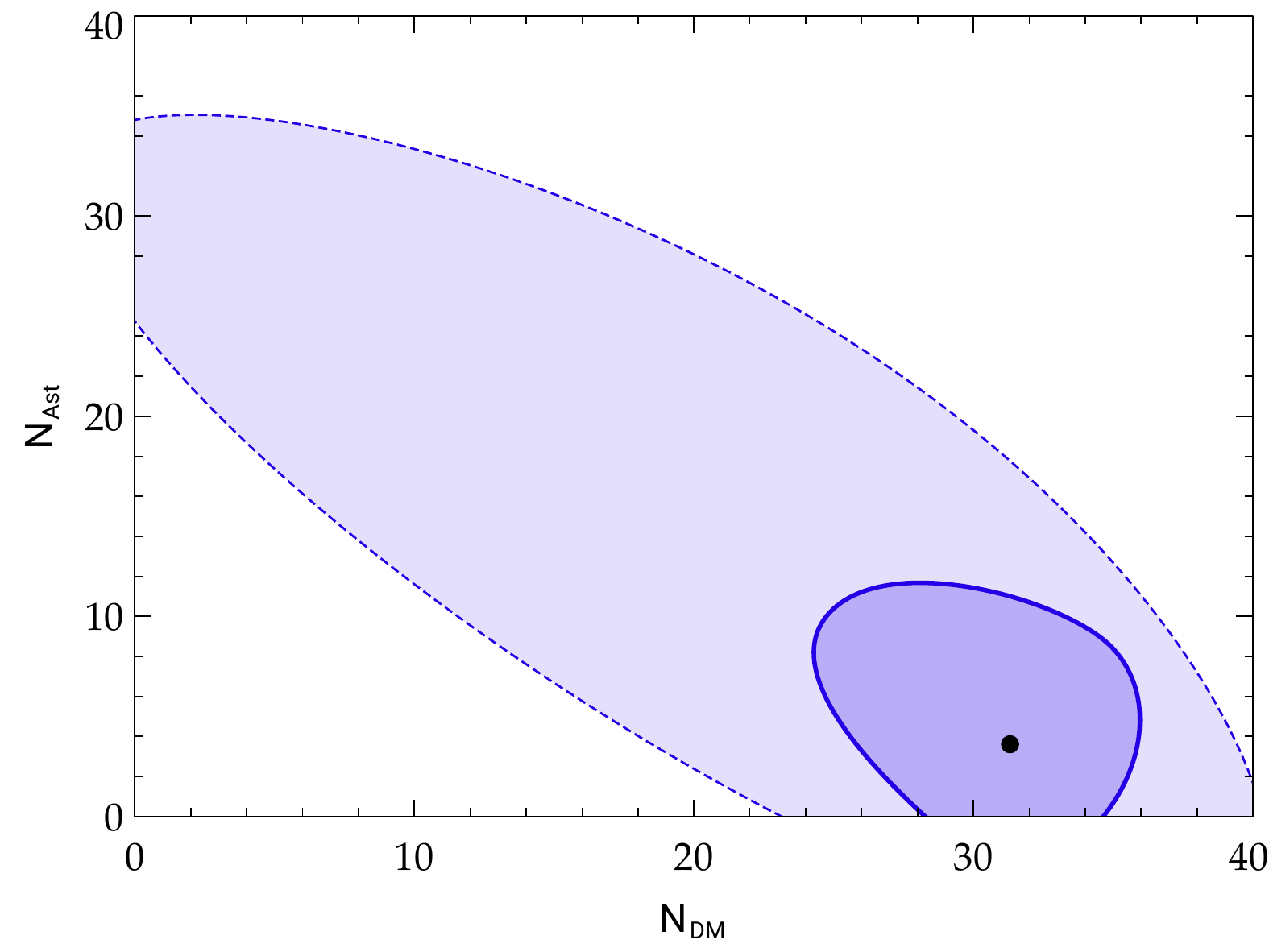}
	\end{subfigure}
	\caption{\label{fig:bbbar-cntr1}$ 1\sigma $ and $ 3\sigma $ allowed regions for
		parameters $ N_{\dm} $ and $ m_a $ (left) and $ N_{\dm} $ and $ N_\text{Ast} $
		(right) for mediator decays to $ b\bar{b} $. The solid dot in each case
		represents the corresponding best-fit point in the parameter subspace.
	}
	\label{cntr2}
\end{figure}

\begin{figure}[htb]
	\centering
	\begin{subfigure}{0.49\textwidth}
		\centering
		\includegraphics[width=0.99\textwidth]{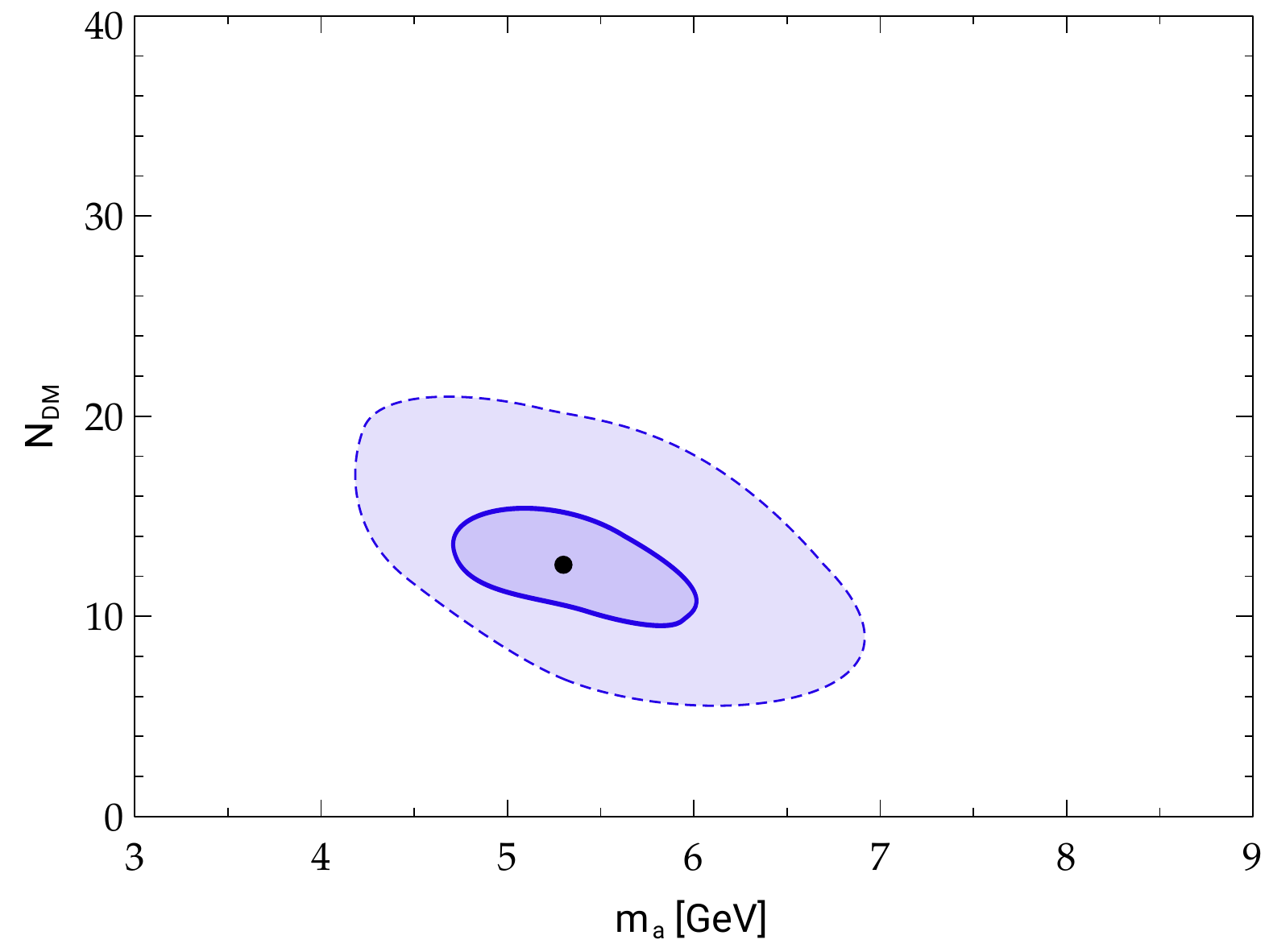}
	\end{subfigure}
	\begin{subfigure}{0.49\textwidth}
		\centering
		\includegraphics[width=0.99\textwidth]{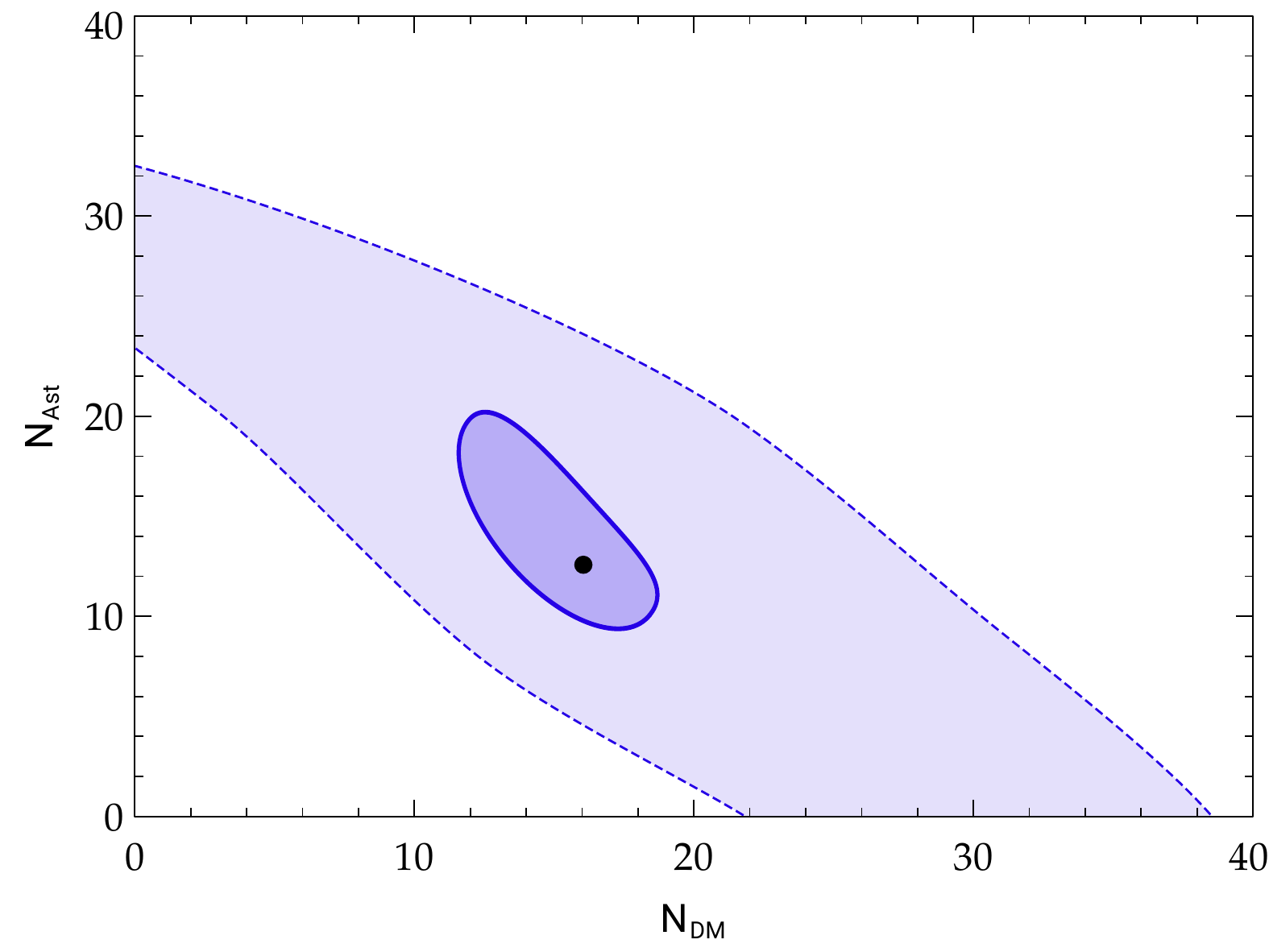}
	\end{subfigure}
	\caption{\label{fig:ccbar-cntr1}$ 1\sigma $ and $ 3\sigma $ allowed regions for
		parameters $ N_{\dm} $ and $ m_a $ (left) and $ N_{\dm} $ and $ N_\text{Ast} $
		(right) for mediator decays to $ c\bar{c} $. The solid dot in each case
		represents the corresponding best-fit point in the parameter subspace.
	}
\label{cntr1}
\end{figure}

\begin{figure}[t]
	\includegraphics[width=0.48\textwidth]{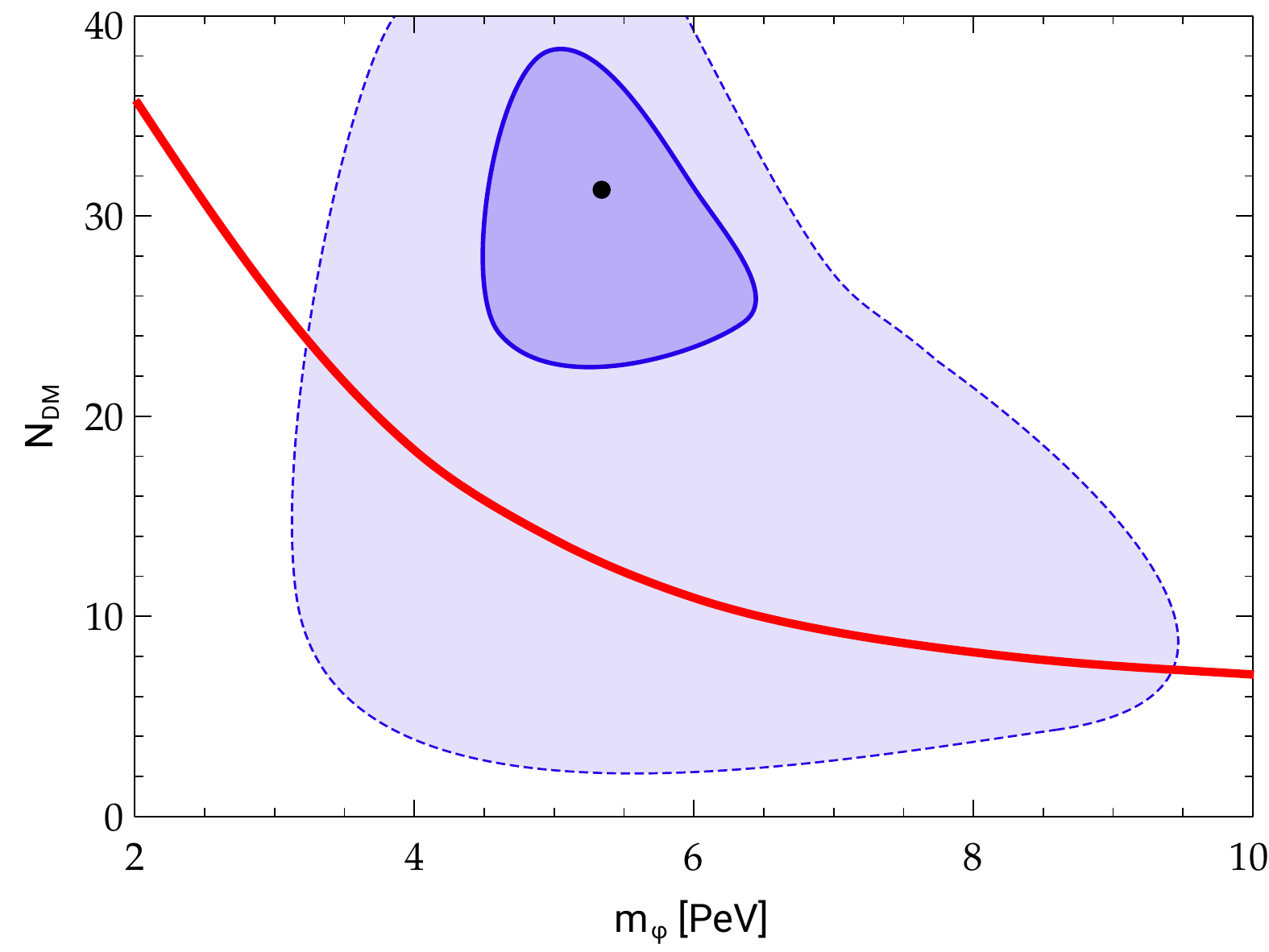}
	\includegraphics[width=0.48\textwidth]{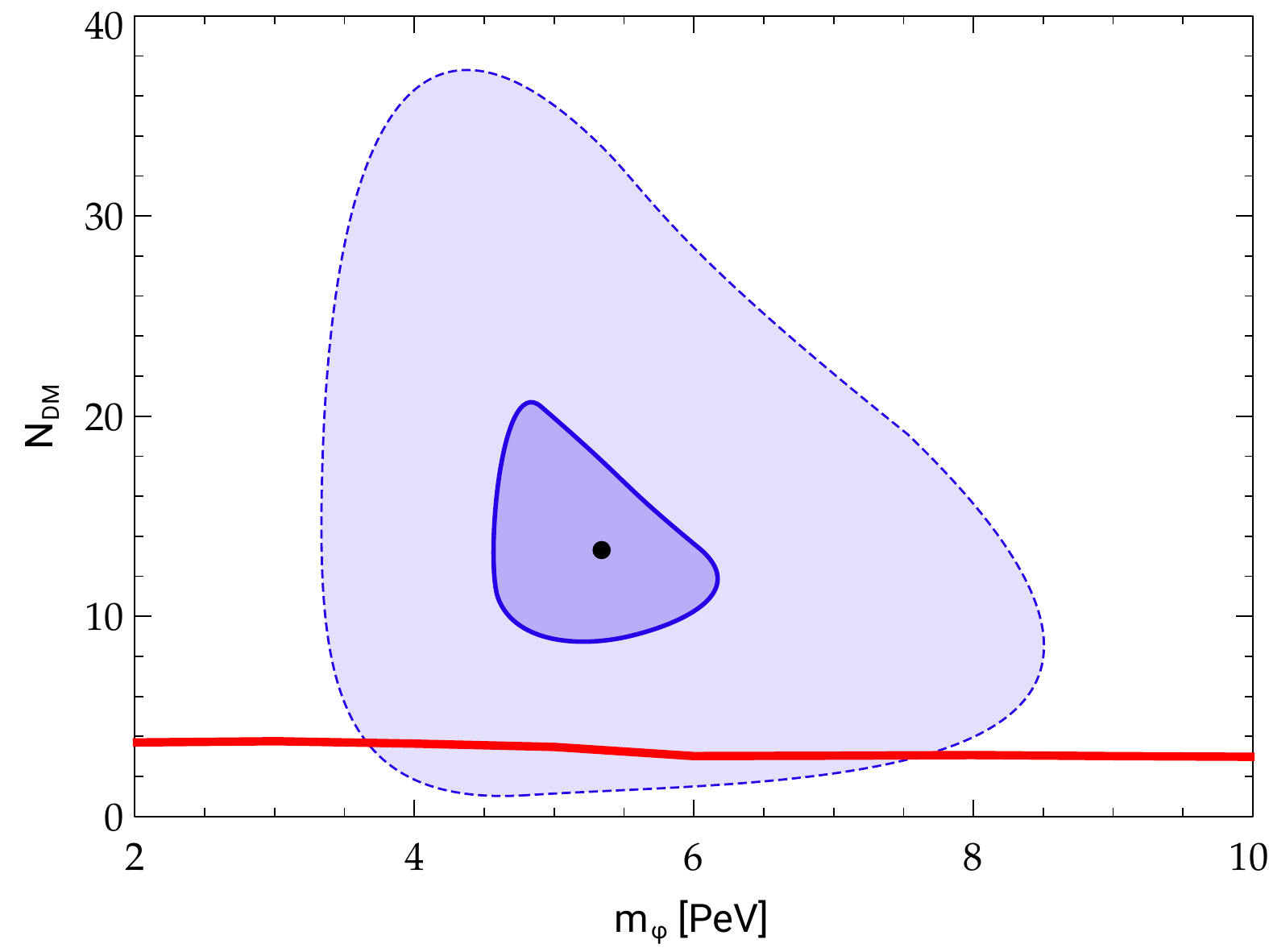}
	\caption{Plot showing allowed regions satisfying gamma ray constraints in the case when pseudoscalar mediator decays to $b\bar{b}$ (Left) and  to $c \bar{c}$ (Right). Regions above the red line are constrained by observations of the diffuse gamma ray flux.}
	\label{gammabnds}
\end{figure}

The total number of signal events observed in the 1347-day  IC sample is 35 at its best-fit value,
with a $ 1\sigma\ (3\sigma) $ variation of 29--42 (20--57).
This assumes the conventional atmospheric background is at the expected best-fit, and
the prompt background is zero.
Selecting two parameters for each analysis, we vary their values 
progressively from their best-fits,  while marginalizing over the other parameters
over their allowed $ 1\sigma $ ($ 3\sigma $) ranges.
For each pair of the chosen two-parameter subset, we compute the
$ \Delta\chisq(p_a, p_b) = \chisq(p_a, p_b) - \chisq_\text{b.f.}$
where $ p_a,p_b $ represent the value of the two chosen parameters in the iteration.
With the resulting $ \Delta\chisq $ we plot $ 1\sigma $ and $ 3\sigma $
contours enclosing the allowed variation of these parameters (Fig.\ \ref{cntr2}, Fig.\ \ref{cntr1} and Fig.\ \ref{gammabnds}). Due to the sparse statistics presently available, the 3$\sigma$ allowed regions in these plots permit the IC data to be fit well for a wide range of values of the chosen variables.

%
\begin{figure}[t]
	\includegraphics[width=0.48\textwidth]{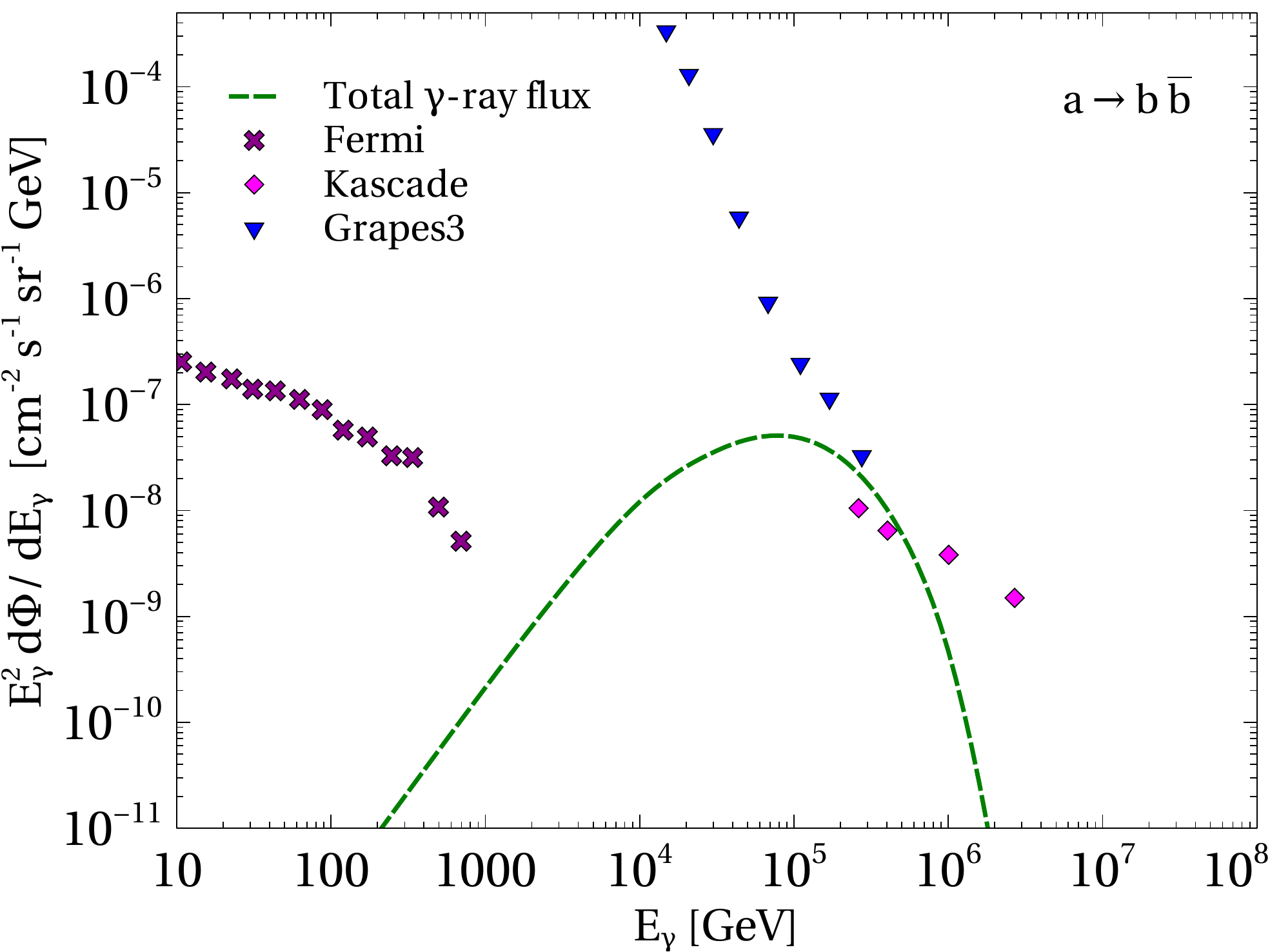}
	\includegraphics[width=0.48\textwidth]{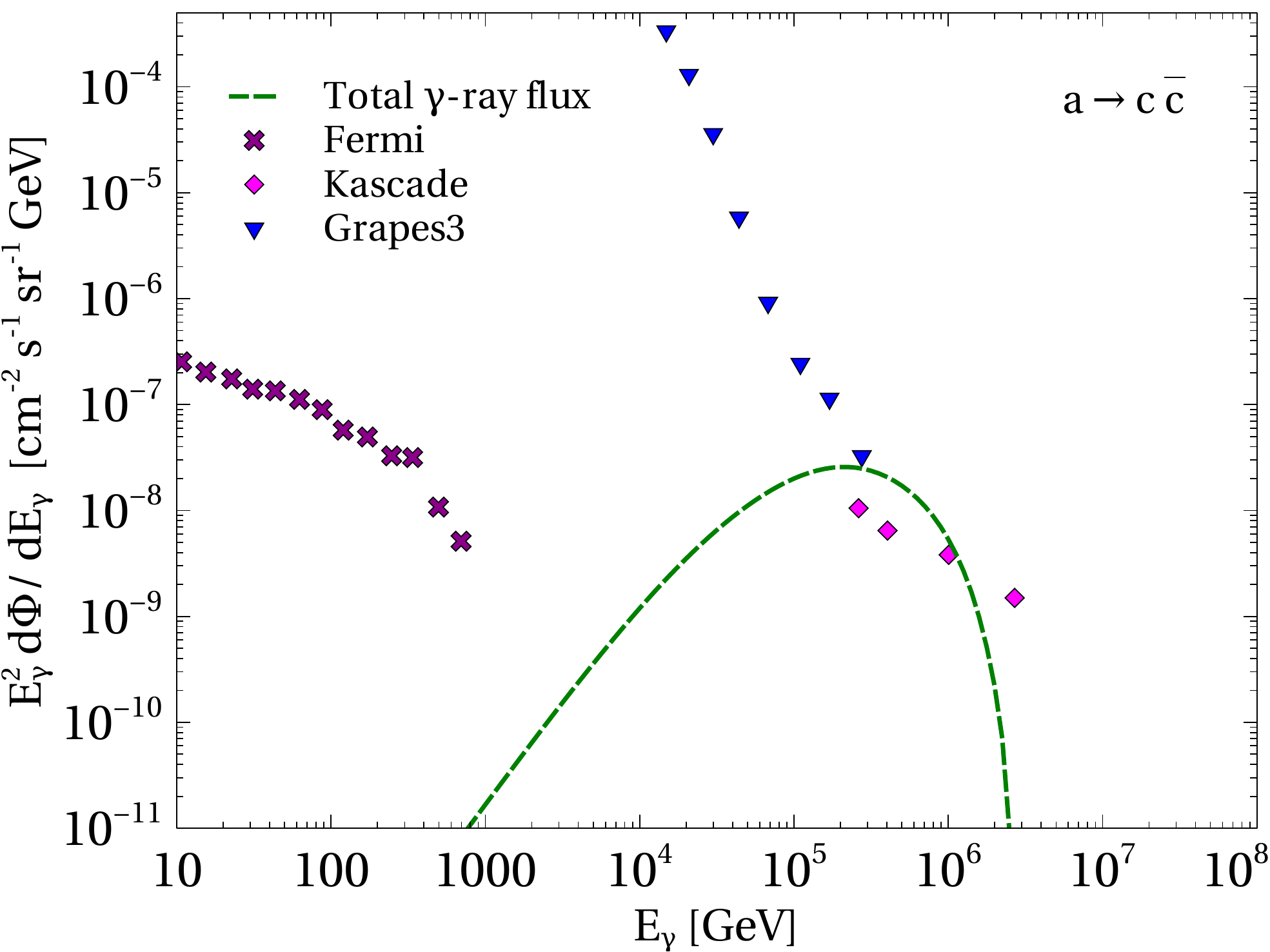}
	\caption{ \small Diffuse gamma-ray flux for the best-fit parameter choice in the pseudo-scalar mediator scenario, where the mediator $a$ dominantly decays to $b \bar{b}$ (left) and $c \bar{c}$ (right). The current constraints from Fermi-LAT data~\cite{Ackermann:2014usa} at lower energies, and cosmic ray air shower experiment (KASCADE~\cite{Feng:2015dye} and GRAPES-3~\cite{Gupta:2009zz}) data at higher energies are also shown.}
	\label{fig:gamma_PS}
\end{figure}

\paragraph*{}
Following the discussion in Sec.\ \ref{ssec:gen-constraints}, the only major constraint on the parameters in the pseudo-scalar mediator scenario stems from the upper bound on diffuse gamma-ray fluxes, while the current collider constraints restrict the values of the couplings to $\mathcal{O}(0.1)$ values.
The sum of the galactic and extragalactic gamma ray fluxes corresponding to the best fit parameter points are shown in Fig.\ \ref{fig:gamma_PS}. 
They are  compared with both the Fermi-LAT data~\cite{Ackermann:2014usa} at lower energies, and cosmic ray air shower experiment (KASCADE~\cite{Feng:2015dye} and GRAPES-3~\cite{Gupta:2009zz})  data at higher energies.
These constraints significantly restrict the available parameter-space,
and, indeed, our best-fit values for the \ndm\ lie in a disfavoured region.
We find, however, that a reasonable region of the allowed $ 3\sigma $ parameters-space is 
nonetheless consistent with these constraints, and that the allowed region for $b \bar{b}$ is larger than that for $c \bar{c}$.
Fig.\ \ref{gammabnds} reflects these conclusions.

\subsection{Scalar mediator}
\label{ssec:scalar}
In this section we explore the case when the mediator $a$ in Scenario I is a scalar. 
The relevant double differential $\chi N$ scattering cross-section in this case is given by:
\begin{align}
\dfrac{d^2\sigma}{dxdy} = &\sum_{q}\dfrac{1}{32\,\pi}\dfrac{E_\chi}{x \, M_N (E_\chi^2 - m_\chi^2)}
\dfrac{(g_\chi \, g_q)^2}{(Q^2 + m_a^2)^2} \nonumber \\
                       & \times \left[16m_\chi^2m_q^2+(Q^2)^2+4Q^2(m_\chi^2+m_q^2)\right]\,f_q(x,Q^2)
\end{align}
where, the various quantities used are as before (Eq.\ \eqref{eq:ps-d2sigma}).

The parameter values at the best-fit point are shown in Table~\ref{tab2}, and we show the corresponding event rates in Fig {\ref{scalarEv}. It is interesting to note that, compared to the pseudo-scalar case, due to the additional terms contributing to the differential $\chi N$ scattering cross-section (in particular, the $4Q^2m_\chi^2$ term), the best fit value for $g_q$ turns out be smaller in the scalar case, while rest of the relevant parameters take similar values.

\begin{table}[htb]
\begin{center}
\begin{tabular} {c c c c c c}
\hline
Best fit parameters & $m_a$ [GeV] & $g_{q}$ & $f_{\phi} g_\chi^2 / \tau_\phi\ \left[\text{s}^{-1}\right]$ & $\gamma$ & $\tilde{N}_{ast}$ (all flavour)\\
\hline
$a \rightarrow c \bar{c}$ & 5.3 & 0.29 & $4.88 \times 10^{-27}$ & 2.63 & $5.41 \times 10^{-9}$\\
\hline
\end{tabular}
\end{center}
\caption{The best fit values of relevant parameters in the case of a scalar mediator $a$, when it decays dominantly to $c\bar{c}$. The best fit value of $m_\phi$ here is $\sim$ 5.3 PeV. \tilnast\ is given in terms of\flxunit.}
\label{tab2}
\end{table}

\begin{figure}[htb]
  \centering
  \includegraphics[width=0.7\textwidth]{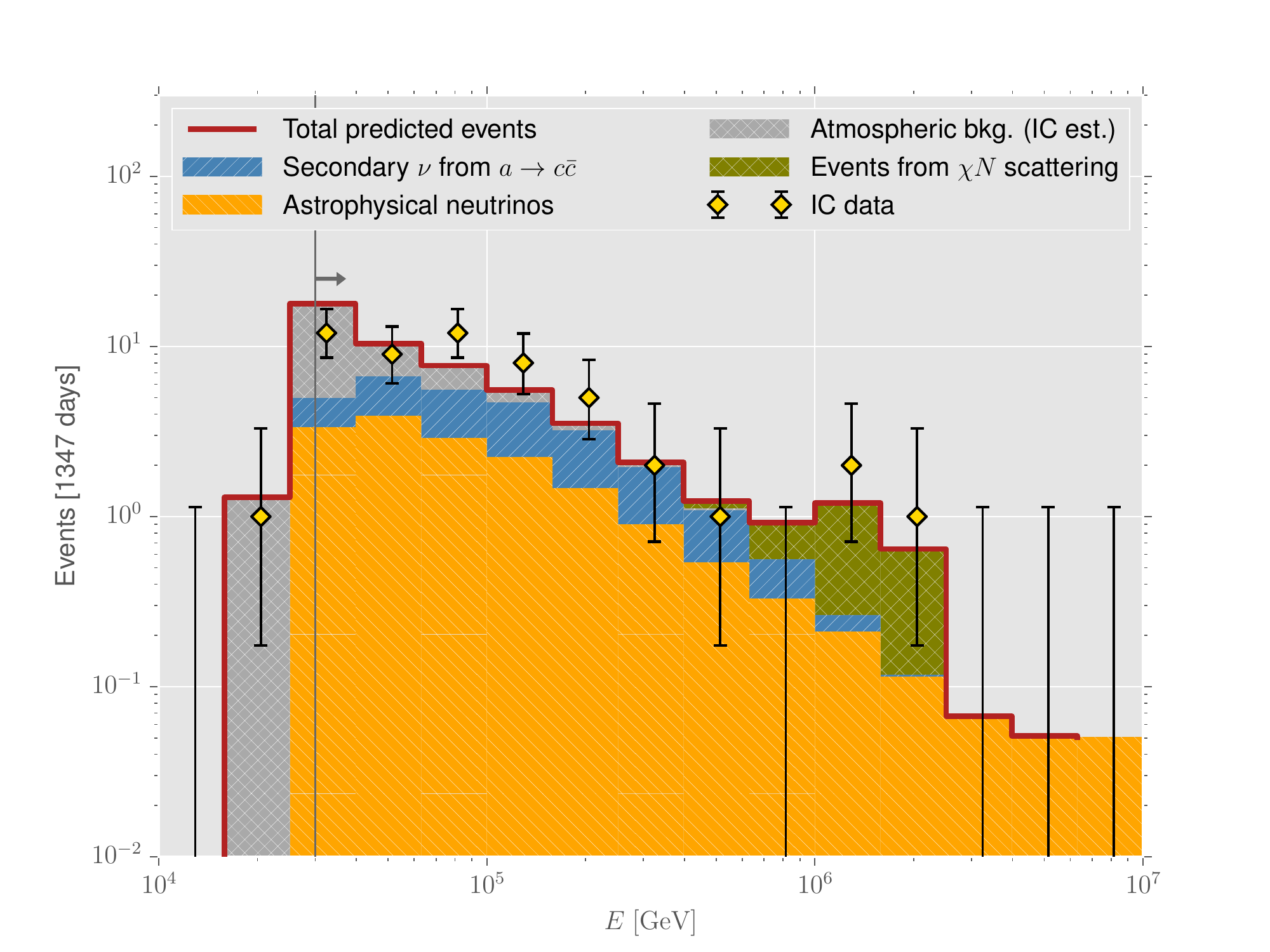}
  \caption{\label{scalarEv}Same as Fig.\ \ref{fig:ps-fixedhdm},
           for the scalar mediator scenario, with the mediator dominantly decaying to $ c\bar{c} $.
          }
\end{figure}

The gamma-ray constraints on the scalar mediator case are found to be similar to the pseudo-scalar case, and as discussed in Sec.~\ref{ssec:gen-constraints}, the collider constraints on the coupling parameters are also of similar magnitude. As further explained in Sec.~\ref{ssec:gen-constraints}, although we restrict ourselves to regions of parameter space where $f_\chi$ is very small, for parameter values where $f_\chi$ becomes appreciable, there are additional constraints from relic density requirements as well as direct detection bounds. The spin-independent direct detection bounds in particular are very stringent in the scalar mediator scenario, unless the DM mass lies below $\mathcal{O}(10 {~\rm GeV})$, where the nuclear-recoil experiments lose sensitivity. Overall, stronger constraints notwithstanding, we find that the best-fit point lies in an allowed region of the parameter space, and provides an excellent fit to the data, with  explanations for the observed features identical to those described in the last subsection.

\subsection{Vector and axial-vector mediators}

The double differential cross section in the case of a vector mediator is given by:
\begin{align}\label{eq:v-d2sigma}
\dfrac{d^2\sigma}{dxdy} =& \sum_{q}\dfrac{1}{32\,\pi}\dfrac{1}{x \, M_N \, E_\chi}
\dfrac{(g_\chi \, g_q)^2}{(Q^2 + m_{Z^\prime}^2)^2} \nonumber \\
                        & \times \Bigg(\dfrac{(Q^2)^2}{2} + s^2-s \, Q^2\Bigg)\,f_q(x,Q^2).
\end{align}
where, $g_q$ is the coupling of $Z^\prime$ to the quark $q$, and $s \approx 2\, x E_{\chi} M_N $.

\paragraph*{}
To evade the strong bounds particular to vector (and axial-vector) mediators coming from dijet resonance searches in collider experiments, as discussed in Sec.\ \ref{sssec:collider-cons}, we impose a penalty on the $ \chisq $ computation whenever the combination of the coupling constant and $ M_{\zp} $ extends into a region disfavoured at more than 90\% confidence level.
Once we have thus determined the allowed region of the parameter space,
we show the results (Fig. \ref{vecEV}) corresponding to a benchmark point in this space, defined by the values in Table\ \ref{tabvec},
that maximises the contribution from
secondary neutrinos from DM decay (Flux-3), and correspondingly deems the
astrophysical neutrino component insignificantly small (which we consequently do not show).
An increased flux for the latter can be accommodated by a corresponding scaling down of the value of $ f_{\phi} g_\chi^2 / \tau_\phi $
and so on.
\begin{table}[htb]
\begin{center}
\begin{tabular} {c c c c c}
\hline
Benchmark Values & $M_{Z^{\prime}}$ [GeV] & $g_{q}$ & $f_{\phi} g_\chi^2 / \tau_\phi\ \left[\text{s}^{-1}\right]$ \\
\hline
$Z^{\prime} \rightarrow q \bar{q}$ & 20 & $3.3\times10^{-3}$ & $2.5 \times 10^{-27}$ \\
\hline
\end{tabular}
\end{center}
\caption{Benchmark values of relevant parameters in the case of a vector mediator $Z^{\prime}$, when it decays to all possible $q\bar{q}$ pairs. The value of $m_\phi$ used here is $\sim$ 5.0 PeV. As noted in the text, we have chosen a benchmark point in the parameter space that maximises the secondary $ \nu $ contribution
from DM decay, and consequently deems the astrophysical flux negligible. The latter has therefore not been shown here.}
\label{tabvec}
\end{table}

As seen in Fig.~\ref{vecEV}, unlike the pseudo-scalar and the scalar cases, we note that the galactic and the extra galactic secondary flux events remain approximately flat with decreasing  energy below $\approx 1$ PeV. This results in the absence of a dip or deficit in the region 400 TeV--1 PeV which is one of the features of the present IC data that we would like to reproduce in Scenario I. This can be mitigated by increasing the mass of the mediator (see Fig \ref{vec2}).
 A comparison with the pseudoscalar mediator event spectrum, where this problem is absent, is shown for a fixed mass, in the right panel Fig. \ref{vec2}.\\

We now turn to the relevant gamma-ray constraints, along the same lines
we studied it for the case of a pseudo-scalar mediator.
While the differential three-body decay width of the HDM follows somewhat different distributions for different choices of mediator spin and CP properties,  the very large boost of the mediator particle washes out these differences to a large extent, and we arrive at a similar spectral shape as discussed for the spin-0 mediators above.   We find that the corresponding constraints are not severe, but may have mild tension in some energy regions. As far as relic density and spin-independent direct detection bounds are concerned, similar considerations as in the scalar mediator case would also apply to the vector mediator scenario, and we refer the reader to the discussion in Sec.~\ref{ssec:scalar}. 

\begin{figure}[htb]
\centering
\includegraphics[scale=0.6]{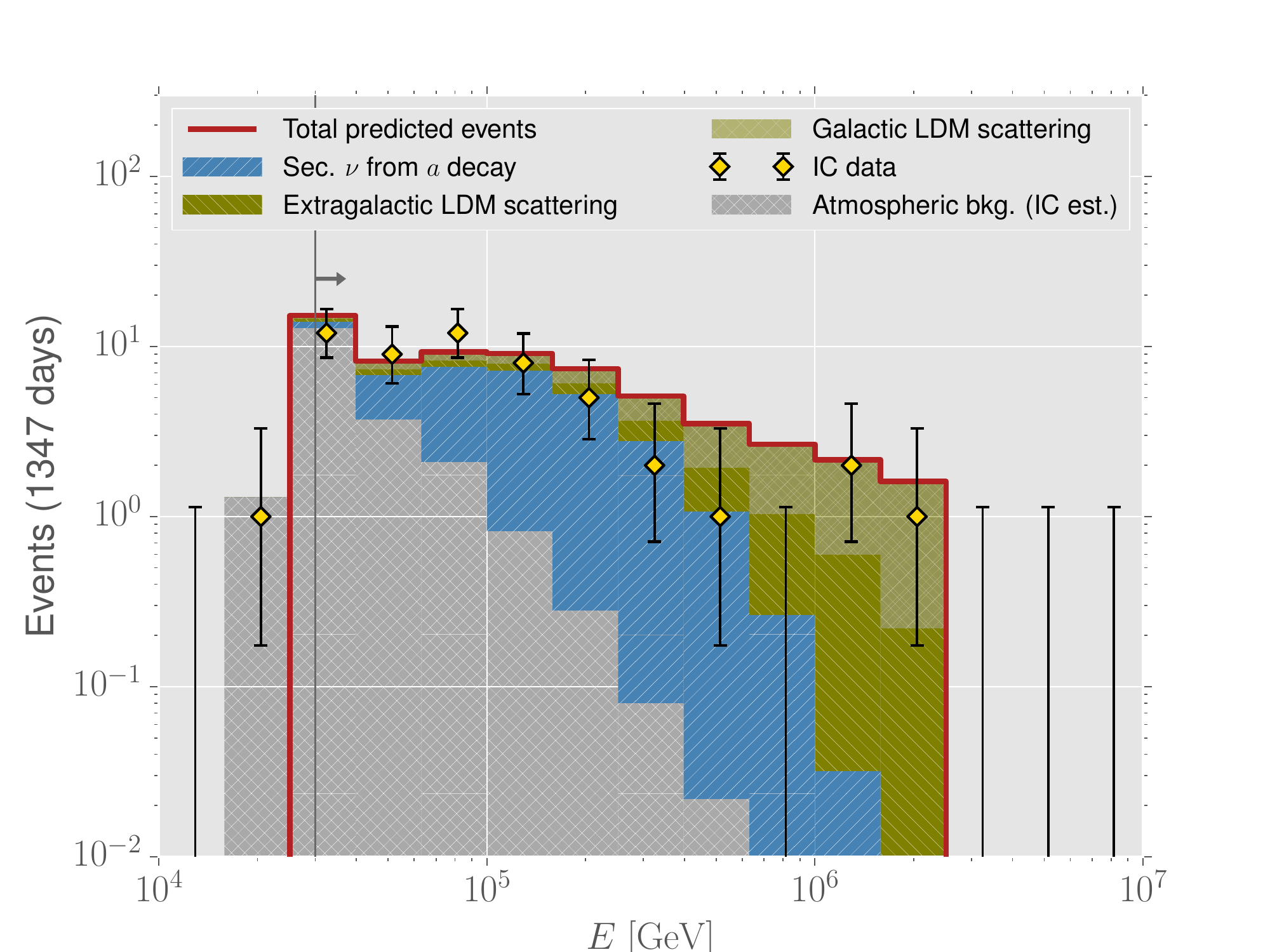}
\caption{\label{vecEV}Event rates for the benchmark parameter values shown in Table\ \ref{tabvec}.
         In keeping with the description in text, the correspondingly tiny number of events from the astrophysical flux have not been shown here.}
\end{figure}

\begin{figure}
\centering
\includegraphics[width=0.48\textwidth]{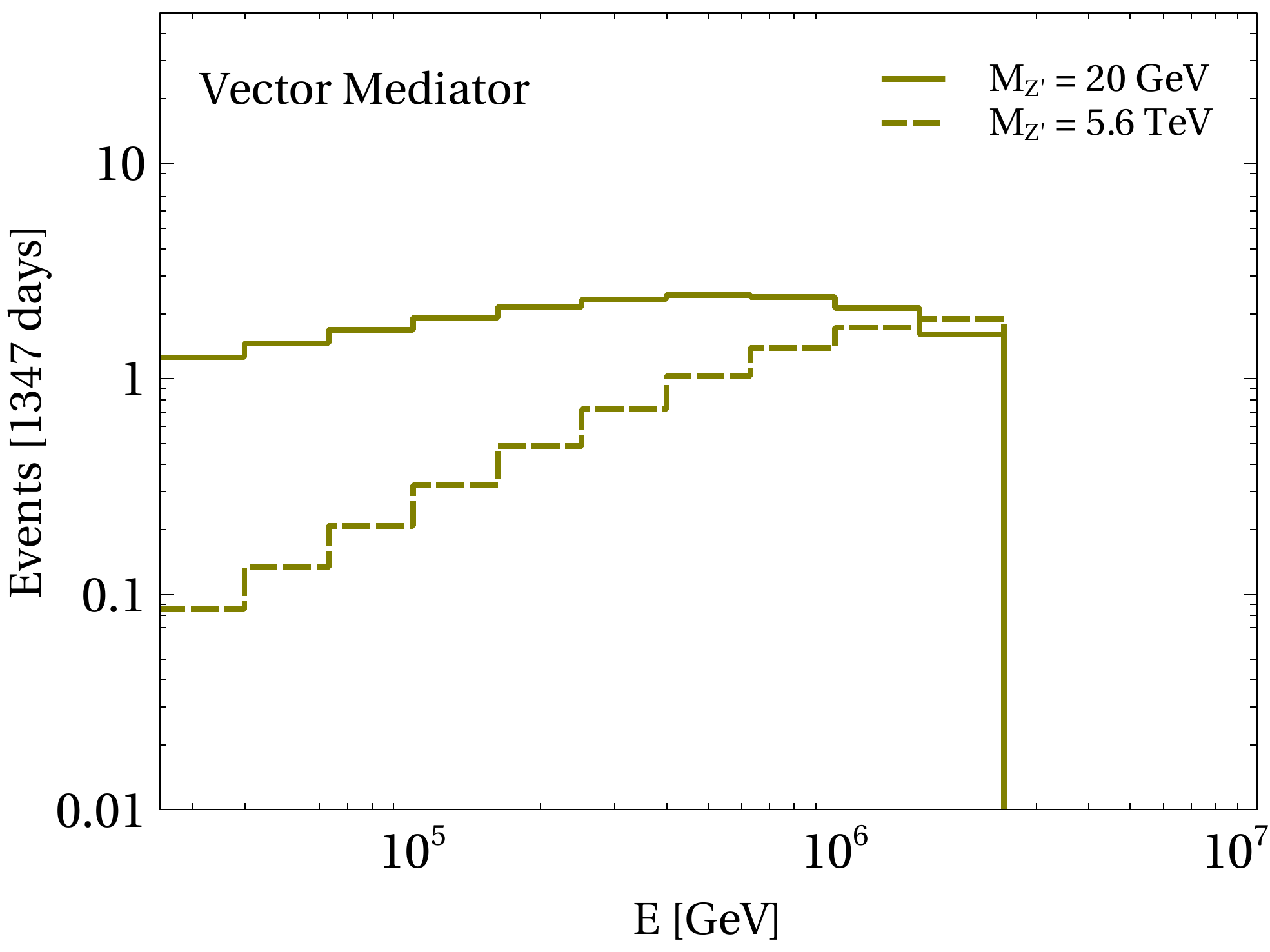}~
\includegraphics[width=0.48\textwidth]{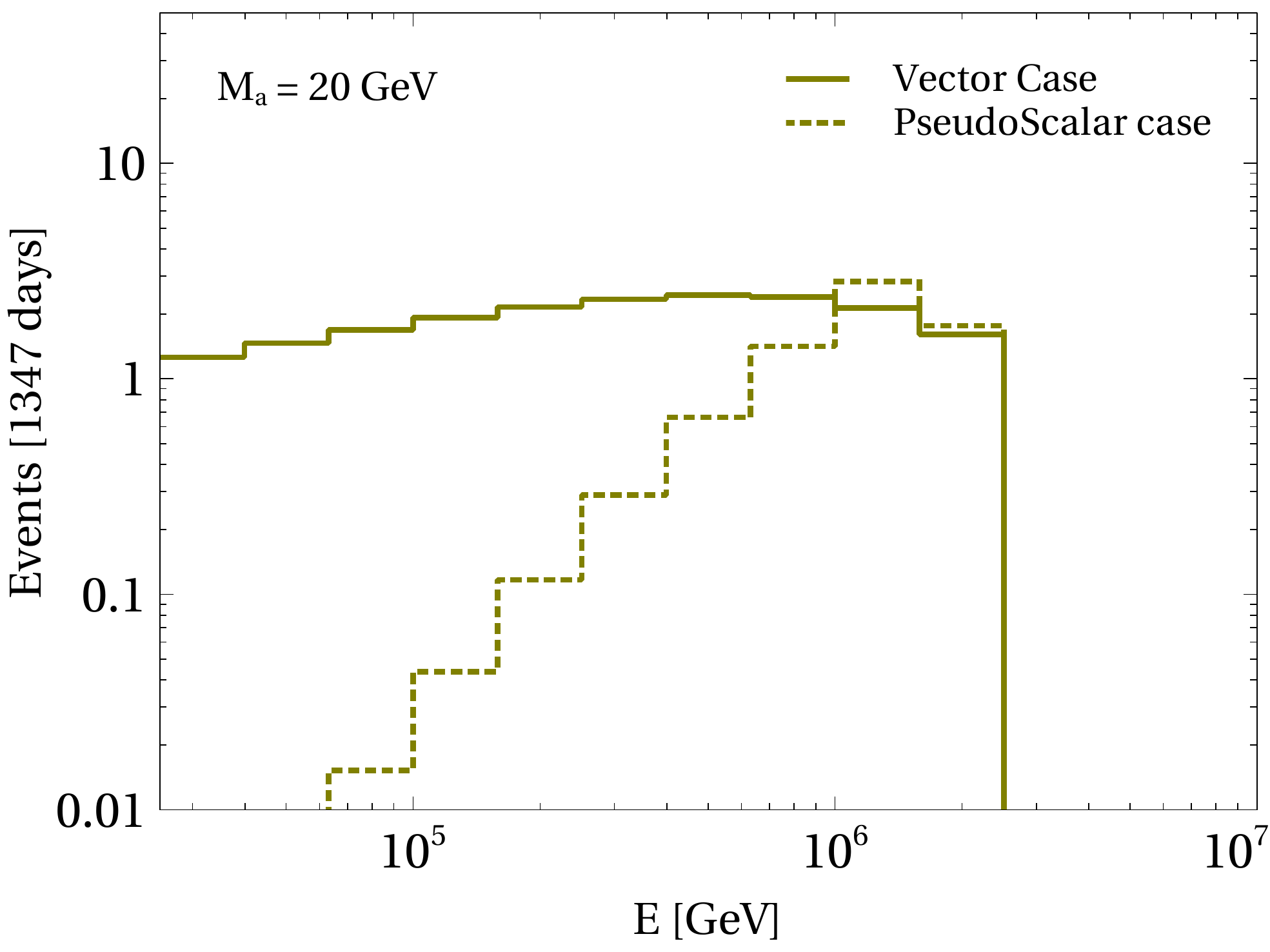}
\caption{{\bf Left:} PeV events in the vector mediator scenario, with different choices for the $Z^\prime$ mass. A larger value of the $Z^\prime$ mass is more likely to explain the {\it dip} at around PeV. {\bf Right:} PeV events in vector and pseudoscalar case with a mediator mass fixed to 20 GeV. The pseudoscalar scenario, as discussed earlier, explains the {\it dip} more accurately because of it's  sharply falling event rates, unlike in the vector scenario.}
\label{vec2}
\end{figure}

\paragraph*{}
Even though the differential $\chi N$ cross-section behaves similarly in the vector and axial-vector scenarios (in small $m_\chi$ and $m_q$ limit), there are additional important considerations particular to the axial-vector case
that limit the available parameter space very stringently. As explained earlier, in order to accommodate the PeV events by $\chi N$ DIS scattering, we require that the three body decay width of the HDM is much smaller than its two body decay width. However, as shown in Fig.~\ref{br3}, the three-body branching ratio starts to dominate for $g_\chi$ values as low as $0.01$ in the axial-vector case, whereas for scalar, pseudo-scalar or vector mediators, the three-body branching ratio becomes large only for $g_\chi \geq 1$. Thus, since the  PeV event rate  is proportional to $ g_\chi^2 g_q^2 f_\phi / \tau_\phi$, to obtain the required number of events in the PeV region, the value of $g_q$ needs to be pushed higher than its perturbative upper bound of $4\pi$. Ultimately, we find that it is not possible  to fit both the PeV and  the sub-PeV events  while simultaneously satisfying the perturbativity requirement for an axial-vector mediator.

\begin{figure}
\centering
\includegraphics[scale=0.4]{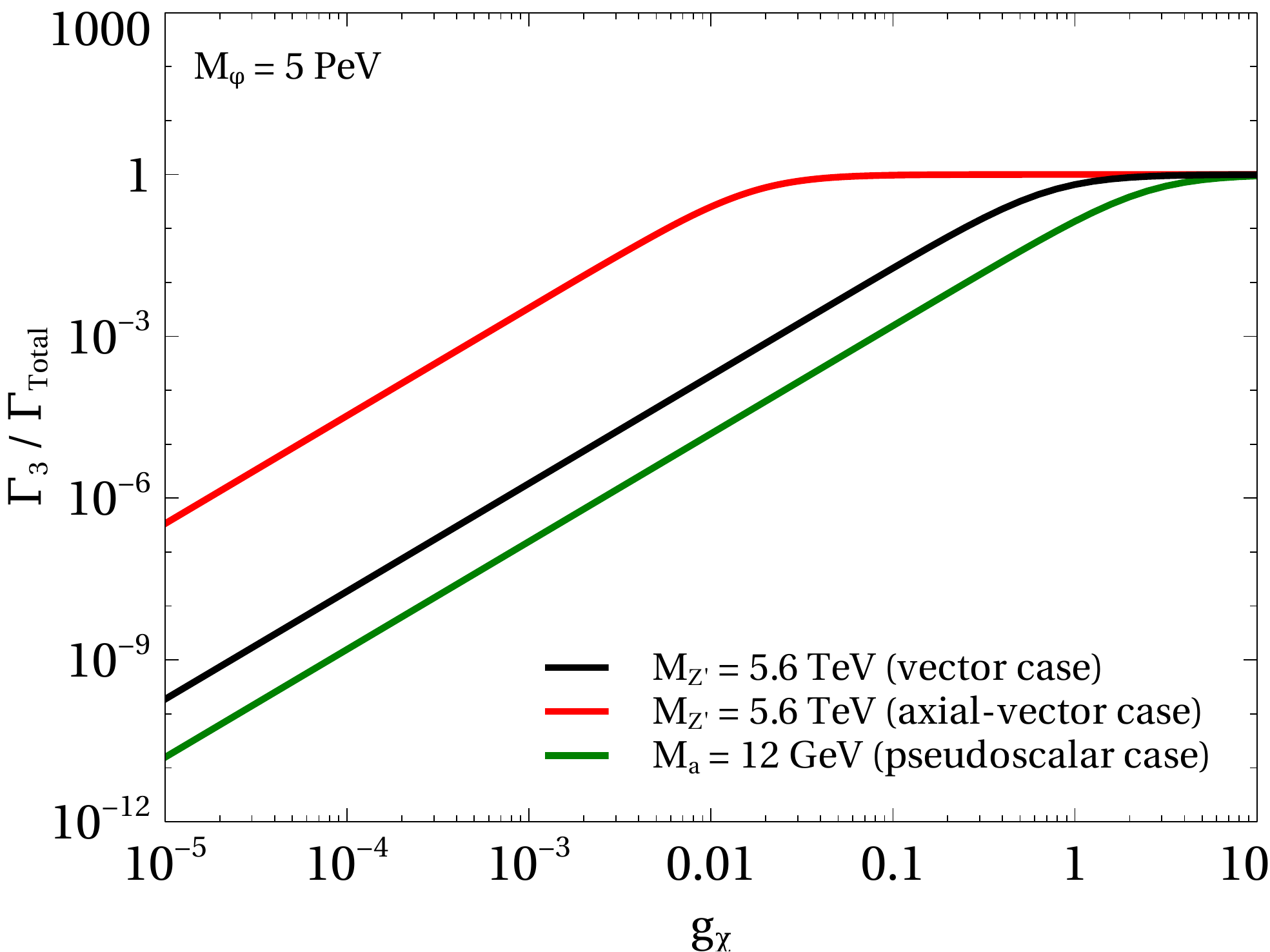}
\caption{\label{br3}Variation of three body branching ratio with $g_\chi$ for the vector, axial-vector and the pseudoscalar mediators. The scalar mediator scenario shows a similar behaviour as the pseudo-scalar one.}
\end{figure}

\section{\label{sec:scenario2}Scenario II: Excess events in the $30$--$100$ TeV region caused by LDM scattering on Ice and its implications}
As discussed in Sec.~\ref{intro2}, in Scenario II, we  relax the assumption made regarding the origin of the three PeV events in Scenario I, and perform a completely general fit to both the PeV and the sub-PeV HESE data, with all four of the flux components taken together. This essentially implies that the HDM mass $m_\phi $ is also left floating in the fit in the entire range $\left[30 {~\rm TeV, 2.5 {~\rm PeV}}\right]$. Therefore, the space of parameters now comprises the set  $m_\phi, F, m_a, \gamma $ and $ N_\text{ast} $. 

We find that  doing this causes the best-fit HDM mass to float to a
value $ \mathcal{O}(500) $ TeV, so that the resulting LDM spectra from its
decay are naturally able to explain the bump, or excess  in the $ \sim$ 50--100 TeV energy range that is seen in the IC data. At the present time, this feature has a statistical significance of about  $2.3 \sigma$.
An important consequence of this is that the flux of secondary neutrinos from  mediator decay, which played an important role in Scenario I, now  populates the low energy bins (between $1$ TeV to $10$
TeV) and falls outside the range relevant to our  fit (the IC threshold for
the HESE events is 30 TeV ). This flux is thus subsumed in the atmospheric background. At energies of around a TeV, where the secondary neutrino 
flux from three-body decays of HDMs in this scenario might have been otherwise important, the atmospheric neutrino flux
 is already about a 1000 times higher, and completely overwhelms it.
  Furthermore, the full-volume IceCube is only sensitive  to contained  events  depositing  at  least  about  10  TeV  in  the  detector, hence this flux is also largely rendered unobservable because it lies outside the HESE sensitivity range.

 Note that  Scenario II also suggests that the other currently emergent features,
the cluster of 3 events close to 1--2 PeV and the dip in the 400 TeV--1 PeV region, which were very
important motivations for Scenario I, may not survive with time. Thus, at  the current level of statistics, this fit gives primacy to the 50--100 TeV excess. In Scenario I, the PeV events, assumed to arise from the 
 two-body decay of HDM,  will (in the form of cascades
resembling NC neutrino events) steadily increase in number and manifest themselves as an excess
or bump, whereas in Scenario II they would just become part of the overall astrophysical power-law neutrino
spectrum without a special origin. The related dip, or deficit, currently seen in the $400$ TeV to $1$ PeV region would gradually become prominent and significant in Scenario I, but would get smoothed over in Scenario II.
Consequently,, in Scenario II the only relevant fluxes are the astrophysical flux and the $\chi$ flux originating from the two body decay of $\phi$, in addition to, of course, the background atmospheric flux.  We show the representative contributing fluxes in Fig.~\ref{flx2}.

The best fit parameters for the fit in Scenario II are given in Table\ \ref{tab3}, and the corresponding results are shown in Fig.\ \ref{fig:multipar}, for the pseudo-scalar mediator scenario (left column), and the axial-vector mediator scenario (right column). As in Scenario I, the scalar and pseudo-scalar mediators lead to similar fits. However, unlike in Scenario I, since the secondary neutrino flux lies outside the energy range under study, both vector and axial-vector mediators lead to similar results for Scenario II. Therefore, we have not shown the scalar and vector cases separately.

\begin{table}[htb]
\begin{center}
\begin{tabular} {c c c c c c}
\hline
Parameter & $m_{a}$ [GeV] & $m_\phi$ [TeV] & $f_{\phi} g_q^2 g_\chi^2/ \tau_\phi \left[\text{s}^{-1}\right]$ & $\gamma$ & $\tilde{N}_{ast}$ (all flavour) \\
\hline
Pseudoscalar        & 16.1 & 680 & $1.15 \times 10^{-27}$ & 2.31 & $1.59 \times 10^{-8}$ \\
 Axial-vector       & $ 5.6 \times 10^{3} $  & 470 & $2.21 \times 10^{-24}$ & 2.30 & $1.59 \times 10^{-8}$  \\
\hline
\end{tabular}
\end{center}
\caption{The best fit values of relevant parameters in case of a pseudoscalar and axial-vector mediator for Scenario II. $\tilde{N}_{ast}$ is given in units of $\flxunit$.}
\label{tab3}
\end{table}

The similarity in the number of events originating from DM and from astrophysical neutrinos
in the two cases is not surprising. In both cases, only the small excess in the vicinity of $ \sim 50-100 $ TeV is due to DM cascades, the remaining events conform to the expected astrophysical neutrino spectrum, which then sets the normalization and the index. 
 Consequently, we also note an important difference between the astrophysical fluxes in Scenario II compared to Scenario I, \ie\ in Scenario I this flux is usually sub-dominant to the secondary neutrino flux,  whereas in Scenario II it accounts for all events except those comprising the excess in the range $ \sim 50$--$100 $ TeV.
The difference in $ m_\phi $ in the two cases is due to the variation in the values of $ \mean{y} $ for the two type of mediators.

\begin{figure}
\centering
\includegraphics[scale=0.5]{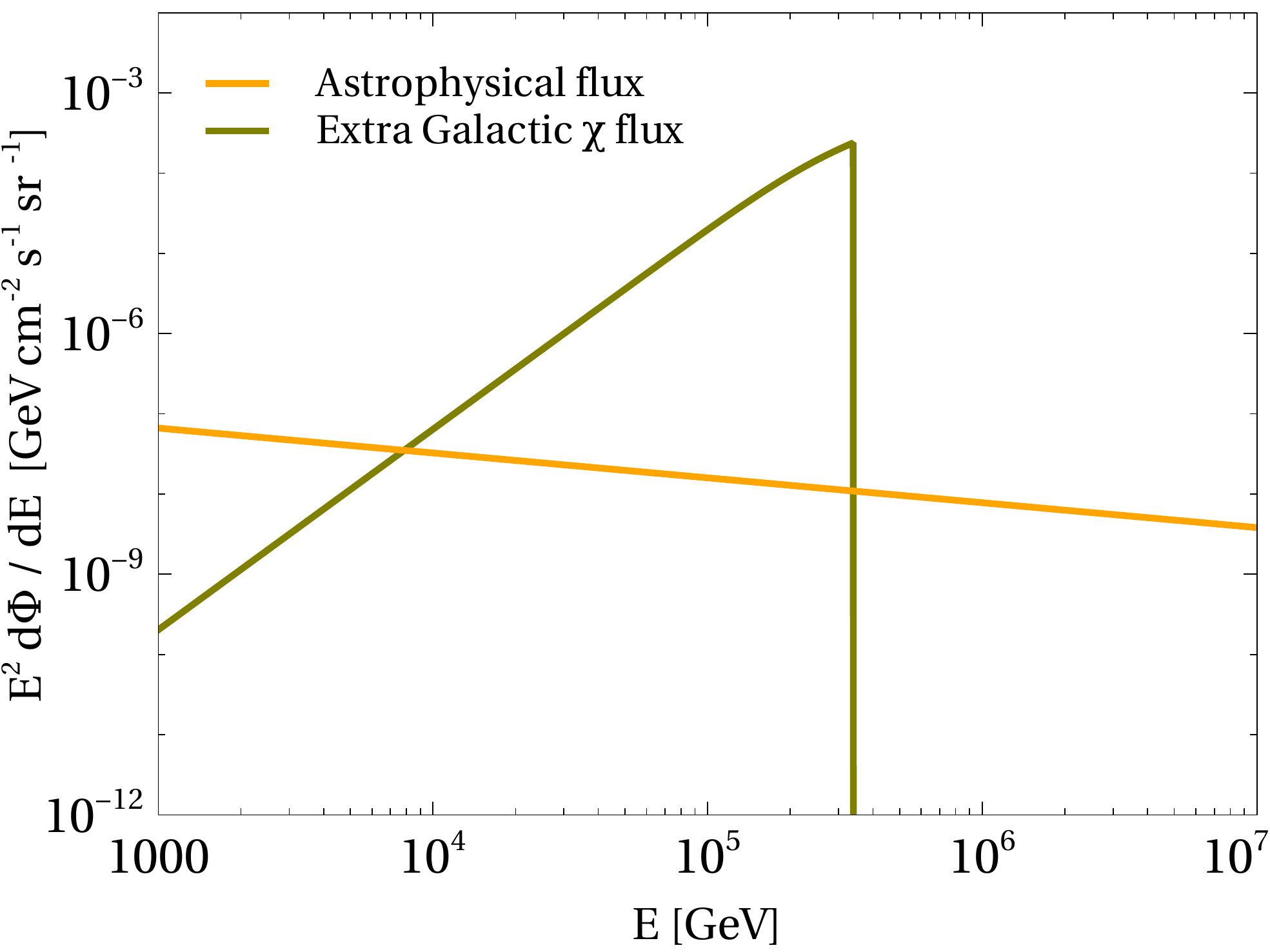}
\caption{Relevant fluxes for Scenario II. The corresponding parameters are given in Table\ \ref{tab3}.
         As before the monochromatic spike at $ m_\phi/2 $ due to the galactic $\chi$ flux is not shown here.}
\label{flx2}
\end{figure}

\begin{figure}[htb]
  \centering
  \includegraphics[width=0.99\textwidth]{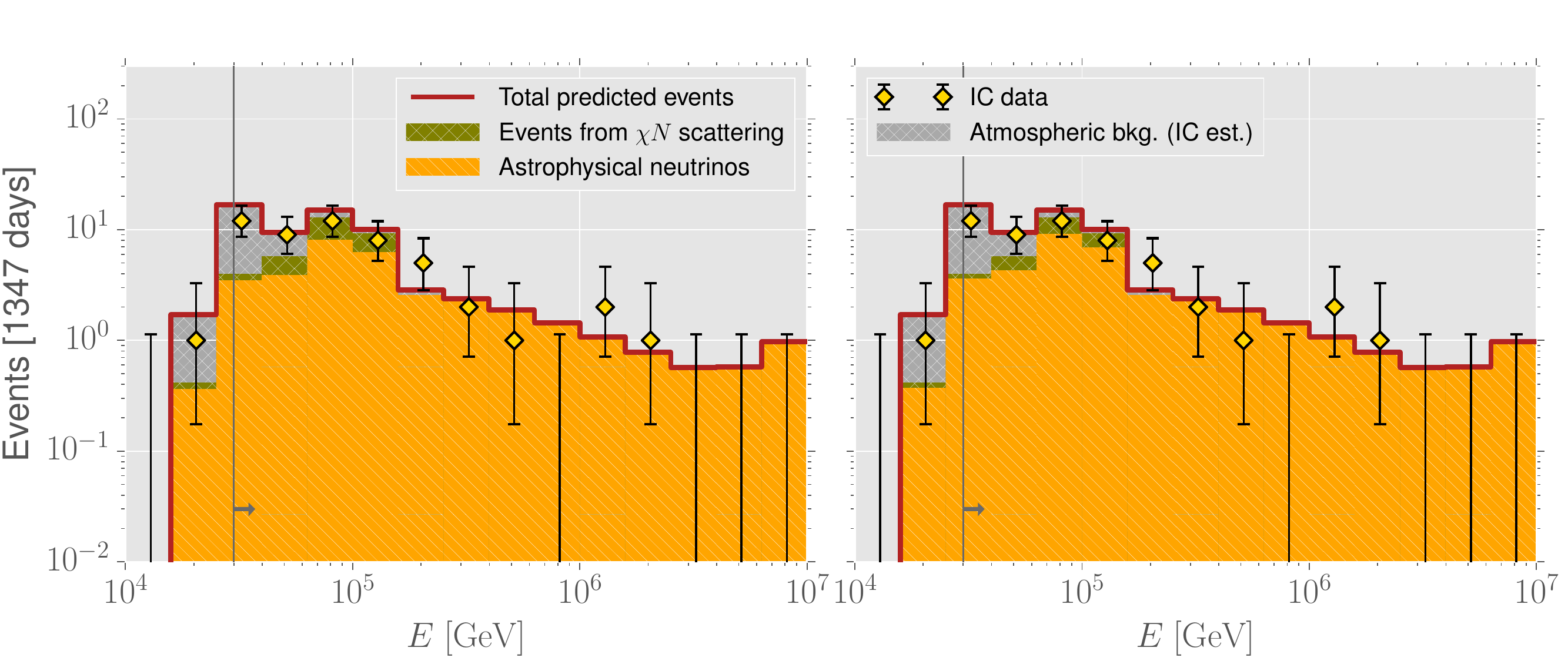}
  \caption{\label{fig:multipar}The total event rate is shown as the red solid curve.
           This comprises events from LDM scattering, astrophysical neutrinos and
           the atmospheric background.
           Events from the astrophysical power-law spectrum are shown as
           orange bars and stacked bars shaded in green show the LDM events over and above
           the astrophysical events.
           The other  events over and above the green/yellow bars are due to atmospheric
           neutrinos and muons. The left hand side shows the pseudo-scalar case while the  right hand side gives the case of an axial-vector type mediator.
          }
\end{figure}

\subsection{Gamma-ray constraints on Scenario II}
As for Scenario I, the diffuse gamma-ray constraints provide the most significant restrictions on our parameter space, and  lead to upper bounds on $ f_\phi g_\chi^2 / \tau_\phi $.
The behaviour of the differential \gray\ flux is sensitive to the mediator mass and the type of mediator under study, as shown in Fig. \ref{gamma_HDM2}.
\begin{figure}[htb]
\centering
\includegraphics[width=0.49\textwidth]{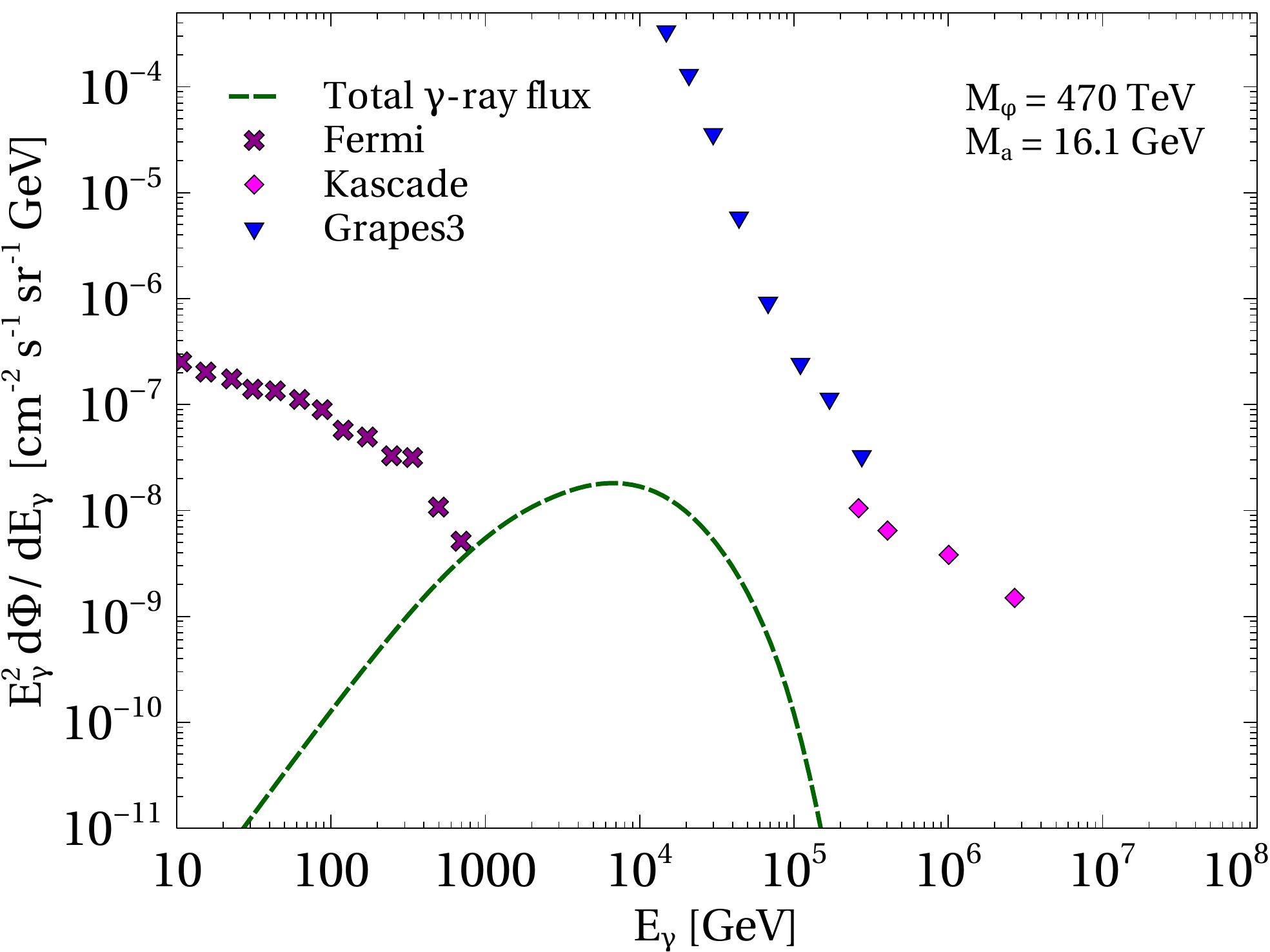}
\includegraphics[width=0.49\textwidth]{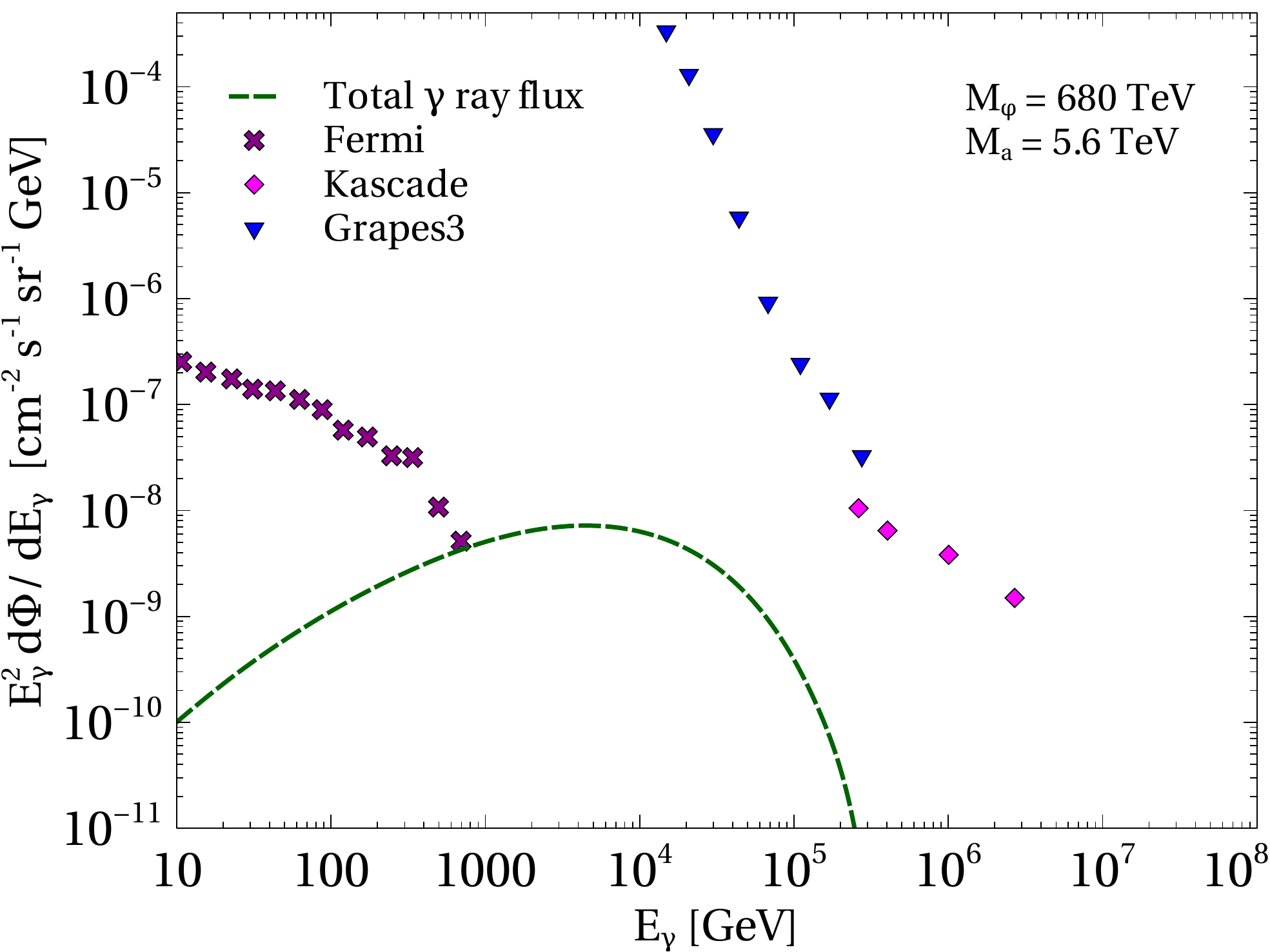}
\caption{Diffuse gamma-ray flux for pseudo-scalar (left) and axial-vector case (right). The maximum allowed values of $(f_\phi g_\chi^2)/\tau_\phi$ have been used for the flux computation here.}
\label{gamma_HDM2}
\end{figure}
Using results on the diffuse gamma ray fluxes from Fermi-LAT, KASCADE and GRAPES3 data, we obtain  upper bounds on $f_\phi g_\chi^2/\tau_\phi$ for the pseudo-scalar and axial-vector cases, respectively, as follows :
\begin{equation}\label{eq:scen2-gamma-bounds}
\dfrac{(g_\chi^2 f_\phi )}{\tau_\phi} \leqslant
\begin{cases}
  5.2\times10^{-27}\,s^{-1}\qquad  \text{for the pseudo-scalar case}  \\
  1.2\times10^{-29}\, s^{-1}\qquad \text{for the axial-vector case}
\end{cases}
\end{equation}
The upper bounds on $F$ that result from the above are significantly more stringent for the axial-vector case, and rule out the best-fit case shown in Fig.~\ref{fig:multipar} for this mediator. The best-fit shown for the pseudo-scalar case is broadly consistent with the current gamma-ray constraints.

\section{\label{sec:track}Muon-Track events}

Our discussion so far has been confined to the HESE events, whose starting vertices are, by definition, contained within the IC instrumented volume.  
More recently, however, a 6-year analysis of through-going muon track events at  IC has been reported  \cite{1607.08006}.
The events in this data sample include those with interaction vertices  outside this volume.
There are events both in the PeV and the sub-PeV regions.
When fit with a uniform astrophysical power-law flux, this sample prefers a stronger astrophysical spectrum, with $ \gamma = 2.13 \pm 0.13 $.
This is notably different from  the conclusion from the HESE analysis, which suggests $\gamma=2.57$, whilst disfavouring
a spectrum with $ \gamma =2.0 $ at more than $ 3\sigma $.
This tension could, perhaps be 
a hint for additional flux components which cannot be accounted for in a simple power-law picture.
Indeed, as pointed out  in \cite{1607.08006}, a 
possible reason for the tension could be a flux component from galactic sources, which becomes sub-dominant as the energy increases.
We note that the secondary neutrino flux from the galaxy, which dominates the sub-PeV contribution in Scenario I, is a possibility that conforms to this requirement.

While we have not attempted a full comparative study of this sample in the context of our scenarios here, we have tried to get an approximate  idea of the track event predictions that Scenario I and II would give.  In Scenario I, for example, contributions to these events would arise from the  secondary neutrino and  astrophysical fluxes. We can then compare the predicted event rates with those predicted by the IC best-fit astrophysical flux (with index 2.13, from \cite{1607.08006}). We show the comparisons in Fig.\ \ref{mtrack1} for the pseudoscalar mediator in Scenario I. The through-going track events span the energy range from 190 TeV to a few PeV \cite{1607.08006} . For both the cases when pseudoscalar $a \rightarrow b \bar{b}$ and $a \rightarrow c \bar{c}$ we have taken a value of $f_\phi \, g_\chi^2 / \tau_\phi$ which satisfies all constraints. For the astrophysical flux,  the values of the index and the normalisations were however fixed to their best-fit values (Fig. \ref{mtrack1}). 

We find good overlap with the IC prediction (\ie, the red and black curves) in the lower part of the energy range of interest, \ie\ 190 TeV to $\sim600$ TeV (where most of the observations lie); however, for higher energies the curves differ, and Scenario I predicts substantially less through-going muon track events. We note that statistics in higher energy region are sparse, making definitive conclusions difficult. In the multi-PeV region, for instance, the highest energy event in this 6-yr sample \cite{1607.08006}, has a deposited energy of $\sim2.6$ PeV, and an estimated muon energy of about $4.5$ PeV.
It is difficult to say if this is an unusually high energy event isolated in origin from the rest; for a detailed discussion of possibilities, see \cite{1605.08781}.

Similarly, we show the IC prediction along with the expectation for  Scenario II  in Fig \ref{mtrack2}.
Although our Scenario II flux is somewhat lower than the IC fit, the agreement overall is reasonable (given the present level of statistics), since the astrophysical power-law flux is a dominant contributor in Scenario II, unlike in Scenario I.
Further confirmation will have to await more data, especially in the high energy region ($E_\nu\geq 3$ PeV). 
\begin{figure}[h]
	\centering
	\includegraphics[width=0.49\textwidth]{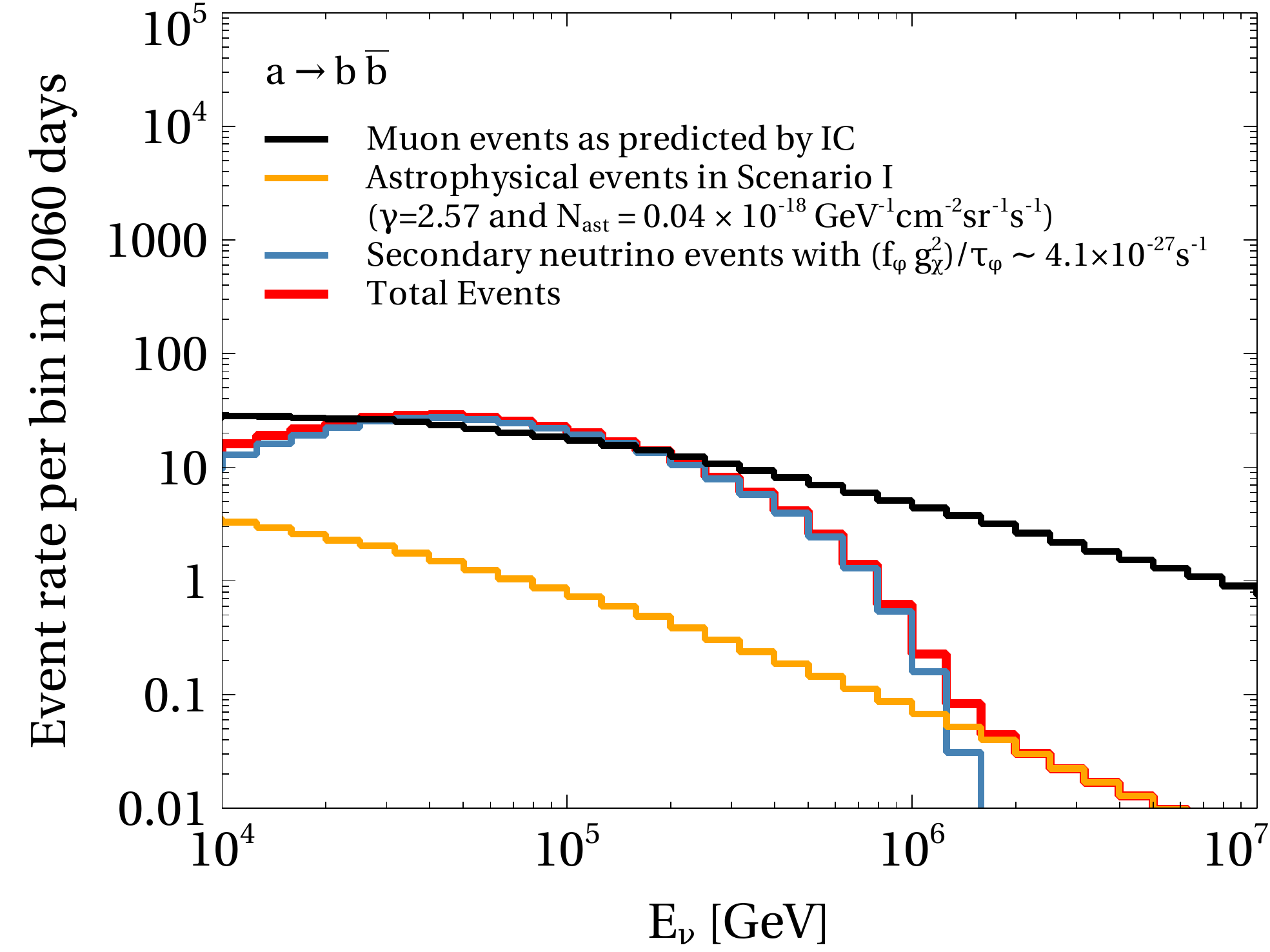}
	\includegraphics[width=0.49\textwidth]{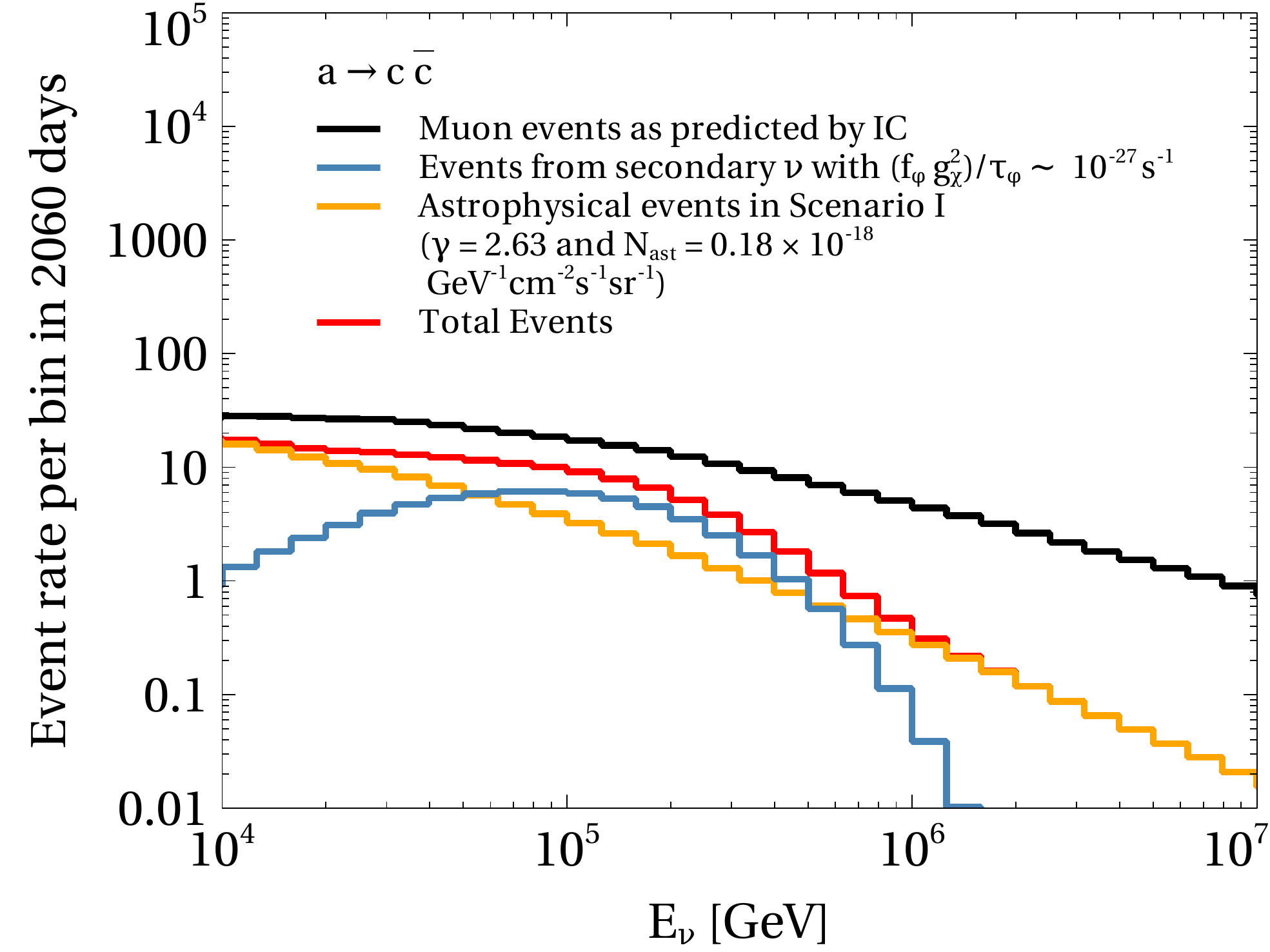}
	\caption{Muon track events for the pseudoscalar case in Scenario I and their comparison with the IC predicted best fit. The black line represents the IC power-law prediction and should be compared to our total prediction for throughgoing track events in the energy region 190 TeV to a few PeV (red line).}
	\label{mtrack1}
\end{figure}

\begin{figure}[h]
	\centering
	\includegraphics[width=0.49\textwidth]{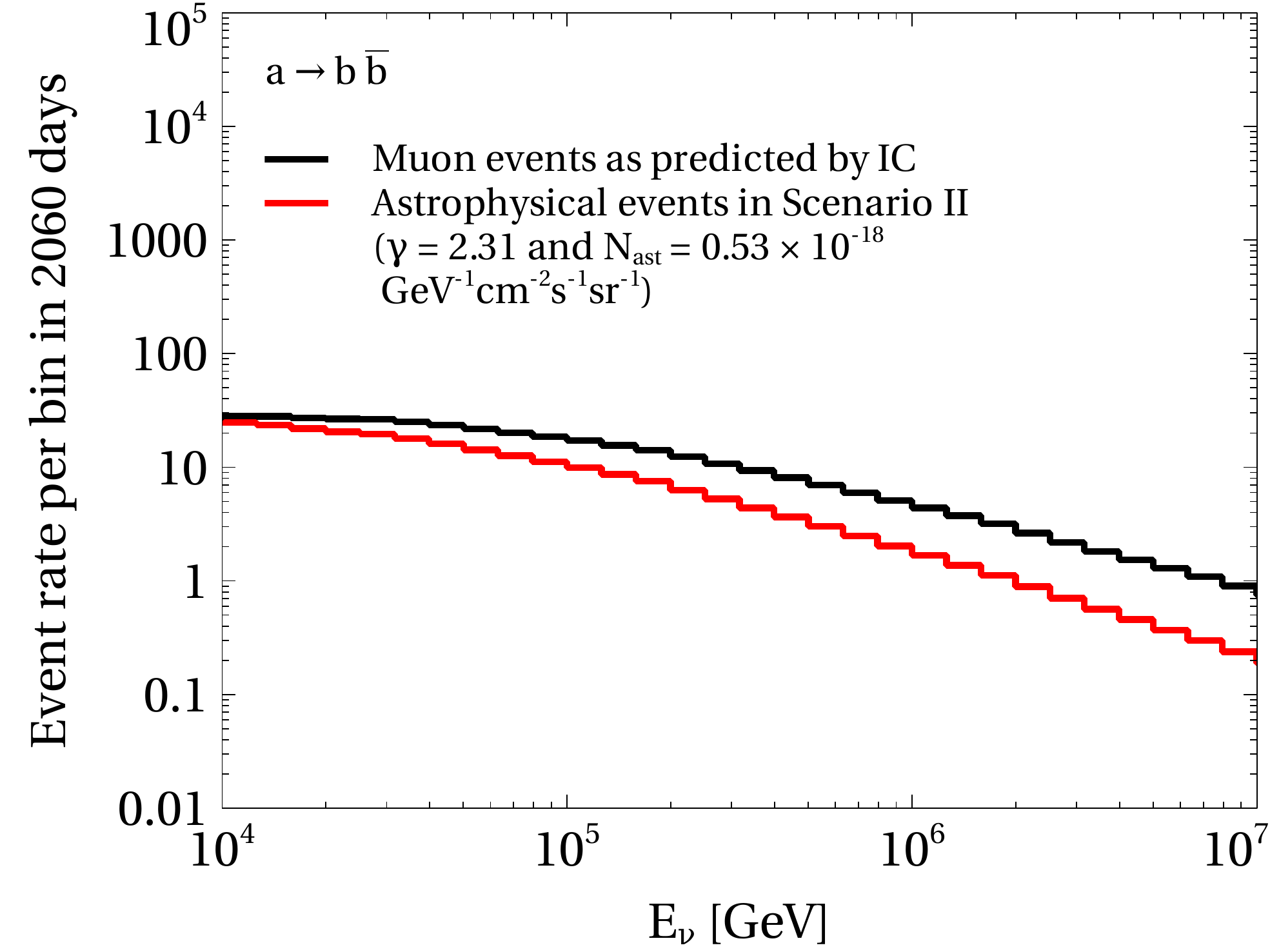}
	\includegraphics[width=0.49\textwidth]{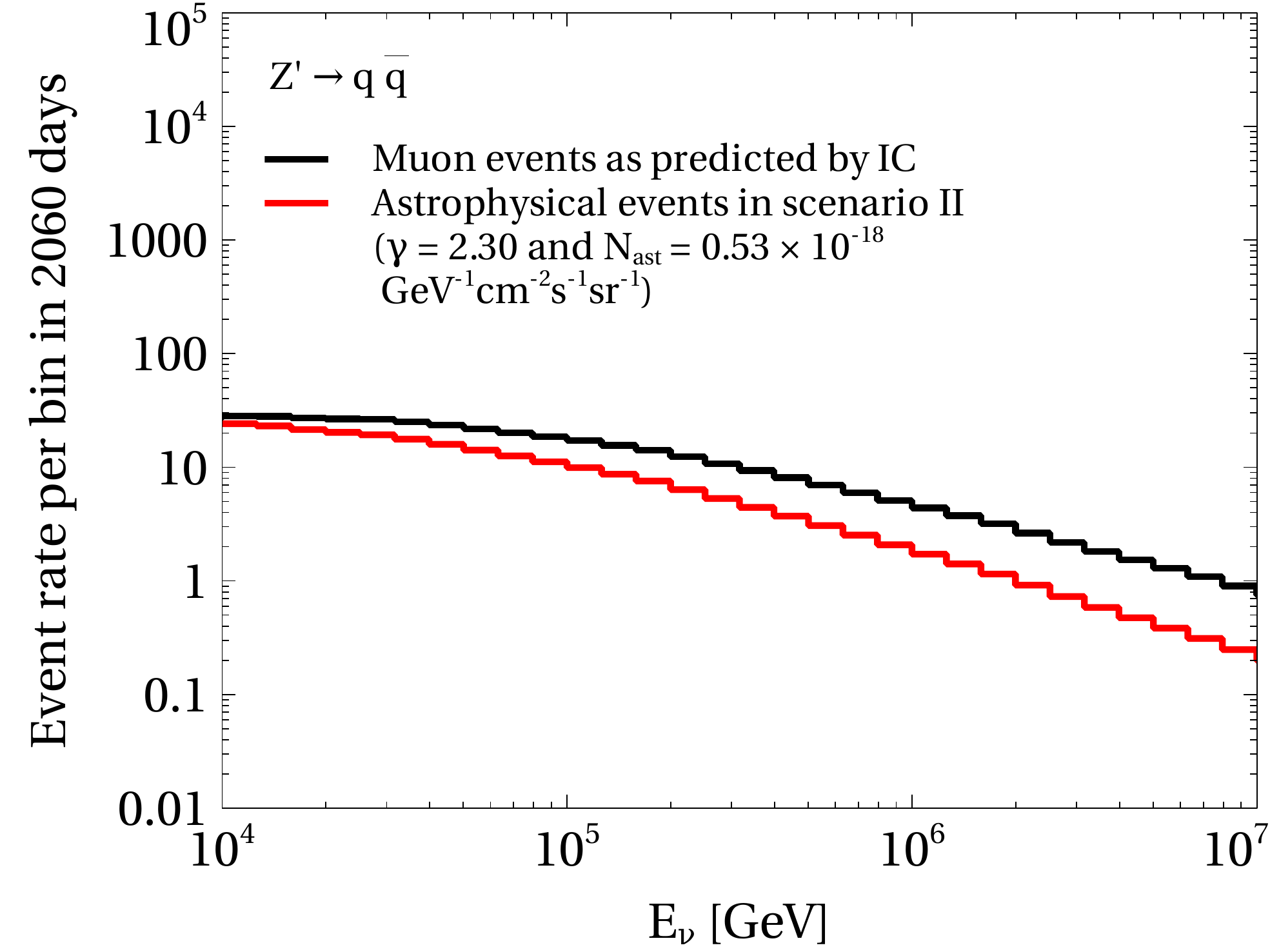}
	\caption{Muon track events in Scenario II. Shown for the case of pseudoscalar (left) and axial-vector type mediators (right). In Scenario II, the astrophysical flux is the main contributor to the track events. In our notation $\Phi_{ast} = N_{ast}\,E^{-\gamma}$. Best fit values of $N_{ast}$ and $\gamma$ are used in the above plot.}
	\label{mtrack2}
\end{figure}

\section{\label{sec:summary}Summary and Conclusions}

By steadily accumulating high energy events over the last four years in the energy range $30$ TeV to $2$ PeV, IC has
conclusively established the presence of a diffuse flux or fluxes which have a non-atmospheric origin and (at least partially) extra-galactic origin, the source(s) of which are at present largely unknown.

 Standard expectations dictate  that this signal is due to a 
flux of astrophysical neutrinos, primarily from sources outside of our galaxy, and that it should correspond to  a uniform power-law flux, characteristic of Fermi shock acceleration, with index approximately $-2$. Features in the data seem to indicate that there are deviations from these expectations, which may signal the presence of one or more additional fluxes.
These  features include a) a  lack of cascade events beyond 2.1 PeV, in spite of both IC's sensitivity in this region, and the presence of the Glashow resonance around 6.3 PeV; b)
a possible  dip  in the spectrum between 400 TeV--1 PeV; c) a low energy excess
of around $ 2.3\sigma $ significance over and above the IC best-fit power-law
spectrum in the energy range $50-100$ TeV. In addition, an overall puzzling feature of the flux is its unexpected proximity to the WB bound, since standard expectations would argue for a neutrino flux that is a factor of $~$a few below this upper limit.

In this work, we have explored the idea that some of the events in IC which cause the overall signal to deviate from the standard power-law originate from the scattering of boosted DM on ice. 
We have considered two scenarios, both  involving the incidence of  such fermionic dark matter (LDM), which is  produced (in the context of a minimal two-component dark matter sector) from the slow decay of
its (significantly) heavier cousin (HDM).
The  LDM, upon scattering off the ice-nuclei inside IC, mimicks standard model neutrino-nucleon
neutral current scattering, but, in general, with weaker interaction strengths.
If the HDM has a mass $ \sim $ 5--10 PeV, the LDM flux can be shown to peak in
a cluster around the 1--2 PeV energies and, with the right parameters, can explain the IC PeV events.
This forms the basis of  Scenario I, which accounts for the rest of the events
(at sub-PeV energies), by a combination of those from astrophysical sources
and a secondary neutrino flux originating from the decay of the mediator involved
in the LDM-nucleon scattering. It is interesting to note that the secondary neutrinos naturally provide a bump in the region $30$--$100$ TeV once the parameters for  the three PeV events from LDM scattering are fixed. 

On the other hand, in Scenario II, for lighter masses of the HDM $ \sim $ 500--800 TeV, the
LDM flux leads to scattering events in the sub-PeV $ \sim $ 30--100 TeV energies and
is helpful in explaining this low-energy excess over and above a 
harder (compared to Scenario I) astrophysical power-law flux. In both scenarios, in order to explain observations, our work incorporates the ${\it{direct}}$ detection of boosted DM by IC, in addition to its detecting UHE neutrinos. This allows the standard astrophysical flux to stay appreciably  below the WB bound for Scenario I, and, to a lesser extent, for Scenario II.

Four different mediators  which connect the SM and DM sectors are considered, specifically, scalar,  pseudo-scalar, vector and axial-vector.
For Scenario I, we find excellent fits to the IC data in both spin-0 mediator cases --- the
LDM scattering explains the three PeV events with a hard cut-off set by
the HDM mass. It has  a soft astrophysical power-law flux that dies out around
energies of 400 TeV, and a small but significant neutrino flux from the decay
of the mediator that helps explain the small bump around 30-100 TeV, making the
full spectrum a  better match to the data than a power-law-only spectrum.
However, for the pseudo-scalar, stringent constraints from \gray\ observations rule out the region
of parameter space where the best-fit itself lies.
The allowed $ 3\sigma $ parameter-space region around the best-fit is quite large, nevertheless,
and we find that a significant portion of this is as yet allowed by the \gray\ bounds.

For spin-1 mediators, in Scenario I we find significantly increased tension between constraints and best-fit parameters
in the vector mediator case, but are, nevertheless, able to fit the IC data well
for specific values of the parameters within the allowed regions.
The case for the axial-vector mediator is, unfortunately, more
pessimistic: we find that perturbativity requirements on the coupling constants 
prevent a simultaneous fit to the observed PeV and sub-PeV data.

If, with future data, Scenario I were to sustain,
we would expect to see a gradual statistical improvement in the evidence for a dip-like structural feature  around  400-800 TeV, since this region marks the interface of fluxes of different origins.
One would see  a paucity of events  beyond  2.1 PeV, due to a significantly lower astrophysical flux compared to current IC predictions.
In addition, a PeV event spectrum predominantly from  LDM scattering (due to HDM decay) 
predicts i) a significantly enhanced ratio of cascade-to-track events  approximately  in the (0.75-2.5 PeV) region, ii) a build-up in the number of such cascade events in this region  as the HDM decay and LDM scattering proceed, and iii)
 a small but non-zero number of up-going cascades in this energy region over time from the northern hemisphere compared
to the case where these events would have been due to a neutrino flux (because of the relatively lower $\chi$-nucleon cross section and consequent reduced screening by the earth.)\footnote{We note that IC has already observed an {\it{upgoing}} cascade in this energy region, with deposited energy $ 0.77\pm0.22$ PeV \cite{Ishihara}}.
Finally, through-going muon track events beyond $\sim 3$ PeV are also expected
to be  lower in number in this scenario than what current IC power-law fit predictions suggest. The overall signal would also exhibit a gradual galactic bias with more statistics, since generically, in DM scenarios, the contributions from our galaxy and from extra-galactic DM are roughly of the same order. Such a directional bias is not expected in a genuinely isotropic flux\footnote{ We stress that in our scenario also,  the events due to the astrophysical neutrino flux (Flux-1) will be isotropic in distribution. The directional bias will be exhibited by only those events that originate in DM( Flux-2 and Flux-3).}.These features would be in contrast to what one would expect to
see if the standard astrophysical power-law flux explanation were indeed responsible
for the observed  events and will be discernable as statistics increase.

Scenario II, on the other hand, is designed to explain only the event excess at 50--100 TeV energies
as being due to DM scattering on ice, with the other events,
including those above 1 PeV,
attributed to an astrophysical  neutrino power-law spectrum. 
It thus assumes that the other features, including the 400-800 TeV dip and the existence of a cut-off beyond $ \sim $ 2 PeV,
which are part of Scenario I, would gradually disappear and smooth out over time.
It is in good agreement with the HESE data, and because it requires a harder astrophysical spectrum to explain the highest energy
events, it is also in better agreement with IC's six-year through-going muon track data.
Indeed, its predictions for both cascade and track events (both starting and through-going) are only slightly below
those of the official IC fits.
The secondary neutrinos produced from the decays of the mediator
in this scenario peak at energies around a TeV and lie in a region
dominated by the conventional atmospheric background.
They are thus not consequential to our considerations here. With respect to the different types of mediators, this scenario is somewhat less constrained over-all compared  to Scenario I. The best-fits we obtain  for the vector and axial-vector cases are disallowed by gamma-ray observations; nevertheless, good fits in the $ 3\sigma $ region are possible.
The scalar and pseudo-scalar mediators make for  better agreement, with their best-fits being allowed.

To conclude, we have shown that present differences in the IC data
in comparison to what is expected from standard astrophysical diffuse neutrino fluxes
may be explained by assuming that the full spectrum is made up of multiple
flux components, with one significant component being the flux of a boosted
DM particle.
Depending on the HDM mass, the LDM flux either peaks at PeV energies (Scenario I)
and explains the PeV events in the 4-yr HESE sample, or at lower energies (Scenario II)
and aids in explaining the 50--100 TeV excess. In Scenario I, the excess at 50-100 TeV is naturally accounted for by a secondary neutrino flux from HDM decay.  In both cases, the different components
conspire in ways that explain the IC data better than any single component
flux can. This is in spite of strong constraints from $\gamma$-ray observations, which limit but do not completely exclude the available 3$\sigma$ parameter space around the corresponding best-fits. On this note, it is worth mentioning that our work skirts the recent strong constraints \cite{1612.05638} on masses and lifetimes of heavy DM decays as explanations of IC events, as they apply to scenarios in which such DM decays directly to SM particles. Finally, we have also discussed signatures that would, with future data, help distinguish each case under consideration from fits with a solely uniform power-law flux.
More data over the next few years should be able to conclusively support or veto such multi-component explanations of high-energy observations at IC compared to other, more standard expectations.

\begin{acknowledgments}
RG thanks Carlos Arguelles, Ben Jones and Joachim Kopp for helpful discussions, and  the Fermilab Theory Group and the Neutrino Division for hospitality while this work was in progress. AG and RG also acknowledge the DAE Neutrino Project Grant for  continued support.
AB is grateful to Jean-Ren\'{e} Cudell, Maxim Laletin and Daniel Wegman Ostrosky for helpful discussions.
AB is supported by the Fonds de la Recherche Scientifique-FNRS, Belgium,
under grant No.~4.4501.15. SM is supported in part by the U.S. Department of Energy under grant No. DE-FG02-95ER40896 and in part by the PITT PACC.
\end{acknowledgments}

\bibliographystyle{jhep}
\bibliography{ICPaper15}

\providecommand{\href}[2]{#2}\begingroup\raggedright\begin{thebibliography}{100}

\bibitem{Aartsen:2015zva}
{\scshape IceCube} collaboration, M.~G. Aartsen et~al., \emph{{The IceCube
  Neutrino Observatory - Contributions to ICRC 2015 Part II: Atmospheric and
  Astrophysical Diffuse Neutrino Searches of All Flavors}},  in
  \emph{{Proceedings, 34th International Cosmic Ray Conference (ICRC 2015)}},
  2015.
\newblock \href{http://arxiv.org/abs/1510.05223}{{\tt 1510.05223}}.

\bibitem{1311.5238}
{\scshape IceCube} collaboration, M.~G. Aartsen et~al., \emph{{Evidence for
  High-Energy Extraterrestrial Neutrinos at the IceCube Detector}},
  \href{http://dx.doi.org/10.1126/science.1242856}{\emph{Science} {\bf 342}
  (2013) 1242856}, [\href{http://arxiv.org/abs/1311.5238}{{\tt 1311.5238}}].

\bibitem{1605.01556}
A.~C. Vincent, S.~Palomares-Ruiz and O.~Mena, \emph{{Analysis of the 4-year
  IceCube HESE data}},  \href{http://arxiv.org/abs/1605.01556}{{\tt
  1605.01556}}.

\bibitem{0710.1557}
J.~K. Becker, \emph{{High-energy neutrinos in the context of multimessenger
  physics}}, \href{http://dx.doi.org/10.1016/j.physrep.2007.10.006}{\emph{Phys.
  Rept.} {\bf 458} (2008) 173--246},
  [\href{http://arxiv.org/abs/0710.1557}{{\tt 0710.1557}}].

\bibitem{astro-ph/0204527}
F.~Halzen and D.~Hooper, \emph{{High-energy neutrino astronomy: The Cosmic ray
  connection}}, \href{http://dx.doi.org/10.1088/0034-4885/65/7/201}{\emph{Rept.
  Prog. Phys.} {\bf 65} (2002) 1025--1078},
  [\href{http://arxiv.org/abs/astro-ph/0204527}{{\tt astro-ph/0204527}}].

\bibitem{1202.0466}
G.~Sigl, \emph{{High Energy Neutrinos and Cosmic Rays}},
  \href{http://dx.doi.org/10.3254/978-1-61499-173-1-145}{\emph{Proc. Int. Sch.
  Phys. Fermi} {\bf 182} (2012) 145--184},
  [\href{http://arxiv.org/abs/1202.0466}{{\tt 1202.0466}}].

\bibitem{1301.1703}
M.~D. Kistler, T.~Stanev and H.~Yüksel, \emph{{Cosmic PeV Neutrinos and the
  Sources of Ultrahigh Energy Protons}},
  \href{http://dx.doi.org/10.1103/PhysRevD.90.123006}{\emph{Phys. Rev.} {\bf
  D90} (2014) 123006}, [\href{http://arxiv.org/abs/1301.1703}{{\tt
  1301.1703}}].

\bibitem{1305.4123}
N.~Gupta, \emph{{Galactic PeV Neutrinos}},
  \href{http://dx.doi.org/10.1016/j.astropartphys.2013.07.003}{\emph{Astropart.
  Phys.} {\bf 48} (2013) 75--77}, [\href{http://arxiv.org/abs/1305.4123}{{\tt
  1305.4123}}].

\bibitem{1306.2309}
R.~Laha, J.~F. Beacom, B.~Dasgupta, S.~Horiuchi and K.~Murase,
  \emph{{Demystifying the PeV Cascades in IceCube: Less (Energy) is More
  (Events)}}, \href{http://dx.doi.org/10.1103/PhysRevD.88.043009}{\emph{Phys.
  Rev.} {\bf D88} (2013) 043009}, [\href{http://arxiv.org/abs/1306.2309}{{\tt
  1306.2309}}].

\bibitem{1306.5021}
L.~A. Anchordoqui, H.~Goldberg, M.~H. Lynch, A.~V. Olinto, T.~C. Paul and T.~J.
  Weiler, \emph{{Pinning down the cosmic ray source mechanism with new IceCube
  data}}, \href{http://dx.doi.org/10.1103/PhysRevD.89.083003}{\emph{Phys. Rev.}
  {\bf D89} (2014) 083003}, [\href{http://arxiv.org/abs/1306.5021}{{\tt
  1306.5021}}].

\bibitem{1312.6587}
L.~A. Anchordoqui et~al., \emph{{Cosmic Neutrino Pevatrons: A Brand New Pathway
  to Astronomy, Astrophysics, and Particle Physics}},
  \href{http://dx.doi.org/10.1016/j.jheap.2014.01.001}{\emph{JHEAp} {\bf 1-2}
  (2014) 1--30}, [\href{http://arxiv.org/abs/1312.6587}{{\tt 1312.6587}}].

\bibitem{1410.3680}
K.~Murase, \emph{{On the Origin of High-Energy Cosmic Neutrinos}},
  \href{http://dx.doi.org/10.1063/1.4915555}{\emph{AIP Conf. Proc.} {\bf 1666}
  (2015) 040006}, [\href{http://arxiv.org/abs/1410.3680}{{\tt 1410.3680}}].

\bibitem{1407.7536}
W.~Winter, \emph{{Describing the Observed Cosmic Neutrinos by Interactions of
  Nuclei with Matter}},
  \href{http://dx.doi.org/10.1103/PhysRevD.90.103003}{\emph{Phys. Rev.} {\bf
  D90} (2014) 103003}, [\href{http://arxiv.org/abs/1407.7536}{{\tt
  1407.7536}}].

\bibitem{1405.5487}
S.~Dado and A.~Dar, \emph{{Origin of the High Energy Cosmic Neutrino
  Background}},
  \href{http://dx.doi.org/10.1103/PhysRevLett.113.191102}{\emph{Phys. Rev.
  Lett.} {\bf 113} (2014) 191102}, [\href{http://arxiv.org/abs/1405.5487}{{\tt
  1405.5487}}].

\bibitem{1211.1974}
I.~Cholis and D.~Hooper, \emph{{On The Origin of IceCube's PeV Neutrinos}},
  \href{http://dx.doi.org/10.1088/1475-7516/2013/06/030}{\emph{JCAP} {\bf 1306}
  (2013) 030}, [\href{http://arxiv.org/abs/1211.1974}{{\tt 1211.1974}}].

\bibitem{1212.1260}
R.-Y. Liu and X.-Y. Wang, \emph{{Diffuse PeV neutrinos from gamma-ray bursts}},
  \href{http://dx.doi.org/10.1088/0004-637X/766/2/73}{\emph{Astrophys. J.} {\bf
  766} (2013) 73}, [\href{http://arxiv.org/abs/1212.1260}{{\tt 1212.1260}}].

\bibitem{1307.7596}
S.~Razzaque, \emph{{Long-lived PeV–EeV neutrinos from gamma-ray burst
  blastwave}}, \href{http://dx.doi.org/10.1103/PhysRevD.88.103003}{\emph{Phys.
  Rev.} {\bf D88} (2013) 103003}, [\href{http://arxiv.org/abs/1307.7596}{{\tt
  1307.7596}}].

\bibitem{1306.2274}
K.~Murase and K.~Ioka, \emph{{TeV–PeV Neutrinos from Low-Power Gamma-Ray
  Burst Jets inside Stars}},
  \href{http://dx.doi.org/10.1103/PhysRevLett.111.121102}{\emph{Phys. Rev.
  Lett.} {\bf 111} (2013) 121102}, [\href{http://arxiv.org/abs/1306.2274}{{\tt
  1306.2274}}].

\bibitem{1310.7061}
N.~Fraija, \emph{{GeV–PeV neutrino production and oscillation in hidden jets
  from gamma-ray bursts}},
  \href{http://dx.doi.org/10.1093/mnras/stt2036}{\emph{Mon. Not. Roy. Astron.
  Soc.} {\bf 437} (2014) 2187--2200},
  [\href{http://arxiv.org/abs/1310.7061}{{\tt 1310.7061}}].

\bibitem{1405.2091}
M.~Petropoulou, D.~Giannios and S.~Dimitrakoudis, \emph{{Implications of a PeV
  neutrino spectral cutoff in GRB models}},
  \href{http://dx.doi.org/10.1093/mnras/stu1757}{\emph{Mon. Not. Roy. Astron.
  Soc.} {\bf 445} (2014) 570--580}, [\href{http://arxiv.org/abs/1405.2091}{{\tt
  1405.2091}}].

\bibitem{1411.7491}
S.~Razzaque and L.~Yang, \emph{{PeV-EeV neutrinos from GRB blast waves in
  IceCube and future neutrino telescopes}},
  \href{http://dx.doi.org/10.1103/PhysRevD.91.043003}{\emph{Phys. Rev.} {\bf
  D91} (2015) 043003}, [\href{http://arxiv.org/abs/1411.7491}{{\tt
  1411.7491}}].

\bibitem{1504.00107}
I.~Tamborra and S.~Ando, \emph{{Diffuse emission of high-energy neutrinos from
  gamma-ray burst fireballs}},
  \href{http://dx.doi.org/10.1088/1475-7516/2015/9/036,
  10.1088/1475-7516/2015/09/036}{\emph{JCAP} {\bf 1509} (2015) 036},
  [\href{http://arxiv.org/abs/1504.00107}{{\tt 1504.00107}}].

\bibitem{1512.08513}
N.~Senno, K.~Murase and P.~Meszaros, \emph{{Choked Jets and Low-Luminosity
  Gamma-Ray Bursts as Hidden Neutrino Sources}},
  \href{http://dx.doi.org/10.1103/PhysRevD.93.083003}{\emph{Phys. Rev.} {\bf
  D93} (2016) 083003}, [\href{http://arxiv.org/abs/1512.08513}{{\tt
  1512.08513}}].

\bibitem{1512.01559}
I.~Tamborra and S.~Ando, \emph{{Inspecting the supernova–gamma-ray-burst
  connection with high-energy neutrinos}},
  \href{http://dx.doi.org/10.1103/PhysRevD.93.053010}{\emph{Phys. Rev.} {\bf
  D93} (2016) 053010}, [\href{http://arxiv.org/abs/1512.01559}{{\tt
  1512.01559}}].

\bibitem{1306.3417}
K.~Murase, M.~Ahlers and B.~C. Lacki, \emph{{Testing the Hadronuclear Origin of
  PeV Neutrinos Observed with IceCube}},
  \href{http://dx.doi.org/10.1103/PhysRevD.88.121301}{\emph{Phys. Rev.} {\bf
  D88} (2013) 121301}, [\href{http://arxiv.org/abs/1306.3417}{{\tt
  1306.3417}}].

\bibitem{1405.7648}
L.~A. Anchordoqui, T.~C. Paul, L.~H.~M. da~Silva, D.~F. Torres and B.~J. Vlcek,
  \emph{{What IceCube data tell us about neutrino emission from star-forming
  galaxies (so far)}},
  \href{http://dx.doi.org/10.1103/PhysRevD.89.127304}{\emph{Phys. Rev.} {\bf
  D89} (2014) 127304}, [\href{http://arxiv.org/abs/1405.7648}{{\tt
  1405.7648}}].

\bibitem{1404.1189}
I.~Tamborra, S.~Ando and K.~Murase, \emph{{Star-forming galaxies as the origin
  of diffuse high-energy backgrounds: Gamma-ray and neutrino connections, and
  implications for starburst history}},
  \href{http://dx.doi.org/10.1088/1475-7516/2014/09/043}{\emph{JCAP} {\bf 1409}
  (2014) 043}, [\href{http://arxiv.org/abs/1404.1189}{{\tt 1404.1189}}].

\bibitem{1406.1099}
X.-C. Chang and X.-Y. Wang, \emph{{The diffuse gamma-ray flux associated with
  sub-PeV/PeV neutrinos from starburst galaxies}},
  \href{http://dx.doi.org/10.1088/0004-637X/793/2/131}{\emph{Astrophys. J.}
  {\bf 793} (2014) 131}, [\href{http://arxiv.org/abs/1406.1099}{{\tt
  1406.1099}}].

\bibitem{1509.00983}
I.~Bartos and S.~Marka, \emph{{Spectral Decline of PeV Neutrinos from Starburst
  Galaxies}},  \href{http://arxiv.org/abs/1509.00983}{{\tt 1509.00983}}.

\bibitem{1303.0300}
O.~E. Kalashev, A.~Kusenko and W.~Essey, \emph{{PeV neutrinos from
  intergalactic interactions of cosmic rays emitted by active galactic
  nuclei}}, \href{http://dx.doi.org/10.1103/PhysRevLett.111.041103}{\emph{Phys.
  Rev. Lett.} {\bf 111} (2013) 041103},
  [\href{http://arxiv.org/abs/1303.0300}{{\tt 1303.0300}}].

\bibitem{1305.7404}
F.~W. Stecker, \emph{{PeV neutrinos observed by IceCube from cores of active
  galactic nuclei}},
  \href{http://dx.doi.org/10.1103/PhysRevD.88.047301}{\emph{Phys. Rev.} {\bf
  D88} (2013) 047301}, [\href{http://arxiv.org/abs/1305.7404}{{\tt
  1305.7404}}].

\bibitem{1406.2633}
C.~D. Dermer, K.~Murase and Y.~Inoue, \emph{{Photopion Production in Black-Hole
  Jets and Flat-Spectrum Radio Quasars as PeV Neutrino Sources}},
  \href{http://dx.doi.org/10.1016/j.jheap.2014.09.001}{\emph{JHEAp} {\bf 3-4}
  (2014) 29--40}, [\href{http://arxiv.org/abs/1406.2633}{{\tt 1406.2633}}].

\bibitem{1406.0645}
F.~Krauß et~al., \emph{{TANAMI Blazars in the IceCube PeV Neutrino Fields}},
  \href{http://dx.doi.org/10.1051/0004-6361/201424219}{\emph{Astron.
  Astrophys.} {\bf 566} (2014) L7}, [\href{http://arxiv.org/abs/1406.0645}{{\tt
  1406.0645}}].

\bibitem{1407.0907}
F.~Tavecchio, G.~Ghisellini and D.~Guetta, \emph{{Structured jets in BL Lac
  objects: efficient PeV neutrino factories?}},
  \href{http://dx.doi.org/10.1088/2041-8205/793/1/L18}{\emph{Astrophys. J.}
  {\bf 793} (2014) L18}, [\href{http://arxiv.org/abs/1407.0907}{{\tt
  1407.0907}}].

\bibitem{1408.3664}
S.~Sahu and L.~S. Miranda, \emph{{Some possible sources of IceCube TeV–PeV
  neutrino events}},
  \href{http://dx.doi.org/10.1140/epjc/s10052-015-3519-1}{\emph{Eur. Phys. J.}
  {\bf C75} (2015) 273}, [\href{http://arxiv.org/abs/1408.3664}{{\tt
  1408.3664}}].

\bibitem{1410.8124}
O.~Kalashev, D.~Semikoz and I.~Tkachev, \emph{{Neutrinos in IceCube from active
  galactic nuclei}},
  \href{http://dx.doi.org/10.1134/S106377611503022X}{\emph{J. Exp. Theor.
  Phys.} {\bf 120} (2015) 541--548},
  [\href{http://arxiv.org/abs/1410.8124}{{\tt 1410.8124}}].

\bibitem{1411.2783}
F.~Tavecchio and G.~Ghisellini, \emph{{High-energy cosmic neutrinos from
  spine-sheath BL Lac jets}},
  \href{http://dx.doi.org/10.1093/mnras/stv1023}{\emph{Mon. Not. Roy. Astron.
  Soc.} {\bf 451} (2015) 1502--1510},
  [\href{http://arxiv.org/abs/1411.2783}{{\tt 1411.2783}}].

\bibitem{1411.3588}
S.~S. Kimura, K.~Murase and K.~Toma, \emph{{Neutrino and Cosmic-Ray Emission
  and Cumulative Background from Radiatively Inefficient Accretion Flows in
  Low-Luminosity Active Galactic Nuclei}},
  \href{http://dx.doi.org/10.1088/0004-637X/806/2/159}{\emph{Astrophys. J.}
  {\bf 806} (2015) 159}, [\href{http://arxiv.org/abs/1411.3588}{{\tt
  1411.3588}}].

\bibitem{1501.07115}
M.~Petropoulou, S.~Dimitrakoudis, P.~Padovani, A.~Mastichiadis and E.~Resconi,
  \emph{{Photohadronic origin of $\boldsymbol {\gamma }$-ray BL Lac emission:
  implications for IceCube neutrinos}},
  \href{http://dx.doi.org/10.1093/mnras/stv179}{\emph{Mon. Not. Roy. Astron.
  Soc.} {\bf 448} (2015) 2412--2429},
  [\href{http://arxiv.org/abs/1501.07115}{{\tt 1501.07115}}].

\bibitem{1310.1263}
R.-Y. Liu, X.-Y. Wang, S.~Inoue, R.~Crocker and F.~Aharonian, \emph{{Diffuse
  PeV neutrinos from EeV cosmic ray sources: Semirelativistic hypernova
  remnants in star-forming galaxies}},
  \href{http://dx.doi.org/10.1103/PhysRevD.89.083004}{\emph{Phys. Rev.} {\bf
  D89} (2014) 083004}, [\href{http://arxiv.org/abs/1310.1263}{{\tt
  1310.1263}}].

\bibitem{1501.02615}
S.~Chakraborty and I.~Izaguirre, \emph{{Diffuse neutrinos from extragalactic
  supernova remnants: Dominating the 100 TeV IceCube flux}},
  \href{http://dx.doi.org/10.1016/j.physletb.2015.04.032}{\emph{Phys. Lett.}
  {\bf B745} (2015) 35--39}, [\href{http://arxiv.org/abs/1501.02615}{{\tt
  1501.02615}}].

\bibitem{1407.2985}
A.~Bhattacharya, R.~Enberg, M.~H. Reno and I.~Sarcevic, \emph{{Charm decay in
  slow-jet supernovae as the origin of the IceCube ultra-high energy neutrino
  events}}, \href{http://dx.doi.org/10.1088/1475-7516/2015/06/034}{\emph{JCAP}
  {\bf 1506} (2015) 034}, [\href{http://arxiv.org/abs/1407.2985}{{\tt
  1407.2985}}].

\bibitem{1410.0348}
L.~A. Anchordoqui, H.~Goldberg, T.~C. Paul, L.~H.~M. da~Silva and B.~J. Vlcek,
  \emph{{Estimating the contribution of Galactic sources to the diffuse
  neutrino flux}},
  \href{http://dx.doi.org/10.1103/PhysRevD.90.123010}{\emph{Phys. Rev.} {\bf
  D90} (2014) 123010}, [\href{http://arxiv.org/abs/1410.0348}{{\tt
  1410.0348}}].

\bibitem{1306.3006}
H.~Gao, B.~Zhang, X.-F. Wu and Z.-G. Dai, \emph{{Possible High-Energy Neutrino
  and Photon Signals from Gravitational Wave Bursts due to Double Neutron Star
  Mergers}}, \href{http://dx.doi.org/10.1103/PhysRevD.88.043010}{\emph{Phys.
  Rev.} {\bf D88} (2013) 043010}, [\href{http://arxiv.org/abs/1306.3006}{{\tt
  1306.3006}}].

\bibitem{1512.08596}
X.-Y. Wang and R.-Y. Liu, \emph{{Tidal disruption jets of supermassive black
  holes as hidden sources of cosmic rays: explaining the IceCube TeV-PeV
  neutrinos}}, \href{http://dx.doi.org/10.1103/PhysRevD.93.083005}{\emph{Phys.
  Rev.} {\bf D93} (2016) 083005}, [\href{http://arxiv.org/abs/1512.08596}{{\tt
  1512.08596}}].

\bibitem{1307.2158}
A.~Neronov, D.~V. Semikoz and C.~Tchernin, \emph{{PeV neutrinos from
  interactions of cosmic rays with the interstellar medium in the Galaxy}},
  \href{http://dx.doi.org/10.1103/PhysRevD.89.103002}{\emph{Phys. Rev.} {\bf
  D89} (2014) 103002}, [\href{http://arxiv.org/abs/1307.2158}{{\tt
  1307.2158}}].

\bibitem{1310.5123}
J.~C. Joshi, W.~Winter and N.~Gupta, \emph{{How Many of the Observed Neutrino
  Events Can Be Described by Cosmic Ray Interactions in the Milky Way?}},
  \href{http://dx.doi.org/10.1093/mnras/stu189,
  10.1093/mnras/stu2132}{\emph{Mon. Not. Roy. Astron. Soc.} {\bf 439} (2014)
  3414--3419}, [\href{http://arxiv.org/abs/1310.5123}{{\tt 1310.5123}}].

\bibitem{1311.0287}
B.~Katz, E.~Waxman, T.~Thompson and A.~Loeb, \emph{{The energy production rate
  density of cosmic rays in the local universe is $\sim10^{44-45}\rm
  erg~Mpc^{-3}~yr^{-1}$ at all particle energies}},
  \href{http://arxiv.org/abs/1311.0287}{{\tt 1311.0287}}.

\bibitem{1404.6237}
K.~Fang, T.~Fujii, T.~Linden and A.~V. Olinto, \emph{{Is the Ultra-High Energy
  Cosmic-Ray Excess Observed by the Telescope Array Correlated with IceCube
  Neutrinos?}},
  \href{http://dx.doi.org/10.1088/0004-637X/794/2/126}{\emph{Astrophys. J.}
  {\bf 794} (2014) 126}, [\href{http://arxiv.org/abs/1404.6237}{{\tt
  1404.6237}}].

\bibitem{1405.3797}
M.~Kachelrieß and S.~Ostapchenko, \emph{{Neutrino yield from Galactic cosmic
  rays}}, \href{http://dx.doi.org/10.1103/PhysRevD.90.083002}{\emph{Phys. Rev.}
  {\bf D90} (2014) 083002}, [\href{http://arxiv.org/abs/1405.3797}{{\tt
  1405.3797}}].

\bibitem{1411.6457}
L.~A. Anchordoqui, \emph{{Neutron $\beta$-decay as the origin of IceCube’s
  PeV (anti)neutrinos}},
  \href{http://dx.doi.org/10.1103/PhysRevD.91.027301}{\emph{Phys. Rev.} {\bf
  D91} (2015) 027301}, [\href{http://arxiv.org/abs/1411.6457}{{\tt
  1411.6457}}].

\bibitem{1412.8590}
Y.~Q. Guo, H.~B. Hu and Z.~Tian, \emph{{On the Contribution of "Fresh" Cosmic
  Rays to the Excesses of Secondary Particles}},
  \href{http://arxiv.org/abs/1412.8590}{{\tt 1412.8590}}.

\bibitem{1403.3206}
A.~M. Taylor, S.~Gabici and F.~Aharonian, \emph{{Galactic halo origin of the
  neutrinos detected by IceCube}},
  \href{http://dx.doi.org/10.1103/PhysRevD.89.103003}{\emph{Phys. Rev.} {\bf
  D89} (2014) 103003}, [\href{http://arxiv.org/abs/1403.3206}{{\tt
  1403.3206}}].

\bibitem{1410.8697}
F.~Zandanel, I.~Tamborra, S.~Gabici and S.~Ando, \emph{{High-energy gamma-ray
  and neutrino backgrounds from clusters of galaxies and radio constraints}},
  \href{http://dx.doi.org/10.1051/0004-6361/201425249}{\emph{Astron.
  Astrophys.} {\bf 578} (2015) A32},
  [\href{http://arxiv.org/abs/1410.8697}{{\tt 1410.8697}}].

\bibitem{1303.7320}
B.~Feldstein, A.~Kusenko, S.~Matsumoto and T.~T. Yanagida, \emph{{Neutrinos at
  IceCube from Heavy Decaying Dark Matter}},
  \href{http://dx.doi.org/10.1103/PhysRevD.88.015004}{\emph{Phys. Rev.} {\bf
  D88} (2013) 015004}, [\href{http://arxiv.org/abs/1303.7320}{{\tt
  1303.7320}}].

\bibitem{1308.1105}
A.~Esmaili and P.~D. Serpico, \emph{{Are IceCube neutrinos unveiling PeV-scale
  decaying dark matter?}},
  \href{http://dx.doi.org/10.1088/1475-7516/2013/11/054}{\emph{JCAP} {\bf 1311}
  (2013) 054}, [\href{http://arxiv.org/abs/1308.1105}{{\tt 1308.1105}}].

\bibitem{1312.3501}
Y.~Ema, R.~Jinno and T.~Moroi, \emph{{Cosmic-Ray Neutrinos from the Decay of
  Long-Lived Particle and the Recent IceCube Result}},
  \href{http://dx.doi.org/10.1016/j.physletb.2014.04.021}{\emph{Phys. Lett.}
  {\bf B733} (2014) 120--125}, [\href{http://arxiv.org/abs/1312.3501}{{\tt
  1312.3501}}].

\bibitem{1410.5979}
A.~Esmaili, S.~K. Kang and P.~D. Serpico, \emph{{IceCube events and decaying
  dark matter: hints and constraints}},
  \href{http://dx.doi.org/10.1088/1475-7516/2014/12/054}{\emph{JCAP} {\bf 1412}
  (2014) 054}, [\href{http://arxiv.org/abs/1410.5979}{{\tt 1410.5979}}].

\bibitem{1403.1862}
A.~Bhattacharya, M.~H. Reno and I.~Sarcevic, \emph{{Reconciling neutrino flux
  from heavy dark matter decay and recent events at IceCube}},
  \href{http://dx.doi.org/10.1007/JHEP06(2014)110}{\emph{JHEP} {\bf 06} (2014)
  110}, [\href{http://arxiv.org/abs/1403.1862}{{\tt 1403.1862}}].

\bibitem{1407.3280}
A.~Bhattacharya, R.~Gandhi and A.~Gupta, \emph{{The Direct Detection of Boosted
  Dark Matter at High Energies and PeV events at IceCube}},
  \href{http://dx.doi.org/10.1088/1475-7516/2015/03/027}{\emph{JCAP} {\bf 1503}
  (2015) 027}, [\href{http://arxiv.org/abs/1407.3280}{{\tt 1407.3280}}].

\bibitem{1408.1745}
Y.~Ema, R.~Jinno and T.~Moroi, \emph{{Cosmological Implications of High-Energy
  Neutrino Emission from the Decay of Long-Lived Particle}},
  \href{http://dx.doi.org/10.1007/JHEP10(2014)150}{\emph{JHEP} {\bf 10} (2014)
  150}, [\href{http://arxiv.org/abs/1408.1745}{{\tt 1408.1745}}].

\bibitem{1411.1071}
J.~F. Cherry, A.~Friedland and I.~M. Shoemaker, \emph{{Neutrino Portal Dark
  Matter: From Dwarf Galaxies to IceCube}},
  \href{http://arxiv.org/abs/1411.1071}{{\tt 1411.1071}}.

\bibitem{1503.02669}
J.~Kopp, J.~Liu and X.-P. Wang, \emph{{Boosted Dark Matter in IceCube and at
  the Galactic Center}},
  \href{http://dx.doi.org/10.1007/JHEP04(2015)105}{\emph{JHEP} {\bf 04} (2015)
  105}, [\href{http://arxiv.org/abs/1503.02669}{{\tt 1503.02669}}].

\bibitem{1503.04663}
K.~Murase, R.~Laha, S.~Ando and M.~Ahlers, \emph{{Testing the Dark Matter
  Scenario for PeV Neutrinos Observed in IceCube}},
  \href{http://dx.doi.org/10.1103/PhysRevLett.115.071301}{\emph{Phys. Rev.
  Lett.} {\bf 115} (2015) 071301}, [\href{http://arxiv.org/abs/1503.04663}{{\tt
  1503.04663}}].

\bibitem{1505.06486}
A.~Esmaili and P.~D. Serpico, \emph{{Gamma-ray bounds from EAS detectors and
  heavy decaying dark matter constraints}},
  \href{http://dx.doi.org/10.1088/1475-7516/2015/10/014}{\emph{JCAP} {\bf 1510}
  (2015) 014}, [\href{http://arxiv.org/abs/1505.06486}{{\tt 1505.06486}}].

\bibitem{1506.08788}
L.~A. Anchordoqui, V.~Barger, H.~Goldberg, X.~Huang, D.~Marfatia, L.~H.~M.
  da~Silva et~al., \emph{{IceCube neutrinos, decaying dark matter, and the
  Hubble constant}},
  \href{http://dx.doi.org/10.1103/PhysRevD.92.061301}{\emph{Phys. Rev.} {\bf
  D92} (2015) 061301}, [\href{http://arxiv.org/abs/1506.08788}{{\tt
  1506.08788}}].

\bibitem{Boucenna:2015tra}
S.~M. Boucenna, M.~Chianese, G.~Mangano, G.~Miele, S.~Morisi, O.~Pisanti
  et~al., \emph{{Decaying Leptophilic Dark Matter at IceCube}},
  \href{http://dx.doi.org/10.1088/1475-7516/2015/12/055}{\emph{JCAP} {\bf 1512}
  (2015) 055}, [\href{http://arxiv.org/abs/1507.01000}{{\tt 1507.01000}}].

\bibitem{1508.02500}
P.~Ko and Y.~Tang, \emph{{IceCube Events from Heavy DM decays through the
  Right-handed Neutrino Portal}},
  \href{http://dx.doi.org/10.1016/j.physletb.2015.10.021}{\emph{Phys. Lett.}
  {\bf B751} (2015) 81--88}, [\href{http://arxiv.org/abs/1508.02500}{{\tt
  1508.02500}}].

\bibitem{1601.02934}
M.~Chianese, G.~Miele, S.~Morisi and E.~Vitagliano, \emph{{Low energy IceCube
  data and a possible Dark Matter related excess}},
  \href{http://dx.doi.org/10.1016/j.physletb.2016.03.084}{\emph{Phys. Lett.}
  {\bf B757} (2016) 251--256}, [\href{http://arxiv.org/abs/1601.02934}{{\tt
  1601.02934}}].

\bibitem{1606.04517}
P.~S.~B. Dev, D.~Kazanas, R.~N. Mohapatra, V.~L. Teplitz and Y.~Zhang,
  \emph{{Heavy right-handed neutrino dark matter and PeV neutrinos at
  IceCube}}, \href{http://dx.doi.org/10.1088/1475-7516/2016/08/034}{\emph{JCAP}
  {\bf 1608} (2016) 034}, [\href{http://arxiv.org/abs/1606.04517}{{\tt
  1606.04517}}].

\bibitem{1607.05283}
M.~Chianese and A.~Merle, \emph{{A Consistent Theory of Decaying Dark Matter
  Connecting IceCube to the Sesame Street}},
  \href{http://arxiv.org/abs/1607.05283}{{\tt 1607.05283}}.

\bibitem{1305.6907}
V.~Barger and W.-Y. Keung, \emph{{Superheavy Particle Origin of IceCube PeV
  Neutrino Events}},
  \href{http://dx.doi.org/10.1016/j.physletb.2013.10.021}{\emph{Phys. Lett.}
  {\bf B727} (2013) 190--193}, [\href{http://arxiv.org/abs/1305.6907}{{\tt
  1305.6907}}].

\bibitem{1402.1681}
A.~N. Akay, U.~Kaya and S.~Sultansoy, \emph{{Color octet neutrino as the source
  of the IceCube PeV energy neutrino events}},
  \href{http://arxiv.org/abs/1402.1681}{{\tt 1402.1681}}.

\bibitem{1402.6678}
I.~Alikhanov, \emph{{The Glashow resonance in neutrino–photon scattering}},
  \href{http://dx.doi.org/10.1016/j.physletb.2014.12.056}{\emph{Phys. Lett.}
  {\bf B741} (2015) 295--300}, [\href{http://arxiv.org/abs/1402.6678}{{\tt
  1402.6678}}].

\bibitem{1404.0622}
L.~A. Anchordoqui, V.~Barger, H.~Goldberg, J.~G. Learned, D.~Marfatia,
  S.~Pakvasa et~al., \emph{{End of the cosmic neutrino energy spectrum}},
  \href{http://dx.doi.org/10.1016/j.physletb.2014.10.037}{\emph{Phys. Lett.}
  {\bf B739} (2014) 99--101}, [\href{http://arxiv.org/abs/1404.0622}{{\tt
  1404.0622}}].

\bibitem{1404.2279}
K.~Ioka and K.~Murase, \emph{{IceCube PeV–EeV neutrinos and secret
  interactions of neutrinos}},
  \href{http://dx.doi.org/10.1093/ptep/ptu090}{\emph{PTEP} {\bf 2014} (2014)
  061E01}, [\href{http://arxiv.org/abs/1404.2279}{{\tt 1404.2279}}].

\bibitem{1404.2288}
K.~C.~Y. Ng and J.~F. Beacom, \emph{{Cosmic neutrino cascades from secret
  neutrino interactions}}, \href{http://dx.doi.org/10.1103/PhysRevD.90.065035,
  10.1103/PhysRevD.90.089904}{\emph{Phys. Rev.} {\bf D90} (2014) 065035},
  [\href{http://arxiv.org/abs/1404.2288}{{\tt 1404.2288}}].

\bibitem{1404.2932}
J.~Zavala, \emph{{Galactic PeV neutrinos from dark matter annihilation}},
  \href{http://dx.doi.org/10.1103/PhysRevD.89.123516}{\emph{Phys. Rev.} {\bf
  D89} (2014) 123516}, [\href{http://arxiv.org/abs/1404.2932}{{\tt
  1404.2932}}].

\bibitem{1404.7025}
F.~W. Stecker and S.~T. Scully, \emph{{Propagation of Superluminal PeV IceCube
  Neutrinos: A High Energy Spectral Cutoff or New Constraints on Lorentz
  Invariance Violation}},
  \href{http://dx.doi.org/10.1103/PhysRevD.90.043012}{\emph{Phys. Rev.} {\bf
  D90} (2014) 043012}, [\href{http://arxiv.org/abs/1404.7025}{{\tt
  1404.7025}}].

\bibitem{1407.2848}
M.~Ibe and K.~Kaneta, \emph{{Cosmic neutrino background absorption line in the
  neutrino spectrum at IceCube}},
  \href{http://dx.doi.org/10.1103/PhysRevD.90.053011}{\emph{Phys. Rev.} {\bf
  D90} (2014) 053011}, [\href{http://arxiv.org/abs/1407.2848}{{\tt
  1407.2848}}].

\bibitem{1409.4180}
T.~Araki, F.~Kaneko, Y.~Konishi, T.~Ota, J.~Sato and T.~Shimomura,
  \emph{{Cosmic neutrino spectrum and the muon anomalous magnetic moment in the
  gauged $L_\mu - L_\tau$ model}},
  \href{http://dx.doi.org/10.1103/PhysRevD.91.037301}{\emph{Phys. Rev.} {\bf
  D91} (2015) 037301}, [\href{http://arxiv.org/abs/1409.4180}{{\tt
  1409.4180}}].

\bibitem{1409.5896}
A.~N. Akay, O.~Cakir, Y.~O. Gunaydin, U.~Kaya, M.~Sahin and S.~Sultansoy,
  \emph{{New IceCube data and color octet neutrino interpretation of the PeV
  energy events}},
  \href{http://dx.doi.org/10.1142/S0217751X15501638}{\emph{Int. J. Mod. Phys.}
  {\bf A30} (2015) 1550163}, [\href{http://arxiv.org/abs/1409.5896}{{\tt
  1409.5896}}].

\bibitem{1410.0408}
E.~Aeikens, H.~Päs, S.~Pakvasa and P.~Sicking, \emph{{Flavor ratios of
  extragalactic neutrinos and neutrino shortcuts in extra dimensions}},
  \href{http://dx.doi.org/10.1088/1475-7516/2015/10/005}{\emph{JCAP} {\bf 1510}
  (2015) 005}, [\href{http://arxiv.org/abs/1410.0408}{{\tt 1410.0408}}].

\bibitem{1410.3208}
J.~I. Illana, M.~Masip and D.~Meloni, \emph{{A new physics interpretation of
  the IceCube data}},
  \href{http://dx.doi.org/10.1016/j.astropartphys.2014.12.004}{\emph{Astropart.
  Phys.} {\bf 65} (2015) 64--68}, [\href{http://arxiv.org/abs/1410.3208}{{\tt
  1410.3208}}].

\bibitem{1411.5318}
C.~S. Fong, H.~Minakata, B.~Panes and R.~Zukanovich~Funchal, \emph{{Possible
  Interpretations of IceCube High-Energy Neutrino Events}},
  \href{http://dx.doi.org/10.1007/JHEP02(2015)189}{\emph{JHEP} {\bf 02} (2015)
  189}, [\href{http://arxiv.org/abs/1411.5318}{{\tt 1411.5318}}].

\bibitem{1411.5889}
F.~W. Stecker, S.~T. Scully, S.~Liberati and D.~Mattingly, \emph{{Searching for
  Traces of Planck-Scale Physics with High Energy Neutrinos}},
  \href{http://dx.doi.org/10.1103/PhysRevD.91.045009}{\emph{Phys. Rev.} {\bf
  D91} (2015) 045009}, [\href{http://arxiv.org/abs/1411.5889}{{\tt
  1411.5889}}].

\bibitem{1507.03015}
A.~DiFranzo and D.~Hooper, \emph{{Searching for MeV-Scale Gauge Bosons with
  IceCube}}, \href{http://dx.doi.org/10.1103/PhysRevD.92.095007}{\emph{Phys.
  Rev.} {\bf D92} (2015) 095007}, [\href{http://arxiv.org/abs/1507.03015}{{\tt
  1507.03015}}].

\bibitem{1507.03193}
G.~Tomar, S.~Mohanty and S.~Pakvasa, \emph{{Lorentz Invariance Violation and
  IceCube Neutrino Events}},
  \href{http://dx.doi.org/10.1007/JHEP11(2015)022}{\emph{JHEP} {\bf 11} (2015)
  022}, [\href{http://arxiv.org/abs/1507.03193}{{\tt 1507.03193}}].

\bibitem{1606.06238}
P.~Di~Bari, P.~O. Ludl and S.~Palomares-Ruiz, \emph{{Unifying leptogenesis,
  dark matter and high-energy neutrinos with right-handed neutrino mixing via
  Higgs portal}},  \href{http://arxiv.org/abs/1606.06238}{{\tt 1606.06238}}.

\bibitem{1606.07903}
U.~K. Dey, S.~Mohanty and G.~Tomar, \emph{{Leptoquarks: 750 GeV Diphoton
  Resonance and IceCube Events}},  \href{http://arxiv.org/abs/1606.07903}{{\tt
  1606.07903}}.

\bibitem{Dev:2016uxj}
P.~S.~B. Dev, D.~K. Ghosh and W.~Rodejohann, \emph{{R-parity Violating
  Supersymmetry at IceCube}},
  \href{http://dx.doi.org/10.1016/j.physletb.2016.08.066}{\emph{Phys. Lett.}
  {\bf B762} (2016) 116--123}, [\href{http://arxiv.org/abs/1605.09743}{{\tt
  1605.09743}}].

\bibitem{Fermi:1949ee}
E.~Fermi, \emph{{On the Origin of the Cosmic Radiation}},
  \href{http://dx.doi.org/10.1103/PhysRev.75.1169}{\emph{Phys. Rev.} {\bf 75}
  (1949) 1169--1174}.

\bibitem{2005PhRvL..95r1101K}
T.~{Kashti} and E.~{Waxman}, \emph{{Astrophysical Neutrinos: Flavor Ratios
  Depend on Energy}},
  \href{http://dx.doi.org/10.1103/PhysRevLett.95.181101}{\emph{Physical Review
  Letters} {\bf 95} (Oct., 2005) 181101},
  [\href{http://arxiv.org/abs/astro-ph/0507599}{{\tt astro-ph/0507599}}].

\bibitem{hep-ph/9405296}
J.~G. Learned and S.~Pakvasa, \emph{{Detecting tau-neutrino oscillations at PeV
  energies}},
  \href{http://dx.doi.org/10.1016/0927-6505(94)00043-3}{\emph{Astropart. Phys.}
  {\bf 3} (1995) 267--274}, [\href{http://arxiv.org/abs/hep-ph/9405296}{{\tt
  hep-ph/9405296}}].

\bibitem{Bhattacharya:2011qu}
A.~Bhattacharya, R.~Gandhi, W.~Rodejohann and A.~Watanabe, \emph{{The Glashow
  resonance at IceCube: signatures, event rates and $pp$ vs. $p\gamma$
  interactions}},
  \href{http://dx.doi.org/10.1088/1475-7516/2011/10/017}{\emph{JCAP} {\bf 1110}
  (2011) 017}, [\href{http://arxiv.org/abs/1108.3163}{{\tt 1108.3163}}].

\bibitem{PhysRev.118.316}
S.~L. Glashow, \emph{Resonant scattering of antineutrinos},
  \href{http://dx.doi.org/10.1103/PhysRev.118.316}{\emph{Phys. Rev.} {\bf 118}
  (Apr, 1960) 316--317}.

\bibitem{Berezinsky:1981bt}
V.~S. Berezinsky and A.~Z. Gazizov, \emph{{Neutrino - electron scattering at
  energies above the W boson production threshold}}, {\emph{Sov. J. Nucl.
  Phys.} {\bf 33} (1981) 120--125}.

\bibitem{Gandhi:1995tf}
R.~Gandhi, C.~Quigg, M.~H. Reno and I.~Sarcevic, \emph{{Ultrahigh-energy
  neutrino interactions}},
  \href{http://dx.doi.org/10.1016/0927-6505(96)00008-4}{\emph{Astropart. Phys.}
  {\bf 5} (1996) 81--110}, [\href{http://arxiv.org/abs/hep-ph/9512364}{{\tt
  hep-ph/9512364}}].

\bibitem{0505017}
H.~Athar, C.~S. Kim and J.~Lee, \emph{{The Intrinsic and oscillated
  astrophysical neutrino flavor ratios}},
  \href{http://dx.doi.org/10.1142/S021773230602038X}{\emph{Mod. Phys. Lett.}
  {\bf A21} (2006) 1049--1066},
  [\href{http://arxiv.org/abs/hep-ph/0505017}{{\tt hep-ph/0505017}}].

\bibitem{Beacom:2004jb}
J.~F. Beacom and J.~Candia, \emph{{Shower power: Isolating the prompt
  atmospheric neutrino flux using electron neutrinos}},
  \href{http://dx.doi.org/10.1088/1475-7516/2004/11/009}{\emph{JCAP} {\bf 0411}
  (2004) 009}, [\href{http://arxiv.org/abs/hep-ph/0409046}{{\tt
  hep-ph/0409046}}].

\bibitem{Glashow:1960zz}
S.~L. Glashow, \emph{{Resonant Scattering of Antineutrinos}},
  \href{http://dx.doi.org/10.1103/PhysRev.118.316}{\emph{Phys. Rev.} {\bf 118}
  (1960) 316--317}.

\bibitem{Barger:2014iua}
V.~Barger, L.~Fu, J.~G. Learned, D.~Marfatia, S.~Pakvasa and T.~J. Weiler,
  \emph{{Glashow resonance as a window into cosmic neutrino sources}},
  \href{http://dx.doi.org/10.1103/PhysRevD.90.121301}{\emph{Phys. Rev.} {\bf
  D90} (2014) 121301}, [\href{http://arxiv.org/abs/1407.3255}{{\tt
  1407.3255}}].

\bibitem{halzen_pheno16}
F.~Halzen, ``Particle physics beyond laboratory energies.'' Phenomenology 2016
  Symposium, University of Pittsburgh, USA, 2016.

\bibitem{1309.2756}
S.~Razzaque, \emph{{The Galactic Center Origin of a Subset of IceCube Neutrino
  Events}}, \href{http://dx.doi.org/10.1103/PhysRevD.88.081302}{\emph{Phys.
  Rev.} {\bf D88} (2013) 081302}, [\href{http://arxiv.org/abs/1309.2756}{{\tt
  1309.2756}}].

\bibitem{1309.4077}
M.~Ahlers and K.~Murase, \emph{{Probing the Galactic Origin of the IceCube
  Excess with Gamma-Rays}},
  \href{http://dx.doi.org/10.1103/PhysRevD.90.023010}{\emph{Phys. Rev.} {\bf
  D90} (2014) 023010}, [\href{http://arxiv.org/abs/1309.4077}{{\tt
  1309.4077}}].

\bibitem{1311.5864}
Y.~Bai, R.~Lu and J.~Salvado, \emph{{Geometric Compatibility of IceCube TeV-PeV
  Neutrino Excess and its Galactic Dark Matter Origin}},
  \href{http://dx.doi.org/10.1007/JHEP01(2016)161}{\emph{JHEP} {\bf 01} (2016)
  161}, [\href{http://arxiv.org/abs/1311.5864}{{\tt 1311.5864}}].

\bibitem{1311.7188}
C.~Lunardini, S.~Razzaque, K.~T. Theodoseau and L.~Yang, \emph{{Neutrino Events
  at IceCube and the Fermi Bubbles}},
  \href{http://dx.doi.org/10.1103/PhysRevD.90.023016}{\emph{Phys. Rev.} {\bf
  D90} (2014) 023016}, [\href{http://arxiv.org/abs/1311.7188}{{\tt
  1311.7188}}].

\bibitem{1406.0376}
P.~Padovani and E.~Resconi, \emph{{Are both BL Lacs and pulsar wind nebulae the
  astrophysical counterparts of IceCube neutrino events?}},
  \href{http://dx.doi.org/10.1093/mnras/stu1166}{\emph{Mon. Not. Roy. Astron.
  Soc.} {\bf 443} (2014) 474--484}, [\href{http://arxiv.org/abs/1406.0376}{{\tt
  1406.0376}}].

\bibitem{1406.2160}
M.~Ahlers and F.~Halzen, \emph{{Pinpointing Extragalactic Neutrino Sources in
  Light of Recent IceCube Observations}},
  \href{http://dx.doi.org/10.1103/PhysRevD.90.043005}{\emph{Phys. Rev.} {\bf
  D90} (2014) 043005}, [\href{http://arxiv.org/abs/1406.2160}{{\tt
  1406.2160}}].

\bibitem{1407.2243}
Y.~Bai, A.~J. Barger, V.~Barger, R.~Lu, A.~D. Peterson and J.~Salvado,
  \emph{{Neutrino Lighthouse at Sagittarius A*}},
  \href{http://dx.doi.org/10.1103/PhysRevD.90.063012}{\emph{Phys. Rev.} {\bf
  D90} (2014) 063012}, [\href{http://arxiv.org/abs/1407.2243}{{\tt
  1407.2243}}].

\bibitem{1501.05158}
R.~Moharana and S.~Razzaque, \emph{{Angular correlation of cosmic neutrinos
  with ultrahigh-energy cosmic rays and implications for their sources}},
  \href{http://dx.doi.org/10.1088/1475-7516/2015/08/014}{\emph{JCAP} {\bf 1508}
  (2015) 014}, [\href{http://arxiv.org/abs/1501.05158}{{\tt 1501.05158}}].

\bibitem{1507.05711}
K.~Emig, C.~Lunardini and R.~Windhorst, \emph{{Do high energy astrophysical
  neutrinos trace star formation?}},
  \href{http://dx.doi.org/10.1088/1475-7516/2015/12/029}{\emph{JCAP} {\bf 1512}
  (2015) 029}, [\href{http://arxiv.org/abs/1507.05711}{{\tt 1507.05711}}].

\bibitem{1509.00517}
{\scshape IceCube, VERITAS} collaboration, M.~Santander, \emph{{Searching for
  TeV gamma-ray emission associated with IceCube high-energy neutrinos using
  VERITAS}}, {\emph{PoS} {\bf ICRC2015} (2016) 785},
  [\href{http://arxiv.org/abs/1509.00517}{{\tt 1509.00517}}].

\bibitem{1509.03522}
A.~Neronov and D.~V. Semikoz, \emph{{Evidence the Galactic contribution to the
  IceCube astrophysical neutrino flux}},
  \href{http://dx.doi.org/10.1016/j.astropartphys.2015.11.002}{\emph{Astropart.
  Phys.} {\bf 75} (2016) 60--63}, [\href{http://arxiv.org/abs/1509.03522}{{\tt
  1509.03522}}].

\bibitem{1510.00048}
L.~S. Miranda, A.~R. de~León and S.~Sahu, \emph{{Blazar origin of some IceCube
  events}}, \href{http://dx.doi.org/10.1140/epjc/s10052-016-4247-x}{\emph{Eur.
  Phys. J.} {\bf C76} (2016) 402}, [\href{http://arxiv.org/abs/1510.00048}{{\tt
  1510.00048}}].

\bibitem{1511.09408}
{\scshape IceCube, Pierre Auger, Telescope Array} collaboration, M.~G. Aartsen
  et~al., \emph{{Search for correlations between the arrival directions of
  IceCube neutrino events and ultrahigh-energy cosmic rays detected by the
  Pierre Auger Observatory and the Telescope Array}},
  \href{http://dx.doi.org/10.1088/1475-7516/2016/01/037}{\emph{JCAP} {\bf 1601}
  (2016) 037}, [\href{http://arxiv.org/abs/1511.09408}{{\tt 1511.09408}}].

\bibitem{1603.06733}
A.~Neronov and D.~Semikoz, \emph{{Galactic and extragalactic contributions to
  the astrophysical muon neutrino signal}},
  \href{http://dx.doi.org/10.1103/PhysRevD.93.123002}{\emph{Phys. Rev.} {\bf
  D93} (2016) 123002}, [\href{http://arxiv.org/abs/1603.06733}{{\tt
  1603.06733}}].

\bibitem{1611.07905}
L.~A. Anchordoqui, M.~M. Block, L.~Durand, P.~Ha, J.~F. Soriano and T.~J.
  Weiler, \emph{{Evidence for a break in the spectrum of astrophysical
  neutrinos}},  \href{http://arxiv.org/abs/1611.07905}{{\tt 1611.07905}}.

\bibitem{1410.1749}
{\scshape IceCube} collaboration, M.~G. Aartsen et~al., \emph{{Atmospheric and
  astrophysical neutrinos above 1 TeV interacting in IceCube}},
  \href{http://dx.doi.org/10.1103/PhysRevD.91.022001}{\emph{Phys. Rev.} {\bf
  D91} (2015) 022001}, [\href{http://arxiv.org/abs/1410.1749}{{\tt
  1410.1749}}].

\bibitem{MESE}
M.~Chianese, G.~Miele and S.~Morisi, \emph{{Dark Matter interpretation of low
  energy IceCube MESE excess}},  \href{http://arxiv.org/abs/1610.04612}{{\tt
  1610.04612}}.

\bibitem{Waxman:1998yy}
E.~Waxman and J.~N. Bahcall, \emph{{High-energy neutrinos from astrophysical
  sources: An Upper bound}},
  \href{http://dx.doi.org/10.1103/PhysRevD.59.023002}{\emph{Phys. Rev.} {\bf
  D59} (1999) 023002}, [\href{http://arxiv.org/abs/hep-ph/9807282}{{\tt
  hep-ph/9807282}}].

\bibitem{Bahcall:1999yr}
J.~N. Bahcall and E.~Waxman, \emph{{High-energy astrophysical neutrinos: The
  Upper bound is robust}},
  \href{http://dx.doi.org/10.1103/PhysRevD.64.023002}{\emph{Phys. Rev.} {\bf
  D64} (2001) 023002}, [\href{http://arxiv.org/abs/hep-ph/9902383}{{\tt
  hep-ph/9902383}}].

\bibitem{1507.01000}
S.~M. Boucenna, M.~Chianese, G.~Mangano, G.~Miele, S.~Morisi, O.~Pisanti
  et~al., \emph{{Decaying Leptophilic Dark Matter at IceCube}},
  \href{http://dx.doi.org/10.1088/1475-7516/2015/12/055}{\emph{JCAP} {\bf 1512}
  (2015) 055}, [\href{http://arxiv.org/abs/1507.01000}{{\tt 1507.01000}}].

\bibitem{1402.2846}
K.~Harigaya, M.~Kawasaki, K.~Mukaida and M.~Yamada, \emph{{Dark Matter
  Production in Late Time Reheating}},
  \href{http://dx.doi.org/10.1103/PhysRevD.89.083532}{\emph{Phys. Rev.} {\bf
  D89} (2014) 083532}, [\href{http://arxiv.org/abs/1402.2846}{{\tt
  1402.2846}}].

\bibitem{1201.3696}
Y.~Kurata and N.~Maekawa, \emph{{Averaged Number of the Lightest Supersymmetric
  Particles in Decay of Superheavy Particle with Long Lifetime}},
  \href{http://dx.doi.org/10.1143/PTP.127.657}{\emph{Prog. Theor. Phys.} {\bf
  127} (2012) 657--664}, [\href{http://arxiv.org/abs/1201.3696}{{\tt
  1201.3696}}].

\bibitem{hep-ph/0203118}
R.~Allahverdi and M.~Drees, \emph{{Production of massive stable particles in
  inflaton decay}},
  \href{http://dx.doi.org/10.1103/PhysRevLett.89.091302}{\emph{Phys. Rev.
  Lett.} {\bf 89} (2002) 091302},
  [\href{http://arxiv.org/abs/hep-ph/0203118}{{\tt hep-ph/0203118}}].

\bibitem{hep-ph/0205246}
R.~Allahverdi and M.~Drees, \emph{{Thermalization after inflation and
  production of massive stable particles}},
  \href{http://dx.doi.org/10.1103/PhysRevD.66.063513}{\emph{Phys. Rev.} {\bf
  D66} (2002) 063513}, [\href{http://arxiv.org/abs/hep-ph/0205246}{{\tt
  hep-ph/0205246}}].

\bibitem{PhysRevLett.64.615}
K.~Griest and M.~Kamionkowski, \emph{{Unitarity Limits on the Mass and Radius
  of Dark Matter Particles}},
  \href{http://dx.doi.org/10.1103/PhysRevLett.64.615}{\emph{Phys. Rev. Lett.}
  {\bf 64} (1990) 615}.

\bibitem{1405.7370}
K.~Agashe, Y.~Cui, L.~Necib and J.~Thaler, \emph{{(In)direct Detection of
  Boosted Dark Matter}},
  \href{http://dx.doi.org/10.1088/1475-7516/2014/10/062}{\emph{JCAP} {\bf 1410}
  (2014) 062}, [\href{http://arxiv.org/abs/1405.7370}{{\tt 1405.7370}}].

\bibitem{1410.2246}
J.~Berger, Y.~Cui and Y.~Zhao, \emph{{Detecting Boosted Dark Matter from the
  Sun with Large Volume Neutrino Detectors}},
  \href{http://dx.doi.org/10.1088/1475-7516/2015/02/005}{\emph{JCAP} {\bf 1502}
  (2015) 005}, [\href{http://arxiv.org/abs/1410.2246}{{\tt 1410.2246}}].

\bibitem{1411.6632}
K.~Kong, G.~Mohlabeng and J.-C. Park, \emph{{Boosted dark matter signals
  uplifted with self-interaction}},
  \href{http://dx.doi.org/10.1016/j.physletb.2015.02.057}{\emph{Phys. Lett.}
  {\bf B743} (2015) 256--266}, [\href{http://arxiv.org/abs/1411.6632}{{\tt
  1411.6632}}].

\bibitem{1611.09866}
H.~Alhazmi, K.~Kong, G.~Mohlabeng and J.-C. Park, \emph{{Boosted Dark Matter at
  the Deep Underground Neutrino Experiment}},
  \href{http://arxiv.org/abs/1611.09866}{{\tt 1611.09866}}.

\bibitem{1607.08006}
{\scshape IceCube} collaboration, M.~G. Aartsen et~al., \emph{{Observation and
  Characterization of a Cosmic Muon Neutrino Flux from the Northern Hemisphere
  using six years of IceCube data}},
  \href{http://arxiv.org/abs/1607.08006}{{\tt 1607.08006}}.

\bibitem{TheFermi-LAT:2015kwa}
{\scshape Fermi-LAT} collaboration, M.~Ajello et~al., \emph{{Fermi-LAT
  Observations of High-Energy $\gamma$-Ray Emission Toward the Galactic
  Center}},
  \href{http://dx.doi.org/10.3847/0004-637X/819/1/44}{\emph{Astrophys. J.} {\bf
  819} (2016) 44}, [\href{http://arxiv.org/abs/1511.02938}{{\tt 1511.02938}}].

\bibitem{Ackermann:2014usa}
{\scshape Fermi-LAT} collaboration, M.~Ackermann et~al., \emph{{The spectrum of
  isotropic diffuse gamma-ray emission between 100 MeV and 820 GeV}},
  \href{http://dx.doi.org/10.1088/0004-637X/799/1/86}{\emph{Astrophys. J.} {\bf
  799} (2015) 86}, [\href{http://arxiv.org/abs/1410.3696}{{\tt 1410.3696}}].

\bibitem{Feng:2015dye}
{\scshape KASCADE-Grande} collaboration, Z.~Feng, D.~Kang and A.~Haungs,
  \emph{{Limits on the isotropic diffuse $\gamma$-rays at ultra high energies
  measured with KASCADE}}, {\emph{PoS} {\bf ICRC2015} (2016) 823}.

\bibitem{Gupta:2009zz}
{\scshape AGRAPES-3} collaboration, S.~K. Gupta et~al., \emph{{The current
  status of the GRAPES-3 extensive air shower experiment}},
  \href{http://dx.doi.org/10.1016/j.nuclphysbps.2009.09.027}{\emph{Nucl. Phys.
  Proc. Suppl.} {\bf 196} (2009) 153--158}.

\bibitem{Ahlers:2013xia}
M.~Ahlers and K.~Murase, \emph{{Probing the Galactic Origin of the IceCube
  Excess with Gamma-Rays}},
  \href{http://dx.doi.org/10.1103/PhysRevD.90.023010}{\emph{Phys. Rev.} {\bf
  D90} (2014) 023010}, [\href{http://arxiv.org/abs/1309.4077}{{\tt
  1309.4077}}].

\bibitem{1612.05638}
T.~Cohen, K.~Murase, N.~L. Rodd, B.~R. Safdi and Y.~Soreq, \emph{{Gamma-ray
  Constraints on Decaying Dark Matter and Implications for IceCube}},
  \href{http://arxiv.org/abs/1612.05638}{{\tt 1612.05638}}.

\bibitem{Izaguirre:2014vva}
E.~Izaguirre, G.~Krnjaic and B.~Shuve, \emph{{The Galactic Center Excess from
  the Bottom Up}},
  \href{http://dx.doi.org/10.1103/PhysRevD.90.055002}{\emph{Phys. Rev.} {\bf
  D90} (2014) 055002}, [\href{http://arxiv.org/abs/1404.2018}{{\tt
  1404.2018}}].

\bibitem{D'Ambrosio:2002ex}
G.~D'Ambrosio, G.~F. Giudice, G.~Isidori and A.~Strumia, \emph{{Minimal flavor
  violation: An Effective field theory approach}},
  \href{http://dx.doi.org/10.1016/S0550-3213(02)00836-2}{\emph{Nucl. Phys.}
  {\bf B645} (2002) 155--187}, [\href{http://arxiv.org/abs/hep-ph/0207036}{{\tt
  hep-ph/0207036}}].

\bibitem{Kahlhoefer:2015bea}
F.~Kahlhoefer, K.~Schmidt-Hoberg, T.~Schwetz and S.~Vogl, \emph{{Implications
  of unitarity and gauge invariance for simplified dark matter models}},
  \href{http://dx.doi.org/10.1007/JHEP02(2016)016}{\emph{JHEP} {\bf 02} (2016)
  016}, [\href{http://arxiv.org/abs/1510.02110}{{\tt 1510.02110}}].

\bibitem{1502.01589}
{\scshape Planck} collaboration, P.~A.~R. Ade et~al., \emph{{Planck 2015
  results. XIII. Cosmological parameters}},
  \href{http://dx.doi.org/10.1051/0004-6361/201525830}{\emph{Astron.
  Astrophys.} {\bf 594} (2016) A13},
  [\href{http://arxiv.org/abs/1502.01589}{{\tt 1502.01589}}].

\bibitem{Buckley:2014fba}
M.~R. Buckley, D.~Feld and D.~Goncalves, \emph{{Scalar Simplified Models for
  Dark Matter}},
  \href{http://dx.doi.org/10.1103/PhysRevD.91.015017}{\emph{Phys. Rev.} {\bf
  D91} (2015) 015017}, [\href{http://arxiv.org/abs/1410.6497}{{\tt
  1410.6497}}].

\bibitem{Chala:2015ama}
M.~Chala, F.~Kahlhoefer, M.~McCullough, G.~Nardini and K.~Schmidt-Hoberg,
  \emph{{Constraining Dark Sectors with Monojets and Dijets}},
  \href{http://dx.doi.org/10.1007/JHEP07(2015)089}{\emph{JHEP} {\bf 07} (2015)
  089}, [\href{http://arxiv.org/abs/1503.05916}{{\tt 1503.05916}}].

\bibitem{ATLAS:2016lvi}
{\scshape ATLAS} collaboration, T.~A. collaboration, \emph{{Search for New
  Phenomena in Dijet Events with the ATLAS Detector at $\sqrt{s}$=13 TeV with
  2015 and 2016 data}}, .

\bibitem{Cirelli:2010xx}
M.~Cirelli, G.~Corcella, A.~Hektor, G.~Hutsi, M.~Kadastik, P.~Panci et~al.,
  \emph{{PPPC 4 DM ID: A Poor Particle Physicist Cookbook for Dark Matter
  Indirect Detection}}, \href{http://dx.doi.org/10.1088/1475-7516/2012/10/E01,
  10.1088/1475-7516/2011/03/051}{\emph{JCAP} {\bf 1103} (2011) 051},
  [\href{http://arxiv.org/abs/1012.4515}{{\tt 1012.4515}}].

\bibitem{Cirelli:2012ut}
M.~Cirelli, E.~Moulin, P.~Panci, P.~D. Serpico and A.~Viana, \emph{{Gamma ray
  constraints on Decaying Dark Matter}},
  \href{http://dx.doi.org/10.1103/PhysRevD.86.083506,
  10.1103/PhysRevD.86.109901}{\emph{Phys. Rev.} {\bf D86} (2012) 083506},
  [\href{http://arxiv.org/abs/1205.5283}{{\tt 1205.5283}}].

\bibitem{astro-ph/9508025}
J.~F. Navarro, C.~S. Frenk and S.~D.~M. White, \emph{{The Structure of cold
  dark matter halos}}, \href{http://dx.doi.org/10.1086/177173}{\emph{Astrophys.
  J.} {\bf 462} (1996) 563--575},
  [\href{http://arxiv.org/abs/astro-ph/9508025}{{\tt astro-ph/9508025}}].

\bibitem{1012.4515}
M.~Cirelli, G.~Corcella, A.~Hektor, G.~Hutsi, M.~Kadastik, P.~Panci et~al.,
  \emph{{PPPC 4 DM ID: A Poor Particle Physicist Cookbook for Dark Matter
  Indirect Detection}}, \href{http://dx.doi.org/10.1088/1475-7516/2012/10/E01,
  10.1088/1475-7516/2011/03/051}{\emph{JCAP} {\bf 1103} (2011) 051},
  [\href{http://arxiv.org/abs/1012.4515}{{\tt 1012.4515}}].

\bibitem{1207.6082}
A.~Belyaev, N.~D. Christensen and A.~Pukhov, \emph{{CalcHEP 3.4 for collider
  physics within and beyond the Standard Model}},
  \href{http://dx.doi.org/10.1016/j.cpc.2013.01.014}{\emph{Comput. Phys.
  Commun.} {\bf 184} (2013) 1729--1769},
  [\href{http://arxiv.org/abs/1207.6082}{{\tt 1207.6082}}].

\bibitem{1303.5076}
{\scshape Planck} collaboration, P.~A.~R. Ade et~al., \emph{{Planck 2013
  results. XVI. Cosmological parameters}},
  \href{http://dx.doi.org/10.1051/0004-6361/201321591}{\emph{Astron.
  Astrophys.} {\bf 571} (2014) A16},
  [\href{http://arxiv.org/abs/1303.5076}{{\tt 1303.5076}}].

\bibitem{1404.0017}
O.~Mena, S.~Palomares-Ruiz and A.~C. Vincent, \emph{{Flavor Composition of the
  High-Energy Neutrino Events in IceCube}},
  \href{http://dx.doi.org/10.1103/PhysRevLett.113.091103}{\emph{Phys. Rev.
  Lett.} {\bf 113} (2014) 091103}, [\href{http://arxiv.org/abs/1404.0017}{{\tt
  1404.0017}}].

\bibitem{Esmaili:2012us}
A.~Esmaili, A.~Ibarra and O.~L.~G. Peres, \emph{{Probing the stability of
  superheavy dark matter particles with high-energy neutrinos}},
  \href{http://dx.doi.org/10.1088/1475-7516/2012/11/034}{\emph{JCAP} {\bf 1211}
  (2012) 034}, [\href{http://arxiv.org/abs/1205.5281}{{\tt 1205.5281}}].

\bibitem{0710.3820}
T.~Sjostrand, S.~Mrenna and P.~Z. Skands, \emph{{A Brief Introduction to PYTHIA
  8.1}}, \href{http://dx.doi.org/10.1016/j.cpc.2008.01.036}{\emph{Comput. Phys.
  Commun.} {\bf 178} (2008) 852--867},
  [\href{http://arxiv.org/abs/0710.3820}{{\tt 0710.3820}}].

\bibitem{Lai:2010vv}
H.-L. Lai, M.~Guzzi, J.~Huston, Z.~Li, P.~M. Nadolsky, J.~Pumplin et~al.,
  \emph{{New parton distributions for collider physics}},
  \href{http://dx.doi.org/10.1103/PhysRevD.82.074024}{\emph{Phys. Rev.} {\bf
  D82} (2010) 074024}, [\href{http://arxiv.org/abs/1007.2241}{{\tt
  1007.2241}}].

\bibitem{Gandhi:1998ri}
R.~Gandhi, C.~Quigg, M.~H. Reno and I.~Sarcevic, \emph{{Neutrino interactions
  at ultrahigh-energies}},
  \href{http://dx.doi.org/10.1103/PhysRevD.58.093009}{\emph{Phys. Rev.} {\bf
  D58} (1998) 093009}, [\href{http://arxiv.org/abs/hep-ph/9807264}{{\tt
  hep-ph/9807264}}].

\bibitem{1106.3723}
A.~Cooper-Sarkar, P.~Mertsch and S.~Sarkar, \emph{{The high energy neutrino
  cross-section in the Standard Model and its uncertainty}},
  \href{http://dx.doi.org/10.1007/JHEP08(2011)042}{\emph{JHEP} {\bf 08} (2011)
  042}, [\href{http://arxiv.org/abs/1106.3723}{{\tt 1106.3723}}].

\bibitem{1605.08781}
M.~D. Kistler and R.~Laha, \emph{{Multi-PeV Signals from a New Astrophysical
  Neutrino Flux Beyond the Glashow Resonance}},
  \href{http://arxiv.org/abs/1605.08781}{{\tt 1605.08781}}.

\bibitem{Ishihara}
A.~Ishihara and I.~Collaboration, \emph{Extremely high energy neutrinos in six
  years of icecube data}, {\emph{Journal of Physics: Conference Series} {\bf
  718} (2016) 062027}.

\end{thebibliography}\endgroup

\end{document}